\documentclass[acmtog,authorversion]{acmart}

\usepackage{booktabs} 

\setcitestyle{square}

\usepackage[ruled]{algorithm2e} 

\SetAlFnt{\small}
\SetAlCapFnt{\small}
\SetAlCapNameFnt{\small}
\SetAlCapHSkip{0pt}
\IncMargin{-\parindent}

\acmJournal{TOG}

\acmSubmissionID{317}

\setcopyright{acmlicensed}

\acmYear{2021}\acmVolume{40}\acmNumber{4}\acmArticle{36}\acmMonth{8} \acmDOI{10.1145/3450626.3459812}

\gdef\cropInsets{0}

\usepackage{animate}

\usepackage{booktabs} 
\usepackage{tabularx}
\newcolumntype{Y}{>{\centering\arraybackslash}X}

\usepackage{xargs}
\usepackage{mathtools}
\usepackage{amsmath}
\usepackage{microtype}
\usepackage{units}
\usepackage[binary-units]{siunitx}
\usepackage{subcaption}
\usepackage{overpic}
\usepackage{cancel}
\usepackage{wrapfig}
\usepackage{placeins}
\usepackage{empheq}
\usepackage{xspace}

\usepackage{multirow}

\usepackage{url} 
\usepackage[utf8]{inputenc}

\def\commentType{0}

\ifnum\commentType=0
    \newcommandx{\customComment}[3]{}
    \newcommandx{\customTODO}[3]{}
\fi
\ifnum\commentType=1
    \newcommandx{\customComment}[3]{\textcolor{#2}{\textsl{#1: #3}}}
    \newcommandx{\customTODO}[3]{\textcolor{#2}{\textsl{#1: #3}}}
\fi
\ifnum\commentType=2
    \usepackage{pdfcomment}
    \newcommandx{\customComment}[3]{\pdfcomment[icon=Comment,opacity=0.5,color=#2,author=#1]{#3}}
    \newcommandx{\customTODO}[3]{\pdfcomment[icon=Note,opacity=0.5,color=#2,author=#1]{#3}}
\fi
\ifnum\commentType=3
    \usepackage{pdfcomment} 
    \usepackage{todonotes} 
    \usepackage{silence}
    \newcommandx{\customComment}[3]{\todo[color=#2!40,size=\small]{\textbf{#1:} #3}}
    \newcommandx{\customTODO}[3]{\todo[color=#2!40,size=\small]{\textbf{#1:} #3}}
\fi

\let\originalleft\left 
\let\originalright\right 
\renewcommand{\left}{\mathopen{}\mathclose\bgroup\originalleft} 
\renewcommand{\right}{\aftergroup\egroup\originalright} 

\definecolor{amber}{rgb}{1.0, 0.49, 0.0}
\definecolor{darkgreen}{rgb}{0.0, 0.5, 0.0}
\newcommandx{\All}[1]{\customComment{All}{red}{#1}}
\newcommandx{\Jan}[1]{\customComment{Jan}{magenta}{#1}}
\newcommandx{\Alex}[1]{\customComment{Alex}{blue}{#1}}
\newcommandx{\Fabrice}[1]{\customComment{Fabrice}{amber}{#1}}
\newcommandx{\Thomas}[1]{\customComment{Thomas}{darkgreen}{#1}}

\newcommandx{\TODO}[1]{\customTODO{TODO}{red}{#1}}
\newcommandx{\JanTODO}[1]{\customTODO{Jan}{magenta}{#1}}
\newcommandx{\FabriceTODO}[1]{\customTODO{Fabrice}{amber}{#1}}
\newcommandx{\ThomasTODO}[1]{\customTODO{Thomas}{darkgreen}{#1}}

\newcommand{\IGNORE}[1]{}

\newcommand{\REMOVE}[1]{} 

\def\equationautorefname~#1\null{%
  Equation~(#1)\null
}

\newcommand{\Sphere}{{S^2}}

\newcommand{\Diff}[1]{\,\mathrm{d}#1}

\newcommand{\PdfMC}{p}

\newcommand{\OneBlob}{\text{ob}}
\newcommand{\PosEnc}{\text{freq}}
\newcommand{\Spherical}{\text{sph}}
\newcommand{\Identity}{\text{id}}
\newcommand{\StopGradient}{\text{sg}}

\newcommand{\Params}{W}

\newcommand{\R}{\mathbb{R}}

\newcommand{\nlayers}{L}
\newcommand{\width}{M}
\newcommand{\batchSize}{N}
\newcommand{\inputWidth}{\width_\mathrm{in}}
\newcommand{\networkWidth}{\width_\mathrm{hidden}}
\newcommand{\outputWidth}{\width_\mathrm{out}}

\newcommand{\pos}{\mathbf{x}}
\newcommand{\posy}{\mathbf{y}}
\newcommand{\diro}{\omega}
\newcommand{\diri}{\omega_\mathrm{i}}

\newcommand{\normal}{\mathbf{n}}

\newcommand{\radiance}{L}
\newcommand{\inRadiance}{\radiance_{\mathrm{i}}}

\newcommand{\reflectedRadiance}{\radiance_{\mathrm{s}}}

\newcommand{\cachedReflectedRadiance}{\widehat{\radiance}_{\mathrm{s}}}

\newcommand{\bsdf}{f_{\mathrm{s}}}

\newcommand{\roughness}{r}
\newcommand{\albedo}{\alpha}
\newcommand{\specAlbedo}{\beta}

\newcommand{\fsangle}{\theta}
\newcommand{\fsanglei}{\fsangle_\mathrm{i}}

\newcommand{\AreaHeuristic}{a}

\newcommand{\Loss}{\mathcal{L}}

\newcommand{\ZeroDay}{\textsc{Zero Day}\xspace}
\newcommand{\Classroom}{\textsc{Classroom}\xspace}
\newcommand{\BistroExterior}{\textsc{Bistro}\xspace}
\newcommand{\Attic}{\textsc{Attic}\xspace}

\newcommand{\LivingRoom}{\textsc{Living Room}\xspace}
\newcommand{\PinkRoom}{\textsc{Pink Room}\xspace}

\newcommand{\FLIP}{\reflectbox{F}LIP\xspace}

\gdef\useCroppedImages{1}

\usepackage[export]{adjustbox}
\usepackage{calc}
\usepackage{tabularx}
\usepackage{tikz}
\usepackage{xstring}

\newlength{\beautyHeight}
\newlength{\beautyPixWidth}
\newlength{\beautyPixHeight}
\newlength{\insetvsep}
\gdef\useInsetA{0}
\gdef\useInsetB{0}
\gdef\useInsetC{0}

\newcommand{\setInset}[6]{%
    \expandafter\gdef\csname useInset#1\endcsname{1}%
    \expandafter\gdef\csname inset#1Color\endcsname{#2}%
    \expandafter\gdef\csname crop#1X\endcsname{#3}%
    \expandafter\gdef\csname crop#1Y\endcsname{#4}%
    \expandafter\gdef\csname crop#1W\endcsname{#5}%
    \expandafter\gdef\csname crop#1H\endcsname{#6}%
}

\newcommand{\unsetInset}[1]{%
    \expandafter\gdef\csname useInset#1\endcsname{0}%
}

\newcommand{\addBeautyCrop}[8]{%
    \pdfpxdimen=\dimexpr 1 in/72\relax
    \def\beauty{%
        \let\cropR\relax%
        \let\cropB\relax%
        \newlength\cropR%
        \newlength\cropB%
        \setlength\cropR{{#3 px}-{#5 px}-{#7 px}}%
        \setlength\cropB{{#4 px}-{#6 px}-{#8 px}}%
        \sbox0{\includegraphics[width=#2\textwidth,trim={#5px {\cropB} {\cropR} #6px},clip]{#1}}%
        \begin{tikzpicture}
            \node[anchor=north west,inner sep=0] at (0,0) {\usebox0};
            \begin{scope}[x=\wd0/#7, y=\ht0/#8]
            \if\useInsetA1{
                \draw[\insetAColor,thick] (\cropAX-#5,-\cropAY+#6) rectangle + (\cropAW,-\cropAH);
            }\fi
            \if\useInsetB1{
                \draw[\insetBColor,thick] (\cropBX-#5,-\cropBY+#6) rectangle + (\cropBW,-\cropBH);
            }\fi
            \if\useInsetC1{
                \draw[\insetCColor,thick] (\cropCX-#5,-\cropCY+#6) rectangle + (\cropCW,-\cropCH);
            }\fi
            \end{scope}
        \end{tikzpicture}
    }%
    \setlength\beautyHeight{\heightof{\beauty}}%
    \setlength\beautyPixWidth{#3px}%
    \setlength\beautyPixHeight{#4px}%
    \global\beautyHeight=\beautyHeight%
    \global\beautyPixWidth=\beautyPixWidth%
    \global\beautyPixHeight=\beautyPixHeight%
    \begin{adjustbox}{valign=t}
        \beauty{}
    \end{adjustbox}
}

\newcommand{\trimInset}[6]{%
    \let\cropR\relax%
    \let\cropB\relax%
    \newlength\cropR%
    \newlength\cropB%
    \setlength\cropR{{\beautyPixWidth}-{#3 px}-{#5 px}}%
    \setlength\cropB{{\beautyPixHeight}-{#4 px}-{#6 px}}%
    \color{#2}%
    \fbox{\includegraphics[width=1\linewidth,trim={{#3 px} {\cropB} {\cropR} {#4 px}},clip]{#1}}%
}

\newcommand{\addInset}[2]{%
    \color{#2}%
    \fbox{\includegraphics[width=1\linewidth]{#1}}%
}

\newcommand{\auxtimes}{x}
\newcommand{\auxplus}{+}
\newcommand{\auxspace}{ }

\newcommand{\addInsets}[1]{%
    \begin{adjustbox}{valign=t}
        \StrSubstitute{#1}{.}{-}[\baseFileName]
        \begin{adjustbox}{totalheight=1\beautyHeight,tabular={c}}
            \if\useInsetA1%
                \def\cropfile{\baseFileName-\cropAW\auxtimes\cropAH\auxplus\cropAX\auxplus\cropAY-crop}
                \if\cropInsets1
                    \immediate\write18{convert #1 -crop \cropAW\auxtimes\cropAH\auxplus\cropAX\auxplus\cropAY\auxspace -filter point -resize 800\% \cropfile.jpg}
                \fi
                \if\useCroppedImages1
                    \addInset{\cropfile.jpg}{\insetAColor}
                \else
                    \trimInset{#1}{\insetAColor}{\cropAX}{\cropAY}{\cropAW}{\cropAH}%
                \fi%
            \fi%
            \if\useInsetB1%
                \if\useInsetA1\\[\insetvsep]\fi%
                \def\cropfile{\baseFileName-\cropBW\auxtimes\cropBH\auxplus\cropBX\auxplus\cropBY-crop}
                \if\cropInsets1
                    \immediate\write18{convert #1 -crop \cropBW\auxtimes\cropBH\auxplus\cropBX\auxplus\cropBY\auxspace -filter point -resize 800\% \cropfile.jpg}
                \fi
                \if\useCroppedImages1
                    \addInset{\cropfile.jpg}{\insetBColor}
                \else
                    \trimInset{#1}{\insetBColor}{\cropBX}{\cropBY}{\cropBW}{\cropBH}%
                \fi%
            \fi%
            \if\useInsetC1%
                \if\useInsetB1\\[\insetvsep]\fi%
                \def\cropfile{\baseFileName-\cropCW\auxtimes\cropCH\auxplus\cropCX\auxplus\cropCY-crop}
                \if\cropInsets1
                    \immediate\write18{convert #1 -crop \cropCW\auxtimes\cropCH\auxplus\cropCX\auxplus\cropCY\auxspace -filter point -resize 800\% \cropfile.jpg}
                \fi
                \if\useCroppedImages1
                    \addInset{\cropfile.jpg}{\insetCColor}
                \else
                    \trimInset{#1}{\insetCColor}{\cropCX}{\cropCY}{\cropCW}{\cropCH}%
                \fi%
            \fi%
        \end{adjustbox}
    \end{adjustbox}
}

\definecolor{mathematicaBlue}{rgb}{0.38, 0.51, 0.71}
\definecolor{mathematicaOrange}{rgb}{0.88, 0.61, 0.14}
\definecolor{mathematicaGreen}{rgb}{0.56, 0.69, 0.19}
\definecolor{mathematicaRed}{rgb}{0.92,0.39, 0.21}
\definecolor{mathematicaPurple}{rgb}{0.53, 0.47, 0.7}

\definecolor{cvintegrand}{rgb}{1.0, 0.65, 0.0}
\definecolor{cvg}{rgb}{0.5, 0.0, 0.5}
\definecolor{cvG}{rgb}{0.67, 0.14, 0.19}

\definecolor{cvdifference}{rgb}{1.0, 0.65, 0.0}
\definecolor{cvpdf}{rgb}{0.5, 0.0, 0.5}

\begin{document}
\title{Real-time Neural Radiance Caching for Path Tracing} 

\author{Thomas M\"uller}
\affiliation{%
  \institution{NVIDIA}
  }
\email{tmueller@nvidia.com}

\author{Fabrice Rousselle}
\affiliation{%
  \institution{NVIDIA}
  }
\email{frousselle@nvidia.com}

\author{Jan Nov\'ak}
\affiliation{%
  \institution{NVIDIA}
  }
\email{jnovak@nvidia.com}

\author{Alexander Keller}
\affiliation{%
  \institution{NVIDIA}
  }
\email{akeller@nvidia.com}

\renewcommand\shortauthors{M\"uller et al.}

\begin{teaserfigure}
    \vspace{6mm}
    \setlength{\fboxrule}{1pt}%
\setlength{\tabcolsep}{1pt}%
\renewcommand{\arraystretch}{1}%
\small%
\centering%
\begin{tabular}{cccccccc}%
    Path tracing & + ReSTIR & + NRC (Ours) & Reference
    &
    Path tracing & + ReSTIR & + NRC (Ours) & Reference
    \\%
    \multicolumn{4}{c}{\fcolorbox{black}{black}{\setlength{\fboxsep}{1.5pt}\begin{overpic}[width=0.491\linewidth]{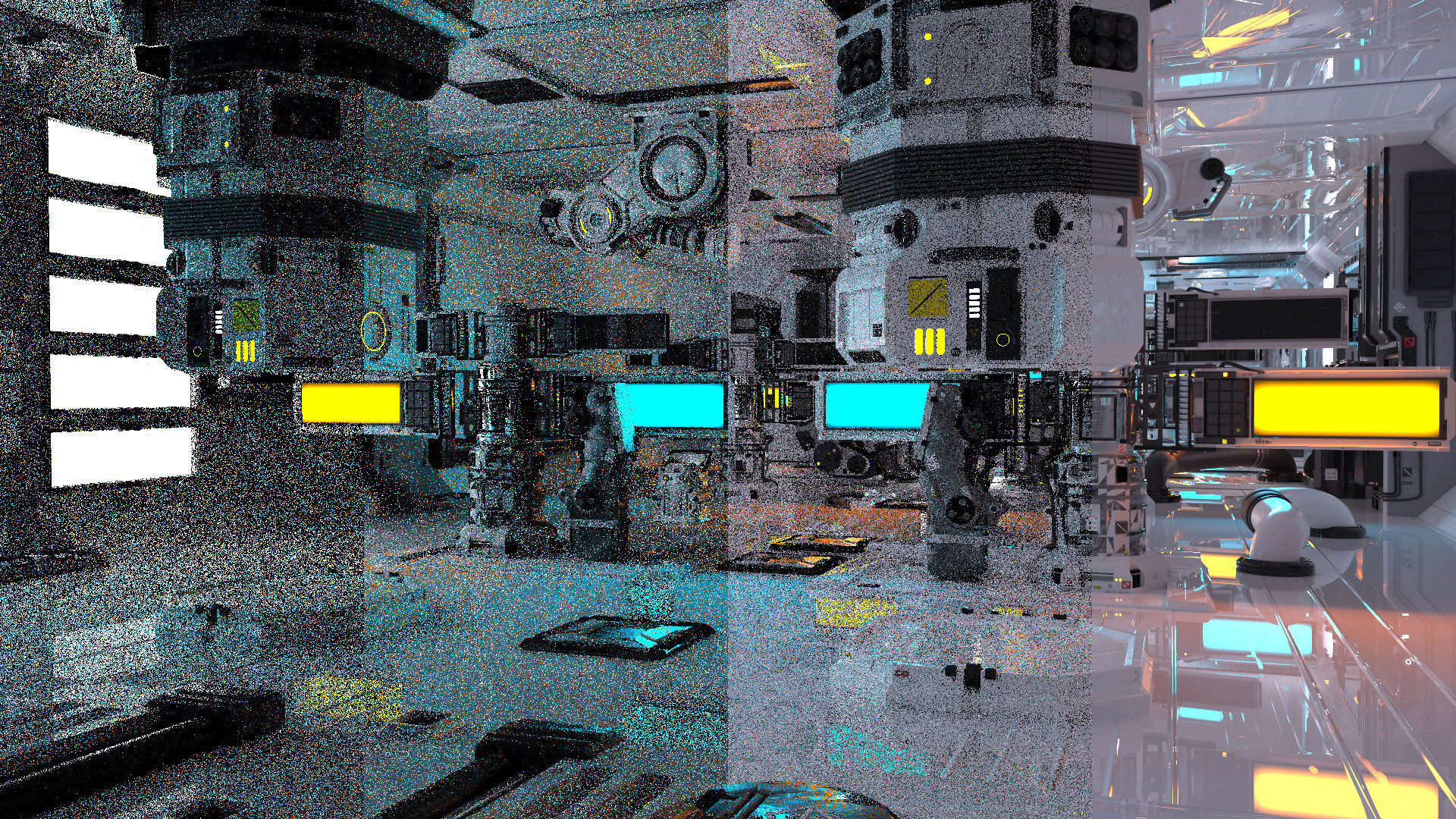}
        \put(45.260416666666664, 41.71875){\makebox(0,0){\tikz\draw[red,ultra thick] (0,0) rectangle (0.051562500000000004\linewidth, 0.020624999999999998\linewidth);}}
        \put(78.125, 9.895833333333332){\makebox(0,0){\tikz\draw[orange,ultra thick] (0,0) rectangle (0.051562500000000004\linewidth, 0.020624999999999998\linewidth);}}
    \end{overpic}}}
    &
    \multicolumn{4}{c}{\fcolorbox{black}{black}{\setlength{\fboxsep}{1.5pt}\begin{overpic}[width=0.491\linewidth]{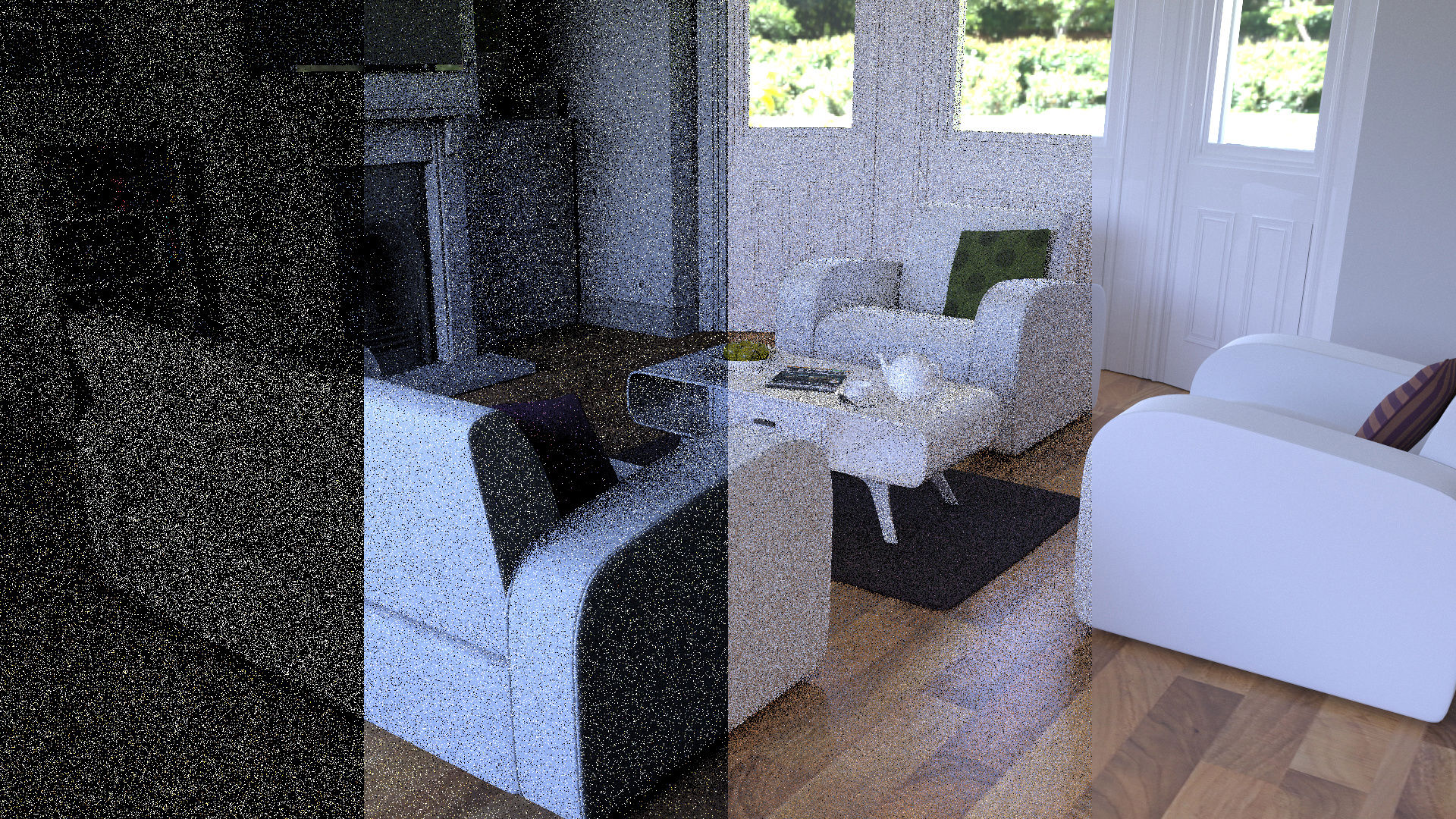}
        \put(59.895833333333336, 29.427083333333332){\makebox(0,0){\tikz\draw[red,ultra thick] (0,0) rectangle (0.051562500000000004\linewidth, 0.020624999999999998\linewidth);}}
        \put(67.70833333333334, 15.104166666666668){\makebox(0,0){\tikz\draw[orange,ultra thick] (0,0) rectangle (0.051562500000000004\linewidth, 0.020624999999999998\linewidth);}}
    \end{overpic}}}
    \\%
    \fcolorbox{red}{red}{\includegraphics[width=0.117\linewidth]{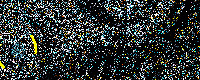}} &
    \fcolorbox{red}{red}{\includegraphics[width=0.117\linewidth]{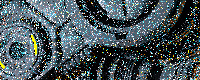}} &
    \fcolorbox{red}{red}{\includegraphics[width=0.117\linewidth]{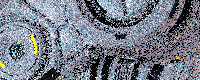}} &
    \fcolorbox{red}{red}{\includegraphics[width=0.117\linewidth]{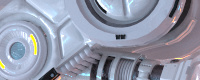}}
    &
    \fcolorbox{red}{red}{\includegraphics[width=0.117\linewidth]{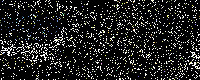}} &
    \fcolorbox{red}{red}{\includegraphics[width=0.117\linewidth]{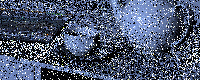}} &
    \fcolorbox{red}{red}{\includegraphics[width=0.117\linewidth]{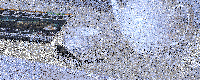}} &
    \fcolorbox{red}{red}{\includegraphics[width=0.117\linewidth]{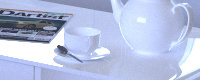}}
    \\%
    \fcolorbox{orange}{orange}{\includegraphics[width=0.117\linewidth]{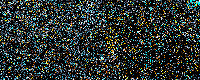}} &
    \fcolorbox{orange}{orange}{\includegraphics[width=0.117\linewidth]{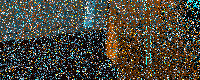}} &
    \fcolorbox{orange}{orange}{\includegraphics[width=0.117\linewidth]{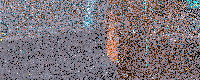}} &
    \fcolorbox{orange}{orange}{\includegraphics[width=0.117\linewidth]{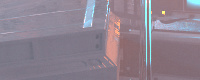}}
    &
    \fcolorbox{orange}{orange}{\includegraphics[width=0.117\linewidth]{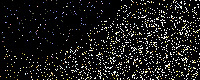}} &
    \fcolorbox{orange}{orange}{\includegraphics[width=0.117\linewidth]{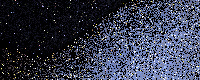}} &
    \fcolorbox{orange}{orange}{\includegraphics[width=0.117\linewidth]{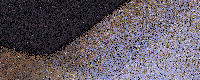}} &
    \fcolorbox{orange}{orange}{\includegraphics[width=0.117\linewidth]{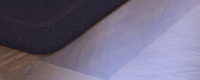}}
    \\%
    Path tracing & + ReSTIR & + NRC (Ours) & Reference
    &
    Path tracing & + ReSTIR & + NRC (Ours) & Reference
    \\%
            97 fps&
            71 fps&
            88 fps&
        &
            186 fps&
            121 fps&
            111 fps&
        \\%
\end{tabular}

    \caption{\label{fig:teaser}
        A path-traced frame from the \ZeroDay{} animation (left) and the \LivingRoom{} scene (right), rendered with 1 sample per pixel each.
        Direct-illumination sampling such as ReSTIR~\citep{bitterli20spatiotemporal} reduces noise at the first path vertex, but does not address noise from indirect illumination.
        We propose to reduce the remaining noise by (online) training a neural network to approximate the radiance field---a new take on classical radiance caching.
        Terminating the paths into the neural cache not only shortens paths---leading to an overall cost \emph{reduction} in the \ZeroDay{} scene---but also removes most of the remaining noise while introducing little bias.
        The images were rendered at a resolution of ${1920\times1080}$ on a high-end desktop machine (i9 9900k and RTX 3090).
        \ZeroDay{} ©beeple
    }\vspace{8mm}
\end{teaserfigure}

\begin{abstract}
We present a real-time neural radiance caching method for path-traced global illumination.
Our system is designed to handle fully dynamic scenes, and makes no assumptions about the lighting, geometry, and materials.
The data-driven nature of our approach sidesteps many difficulties of caching algorithms, such as locating, interpolating, and updating cache points.
Since pretraining neural networks to handle novel, dynamic scenes is a formidable generalization challenge,
we do away with pretraining and instead achieve \emph{generalization via adaptation}, i.e.\ we opt for training the radiance cache while rendering.
We employ self-training to provide low-noise training targets and simulate infinite-bounce transport by merely iterating few-bounce training updates.
The updates and cache queries incur a mild overhead---about 2.6ms on full HD resolution---thanks to a streaming implementation of the neural network that fully exploits modern hardware.
We demonstrate significant noise reduction at the cost of little induced bias, and report state-of-the-art, real-time performance on a number of challenging scenarios.
\end{abstract}



\maketitle

\section{Introduction}%
\label{sec:introduction}

Path-traced global illumination is a long standing challenge in real-time rendering~\cite{AreWeDoneWithRayTracing}.
The problem remains onerous even in offline rendering, especially when considering high-order indirect illumination.
Fortunately, radiative quantities feature significant spatial, directional, and temporal correlations, which can be exploited in various ways to accelerate rendering.

One particularly appealing approach is to cache radiance samples for later reuse.
This can be done in a precomputation step~\cite{Seyb:2020}, or while rendering~\cite{Majercik:2019:ddgi}.
Such systems, however, can be difficult to harness, as they often rely on human intervention or involved heuristics to minimize rendering artifacts~\cite{Hooker:2016}.
We propose to alleviate these difficulties through the use of a \emph{neural} radiance cache, as neural networks are known to be particularly apt at replacing complex heuristics.
Our system is designed according to the following principles:
\begin{itemize}
\item {\bf Dynamic content.}
To handle fully interactive content,
the system must support arbitrary dynamics of the camera, lighting, geometry, and materials.
We strive for a solution that \emph{does not require} precomputation.

\item {\bf Robustness.}
Case-specific handling eventually leads to complex, brittle systems.
Hence, the cache should  be agnostic of materials and scene geometry.
This is particularly important for user-generated content, where design assumptions cannot be easily enforced.
Yet, additional attributes may be provided to the system to improve its rendering quality.

\item{\bf Predictable performance and resource consumption.}
Fluctuations in work load and memory usage lead to unstable framerates.
We seek a solution with stable runtime overhead and memory footprint, both of which should be independent of scene complexity.
The rendering cost must scale at worst linearly with the number of pixels.
\end{itemize}
The first two principles---dynamic content and robustness---present a major challenge for \emph{pre-trained} networks:
the trained model must generalize to novel configurations and, worse, content possibly never observed before.
This form of generalization has not been demonstrated with previous approaches in the context of learning radiative quantities ~\cite{Ren:2013:global,mildenhall2020nerf,hermosilla2019},
and it is unclear if it can ever be achieved.

Instead, we build on the simple but powerful realization that the generalization challenge can be completely sidestepped by fast adaptation.
We rely solely on optimizing the model \emph{online}, during rendering.
Online learning of neural networks has so-far only been used in offline and interactive rendering~\cite{mueller2019nis,mueller2020neural,Lehtinen:2018}.
Fitting the optimization and inference inside the tight rendering loop of real-time applications is a non-trivial task that remains to be tackled.

We present two key contributions that enable \emph{generalization via adaptation} in real time.
First, we describe an efficient mechanism for optimizing the network using (relatively) inexpensive radiance estimates.
The core of this mechanism is self-training of the neural network from its own prediction at a later vertex of the path, providing multi-bounce illumination at the cost of tracing single rays or very short paths.

Second, we propose a streamlined network architecture designed to maximize the quality-cost tradeoff when rendering fully dynamic scenes.
This architecture is key to our system, as its simplicity not only leads to extremely fast convergence but also enables extensive optimizations.
To fully exploit these opportunities, we propose a fully fused implementation tailored to modern GPUs;
the underlying principles however are general and could be adapted to a range of platforms.
We also show how recently proposed input encodings~\citep{mueller2019nis,mildenhall2020nerf} can be combined to greatly enhance fidelity even with our severely constrained budget.

As shown in \autoref{fig:teaser}, our system achieves real-time framerates on current hardware and handles a wide range of material and lighting configurations.
Our accompanying video demonstrates the temporal adaptation to dynamic geometry and lighting.
Lastly, we report preliminary results with an off-the-shelf denoiser demonstrating significantly improved temporal coherence when using our cache.

\section{Related Work}%
\label{sec:related-work}

Reviewing techniques for accelerating global illumination by radiance caching, we focus on precomputation-based techniques, fully dynamic algorithms, and approaches based on artificial neural networks that are most related to our work.
For an extensive survey we refer to \citet{Ritschel:2012:star}.

\paragraph{Radiance caching}
Most real-time global illumination techniques can be traced back to the seminal work of \citet{Ward:1988:caching} on irradiance caching.
Modern techniques for modeling diffuse interreflections follow the same assumption that irradiance tends to vary smoothly across the scene, and texture detail can be recovered using albedo modulation.
Later, \citet{Greger:1998:irradiance} introduced the irradiance probe volume, which became ubiquitous in modern game engines.
The interpolation and location of the various cache records is a key challenge in these techniques,
especially when the aforementioned assumptions on smoothness do not hold.
While robust, principled solutions exist~\cite{Jarosz:2008:caching,Krivanek:2009:PGI}, real-time applications often have to resort to clever heuristics and impose restrictions on scene design to fit their harsh constraints.
In order to handle glossy surfaces, which invalidate the Lambertian assumption at the core of irradiance caching algorithms, \citet{Krivanek:2005:caching} proposed the use of a radiance cache, representing the directional domain with spherical harmonics.
A wealth of recent works further explored the use of radiance caching in offline~\cite{Marco:2018,Zhao:2019,Dubouchet:2017} and real-time rendering, where advances in real-time rendering were enabled for example by compression~\cite{Vardis:2014:radiance}, sparse interpolation~\cite{silvennoinen2017radianceprobes}, pre-convolved environment maps~\cite{Scherzer:2012,Rehfeld:2014}, and spatial hashing~\cite{Pantaleoni:2020,Binder:2018}.
In contrast, our own work achieves robustness through online deep learning.

\paragraph{Precomputation-based techniques}
The high computational cost of simulating global illumination spurred the development of precomputation techniques~\citep{Arvo86backwardray,heckbert1990adaptive}, which have been further developed to address the stringent constraints of real-time applications.
Assuming both the scene lighting and geometry are fixed, irradiance can be computed and then stored in texture space using lightmaps~\cite{Abrash:1997:quake} and in world space using light probes~\cite{Oat:2005:irradiance}.
Offering both approaches in one system, \citet{GeomericsEnlighten} introduced iterated dynamic lighting updates.
These techniques are widespread in modern game engines~\cite{Barre:2017:certain}.
Light probes can be combined with precomputed radiance transfer~\cite{Sloan:2002:precomputed} or visibility~\cite{Iwanicki:2017:precomputed,McGuire:2017:probes}, to account for (self) occlusion when shading scene objects.
While precomputation-based solutions offer a number of indisputable advantages~\cite{Seyb:2020}, we embrace online caching mechanisms that facilitate dynamically changing scenes without assuming (parts of) the scene to be static or known in advance.

\paragraph{Fully dynamic techniques}
Dynamic real-time global illumination methods build upon efficient rendering algorithms that reuse shading and visibility computation across pixels, such as photon mapping~\cite{Jensen:1996:global}, many-light rendering~\cite{Keller:1997:instant} and radiosity maps~\cite{Tabellion:2004:approximate}, extracting further efficiency through various approximations.
Some approximate the scene geometry using a (hierarchical) point cloud, which is then efficiently rasterized into shadow maps~\cite{Ritschel:2008:imperfect} or micro-rendering buffers~\cite{Ritschel:2009:micro}.
Volumetric approximations of the scene lighting and geometry~\cite{Kaplanyan:2010:cascaded,Crassin:2011:interactive}, bootstrapped with large numbers of virtual point lights from reflective shadow maps~\cite{Dachsbacher:2005:reflective}, allowed to scale to larger scenes.
These approaches are sometimes combined with very efficient screen-space approximations of ambient occlusion~\cite{Mittring:2007:ssao}, directional occlusion~\cite{Ritschel:2009:ssdo}, or reflections~\cite{Sousa:2011:ssr}.
Recently, ray-tracing hardware has been used to compute specific components of light transport online, such as diffuse interreflections~\cite{Majercik:2019:ddgi} or glossy reflections~\cite{Deligiannis:2019:battlefield}.
Aside of accuracy limitations inherent to the approximations, such as blurring, missing interactions, or assumptions about the material model, a key limitation of many of these techniques is the reliance on a dual representation of the scene which must be continuously refreshed.
Our neural radiance cache sidesteps the need for an approximate scene representation by operating on a set of sampled path contributions, which effectively decouples the algorithm from the scene lighting and geometric complexity.

\paragraph{Path guiding}
The family of path guiding techniques, originating from \citet{Lafortune:1995:A5T} and \citet{Jensen1995}, is closely related to that of radiance caching in that they often learn an approximation of incident radiance that is amenable to importance sampling.
Recent incident-radiance models tend to be either parametric mixtures~\citep{Vorba:2014:OnlineLearningPMMinLTS} or probability trees~\citep{mueller2017practical}, but (neural) normalizing flows are also possible~\citep{mueller2019nis}.
While these techniques are highly successful in offline rendering of mostly static scenes~\citep{vorba19guiding}, adapting them to the constraints of animated real-time rendering is non-trivial ongoing work~\citep{2020_tsr_nee_pg_rtpt}.
Methods that use an explicit model of the BRDF~\citep{herholz2016,Herholz:2018} or the product integrand~\citep{mueller2019nis,mueller2020neural} are likely the most promising to repurpose as radiance caches, as they yield a more accurate scattered-radiance estimate than methods approximating the incident radiance at the previous camera-path vertex.

\paragraph{Neural techniques}
Neural networks are capable of approximating various visual phenomena remarkably well, whether they operate in screen space~\citep{Nalbach2017b} or in world space, whether they are pre-trained over multiple scenes~\citep{Kallweit2017DeepScattering,hermosilla2019,Nalbach2017b,Jiang:2021} or fit to single scene~\citep{Ren:2013:global,mueller2019nis,mildenhall2020nerf,mueller2020neural,MLandIEQ}.
The latter approaches are most closely related to ours.
\citet{Ren:2013:global} propose to train a set of local neural radiance caches, conditioned on the position of a single point light source.
While lighting can be changed dynamically and area lighting can be approximated using a set of point lights at the cost of multiple cache queries,
geometry and materials have to remain static as a consequence of the cost of the training procedure.
Our technique differs on two important aspects:
(i) we use a single neural radiance cache leveraging recently proposed encodings~\cite{vaswani2017attention,mueller2019nis} to adapt to local scene variations,
and (ii) we train our model online which allows for fully dynamic scenes and readily accounts for all lighting in the scene in a single query.
Neural control variates~\cite{mueller2020neural} and NeRF~\cite{mildenhall2020nerf}, developed in the context of offline rendering, encompass a radiance cache that is parameterized similarly.
The key differences of our work are:
(i) a network architecture and implementation designed for a rendering budget on the order of milliseconds instead of minutes,
and (ii) integration in a renderer using self-training, which has been connected with Q-learning~\citep{Dahm16}, to account for infinite bounces of indirect illumination despite tracing paths of finite length.

\section{Neural Radiance Caching}
\label{sec:radiance-caching}

Our goal is to cache radiance using \emph{one single} neural network that maps spatio-directional coordinates to radiance values and is trained in real-time to support dynamic scenes.
We opt for approximating the \emph{scattered radiance} as it is the most computationally expensive part of the rendering equation~\cite{Kajiya:1986:TRE}.
The \emph{scattered radiance}
\begin{align} \label{eqn:radiance}
\reflectedRadiance(\pos,\diro) := \int_\Sphere \bsdf(\pos,\diro,\diri) \, \inRadiance(\pos,\diri) \, |\cos\fsanglei| \, \Diff{\diri}
\end{align}
represents the radiative energy leaving point $\pos$ in direction $\diro$ after being scattered at $\pos$.
For a given direction of incidence $\diri$, the integrand is
the product of the bidirectional scattering distribution function (BSDF) $\bsdf(\pos,\diro,\diri)$, the incident radiance $\inRadiance(\pos,\diri)$, and the foreshortening term $|\cos \fsanglei|$, where $\fsanglei$ is the angle between $\diri$ and the surface normal at $\pos$.
Our neural network approximates $\reflectedRadiance$ by the cached radiance $\cachedReflectedRadiance$.

In this section, we discuss the algorithmic choices for building a neural radiance cache that are key to satisfy the design principles outlined in \autoref{sec:introduction}.
Real-time performance is enabled by an optimized fully fused network, which is discussed in \autoref{sec:neuralnetwork}.

\subsection{Algorithm Overview}

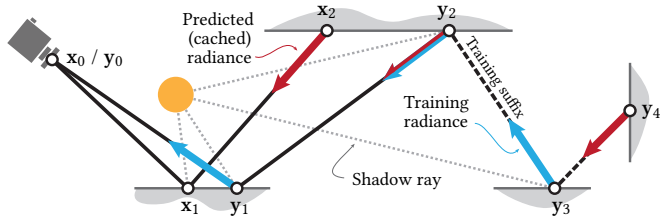
\begin{figure}
    \small
    \begin{overpic}[width=\linewidth]{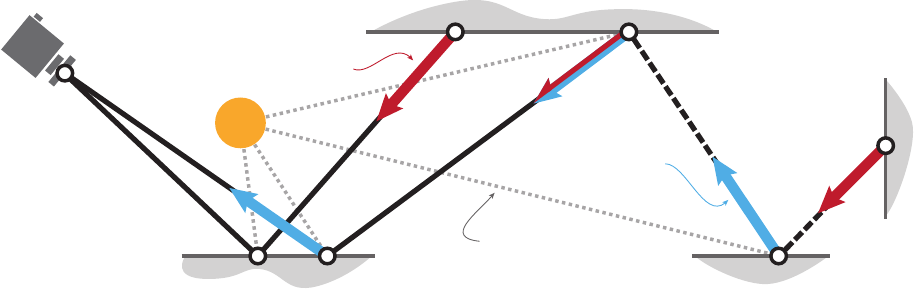}
        \put(8.5, 23) {  $\pos_0$ / $\posy_0$ }
        \put(26.3, 0) { $\pos_1$ }
        \put(34, 0) { $\posy_1$ }
        \put(47.5, 30.5) { $\pos_2$ }
        \put(66.2, 30.5) { $\posy_2$ }
        \put(84, 0) { $\posy_3$ }
        \put(98, 15) { $\posy_4$ }
        \put(26.6, 29) { \footnotesize Predicted }
        \put(27.5, 26) { \footnotesize (cached) }
        \put(27.3, 23) { \footnotesize radiance }
        \put(53, 4) { \footnotesize Shadow ray }
        \put(61, 16) { \footnotesize Training }
        \put(61, 13) { \footnotesize radiance }
        \put(70, 26.5) { \scriptsize \rotatebox{-56}{Training suffix} }
    \end{overpic}
    \caption{\label{fig:path-illustration}
        For rendering, we trace short ``rendering'' paths (e.g.\ $\pos_0\cdots\pos_2$) and terminate them into the neural radiance cache; queries of cached radiance $\cachedReflectedRadiance$ are highlighted by red arrows.
        To optimize the cache, we extend a small subset of the rendering paths by a few vertices (called ``training suffix'', e.g.\ $\posy_2\cdots\posy_4$).
        We collect radiance estimates (blue arrows) to update the neural radiance cache along the vertices of the longer training path,
        reusing the initial path segment that was already traced for rendering.
        Furthermore, the online training together with the termination of training paths into the cache progressively increases the number of simulated light bounces.
    }
\end{figure}

Rendering a single frame consists of computing pixel colors and updating the neural radiance cache; see \autoref{fig:path-illustration} for an illustration.

First, we trace short rendering paths, one for each pixel, and terminate them as soon as the approximation provided by the radiance cache is deemed sufficiently accurate.
We use the heuristic by \citet{bekaert2003}, that was originally developed in the context of photon density estimation, to only query the cache once the spread of the path is sufficiently large to blur small inaccuracies of the cache (more detail in \autoref{sec:termination}).
At each intermediate vertex, we use next-event estimation to integrate light from emitters.
To this end, we use screen-space ReSTIR~\citep{bitterli20spatiotemporal} at the primary vertex and a LightBVH~\citep{moreau19lightbvh}, combined with the BSDF via multiple importance sampling~\citep{Veach:1995:MIS}, at the subsequent vertices.
Truncating the path at the terminal vertex $\pos_k$, we evaluate the neural radiance cache to approximate $\reflectedRadiance(\pos_k, \diro_k)$.

Second, to train the radiance cache, we extend a fraction (typically under $3\%$) of the short rendering paths by a few vertices---a \emph{training suffix}.
As before, we terminate these longer training paths once the area spread of their suffix is sufficiently large; for that purpose we consider the query vertex $\pos_k$ as a primary vertex (see \autoref{fig:path-heuristic}).
In the majority of cases, the suffix consists of one vertex.
The radiance estimates collected along \emph{all} vertices of the longer training paths are used as reference values for training the radiance cache.

\paragraph{Discussion}
Terminating the paths into the radiance cache saves computation and, importantly, replaces a one-sample estimate with an approximation that aggregates samples from spatially and temporally nearby locations.
The variance is thus reduced, however, the viability of caching for real-time applications is still conditioned on how efficiently and quickly we update and query the cache.

\subsection{Fast Adaptation of the Cache by Self-training} \label{sec:qlearning}

As in any data-driven approach, the quality of the approximation depends on the accuracy of the target values $\reflectedRadiance$ that the network is trained on.
The distinct challenge of rendering dynamic scenes in real time requires to continuously adapt the neural radiance cache according to the changing radiance field, for example, due to moving lights or geometry.
This means we neither have the luxury of precomputing precise target values $\reflectedRadiance$, nor can we tolerate noisy
estimates that would slow down convergence.

Instead of estimating target values via Monte Carlo path tracing~\cite{mueller2019nis,mueller2020neural}, we leverage the neural radiance cache itself by evaluating it at the terminal vertices of the longer training paths.
The collected radiance is transported to the preceeding vertices, at each one generating a target value for training the neural network.
Updating the neural radiance cache using its own values resembles the concept of Q-learning~\citep{MLandIEQ,Dahm16}.

The self-training approach has two distinct advantages over fully path-traced estimates: it trades a large fraction of the undesired noise for (potential) bias when estimating $\reflectedRadiance$.
It also allows for capturing global illumination
as long as the training procedure is iterated:
the radiance learned by one training path is transported using multiple other training paths in the next iteration.
Hence, each iteration increases the number of simulated light bounces.
This is reminiscent of progressive radiosity algorithms \cite{GeomericsEnlighten} that simulate multi-bounce diffuse transport by iterating single-bounce radiative transfer.

However, self-training the neural radiance cache also has two caveats: first, the last vertex of the training path may reach scene locations that the radiance cache has not been trained for, which may incur a larger approximation error. The second drawback is that the iterated optimization may simulate only a subset of multi-bounce illumination rather than all light transport.
Specifically, only the transport from emitters that can be reached by training paths will be further bounced around, and only so, if tails of training paths in subsequent frames land near the current optimization points (i.e.\ $\posy_4$ needs to be close to $\posy_2$ or $\posy_3$ in \autoref{fig:path-illustration}).
Both caveats can be alleviated almost for free by making a small fraction $u$ of the training paths truly unbiased, thereby injecting correct source values to be propagated by the self-training mechanism.
We use ${u = 1/16}$, i.e.\ every $16$\textsuperscript{th} training suffix is only terminated by Russian roulette.

\subsection{Temporally Stable Cache Queries}
\label{sec:onlinelearning}

When rendering dynamic content, for example changing camera position or animated geometry, the neural radiance cache continuously needs to adapt, forcing us to use a high learning-rate when optimizing the network by gradient descent.
In addition, we also perform \emph{multiple} (in our case 4) gradient descent steps per frame\footnote{Each step uses a disjoint random subset of the training data that was gathered while rendering the frame to prevent the same data from being seen twice.}, which leads to even faster adaptation.

However, a side effect of such an aggressive optimization schedule are temporal artifacts like flickering and oscillations across the rendered frames---even when the scene and camera are static, because there is noise in the estimated radiance targets.

\begin{figure}
    \setlength{\tabcolsep}{1pt}%
\renewcommand{\arraystretch}{0.75}%
\small%
\begin{tabular}{cccc}%
    &%
    EMA weight $\alpha = 0.00$&%
    EMA weight $\alpha = 0.90$&%
    EMA weight $\alpha = 0.99$\\%
    \hspace{-2.5mm}\rotatebox{90}{\hspace{10mm}\ZeroDay}%
    &\includegraphics[width=0.325\linewidth]{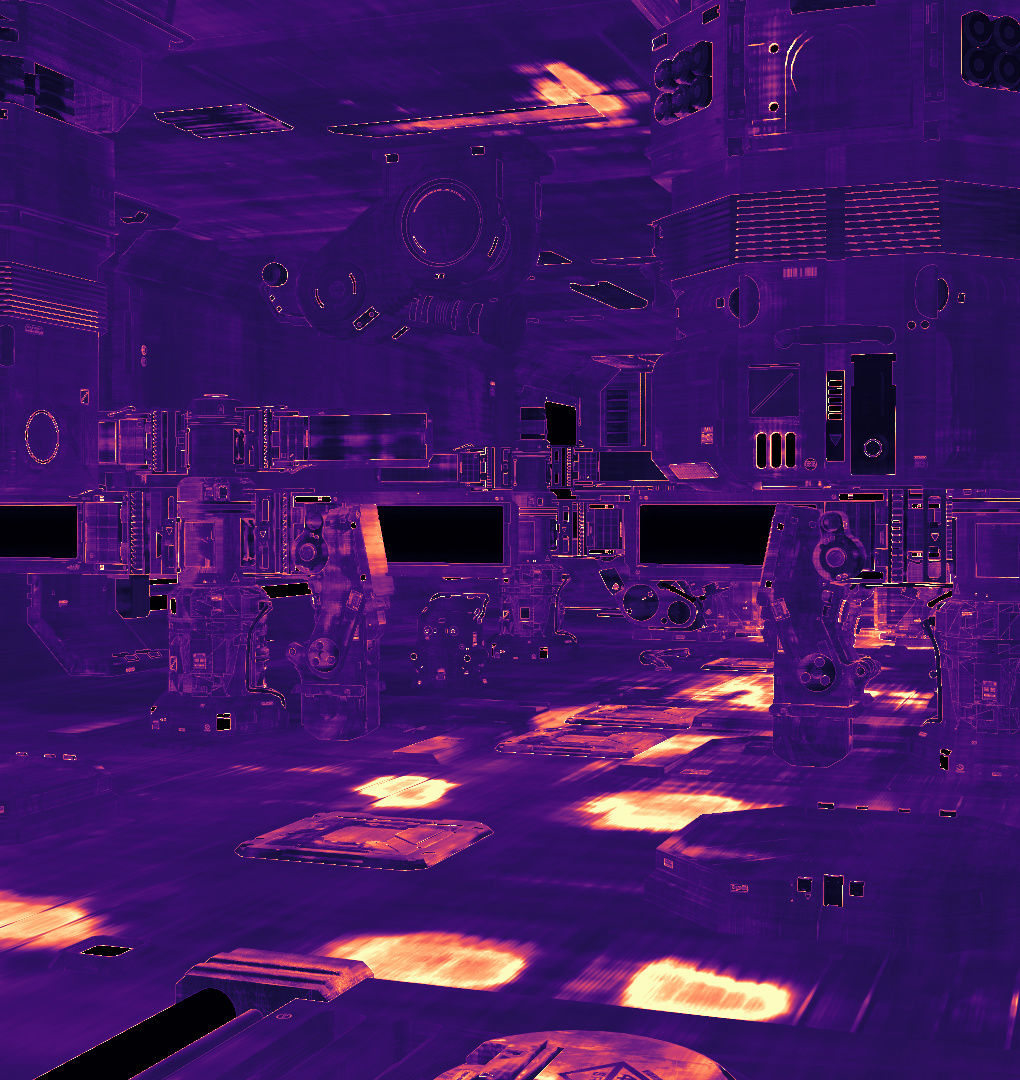}%
    &\includegraphics[width=0.325\linewidth]{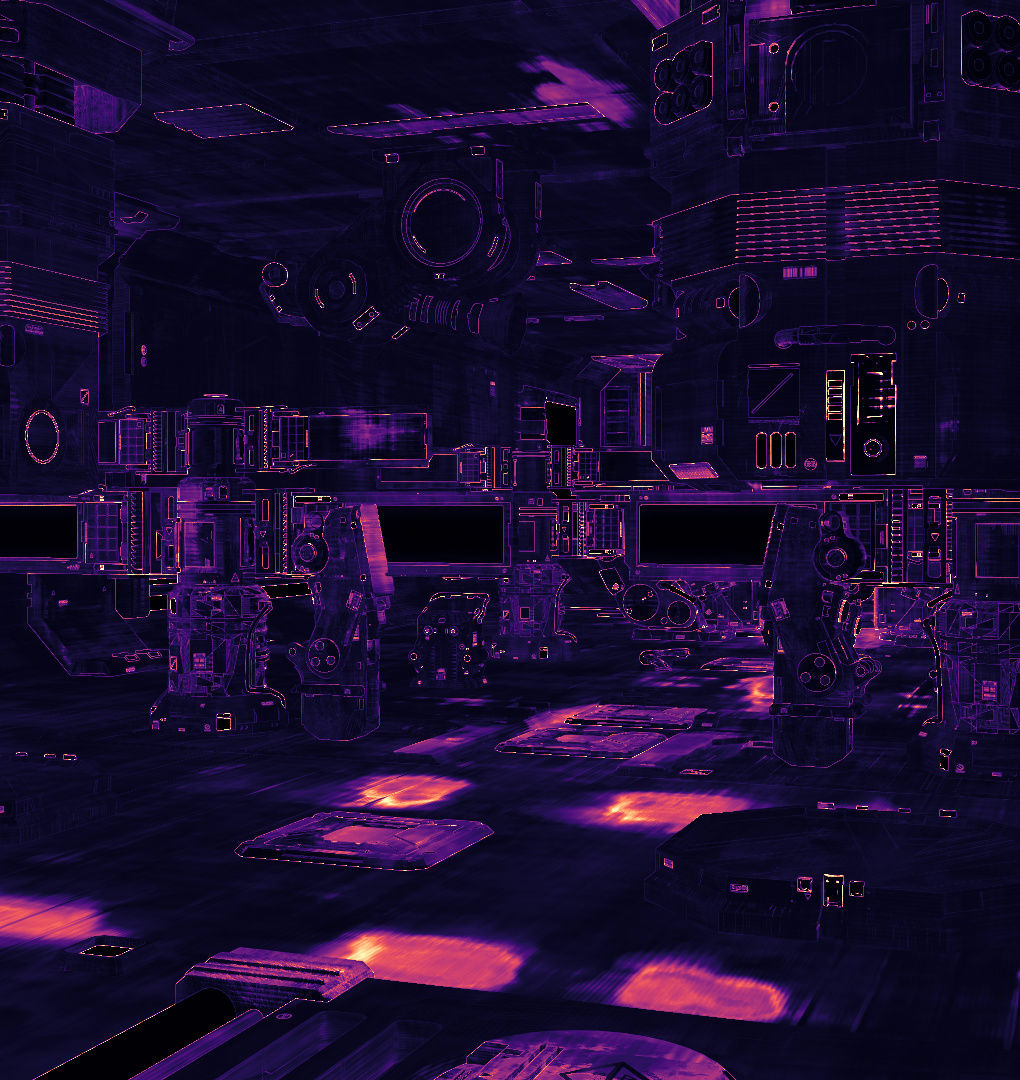}%
    &\includegraphics[width=0.325\linewidth]{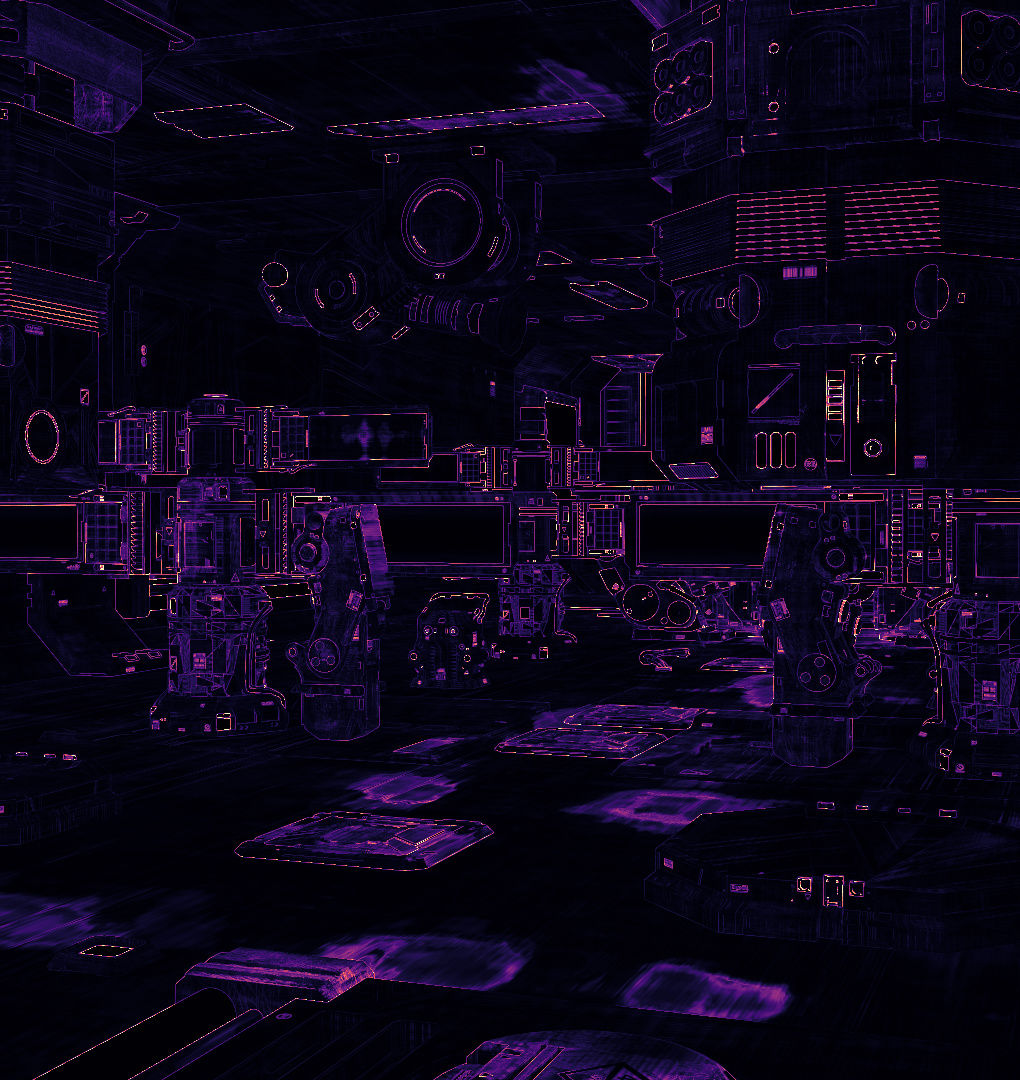}%
    \\%
\end{tabular}
    \vspace{-3mm}
    \caption{
        \label{fig:temporal-stability}
        Temporal stability of online learning with three weight-averaging strategies:
        no averaging (left) and an exponential moving average (EMA) with weights $0.90$ and $0.99$.
        The false color images depict temporal stability across 100 frames, measured as the symmetric mean absolute percentage error (SMAPE) averaged over all consecutive pairs of frames; i.e.\ darker means less variation across frames.
        To be more meaningful in print, the scene and the camera have been fixed.
        Please see the accompanying video to best assess the temporal stability.
    }
\end{figure}

We therefore propose to dampen such oscillations by averaging the network weights produced by the optimization.
More specifically, we compute an exponential moving average (EMA) of the network weights $\Params_t$ produced by the $t$\textsuperscript{th} gradient descent step, which creates a \emph{second} set of weights $\overline{\Params}_t$ that we use when evaluating the cache for rendering.
The exponential moving average reads
\begin{align} \label{eqn:weights}
    \overline{\Params}_t := \frac{1 - \alpha}{\eta_t} \cdot \Params_t + \alpha \cdot \eta_{t-1} \cdot \overline{\Params}_{t-1} \,, \text{ where }
    \eta_t := 1 - \alpha^t
\end{align}
corrects the bias of the average for small $t$ and $\alpha \in [0, 1]$ controls the strength of exponential averaging.
We use ${\alpha = 0.99}$ for a good trade-off between fast adaptation yet temporally stable evolution of the weights $\overline{\Params}_t$; we illustrate the temporal stability in \autoref{fig:temporal-stability}.

Note that the averaging process does not feed back into the training loop; $\overline{\Params}_t$ depends on $\Params_t$, but not the other way around.
Still, recent work suggests that the EMA filtered weights $\overline{\Params}_t$ may be closer to the optimum than any of the raw weights $\Params_t$ produced by the optimizer~\citep{izmailov2019averaging}.

\begin{figure}
    \setlength{\fboxrule}{1pt}%
\setlength{\tabcolsep}{1pt}%
\renewcommand{\arraystretch}{1}%
\small%
\centering%
\begin{tabularx}{\columnwidth}{YYYYYYYYYYY}%
    \multicolumn{11}{c}{ Visualization at first non-specular vertex } \\
    \multicolumn{11}{c}{\setlength{\fboxsep}{1.5pt}\begin{overpic}[width=\linewidth]{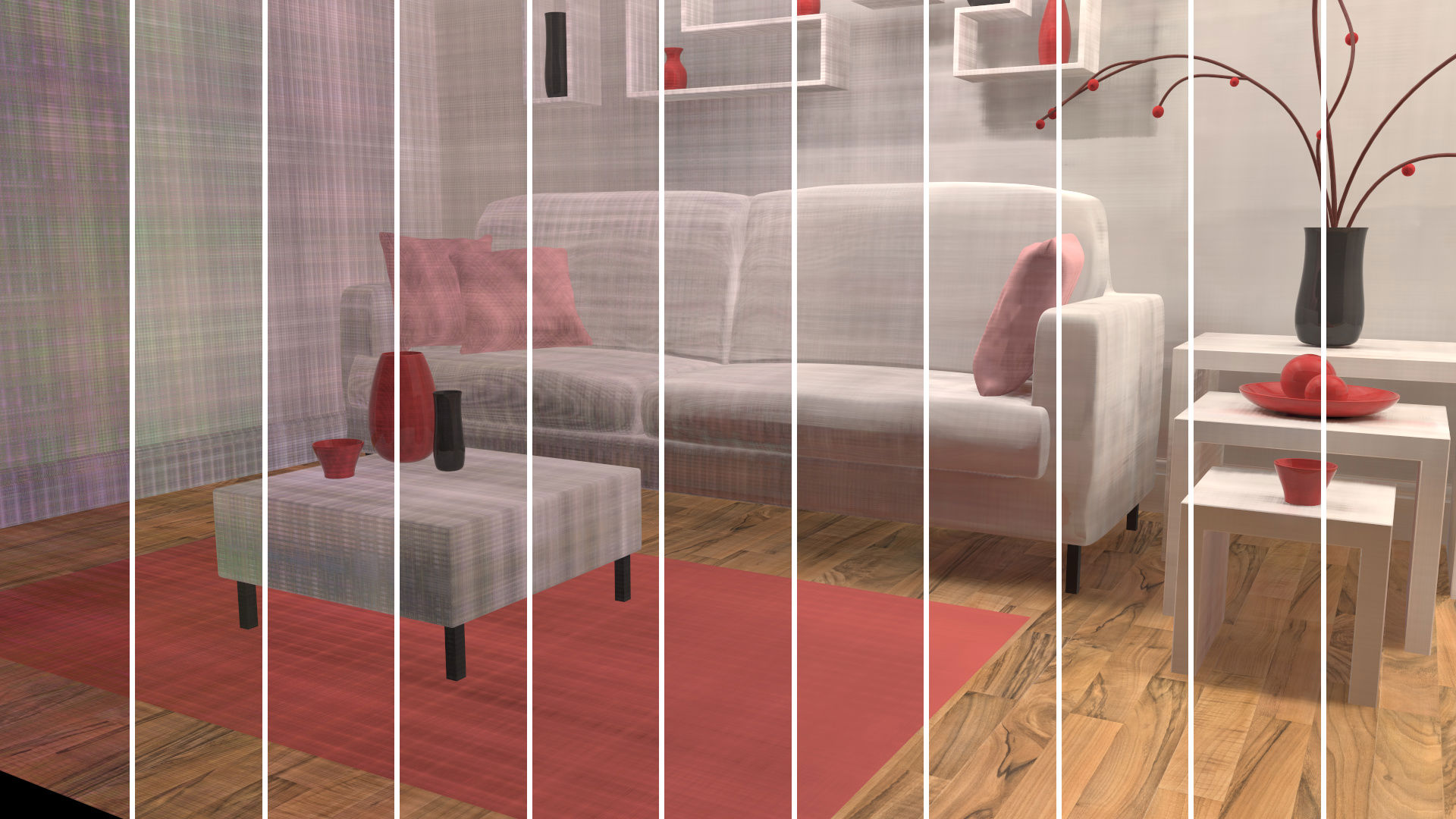}
    \end{overpic}}\\
    \multicolumn{11}{c}{ Proposed use of the cache at the end of short paths } \\
    \multicolumn{11}{c}{\setlength{\fboxsep}{1.5pt}\begin{overpic}[width=\linewidth]{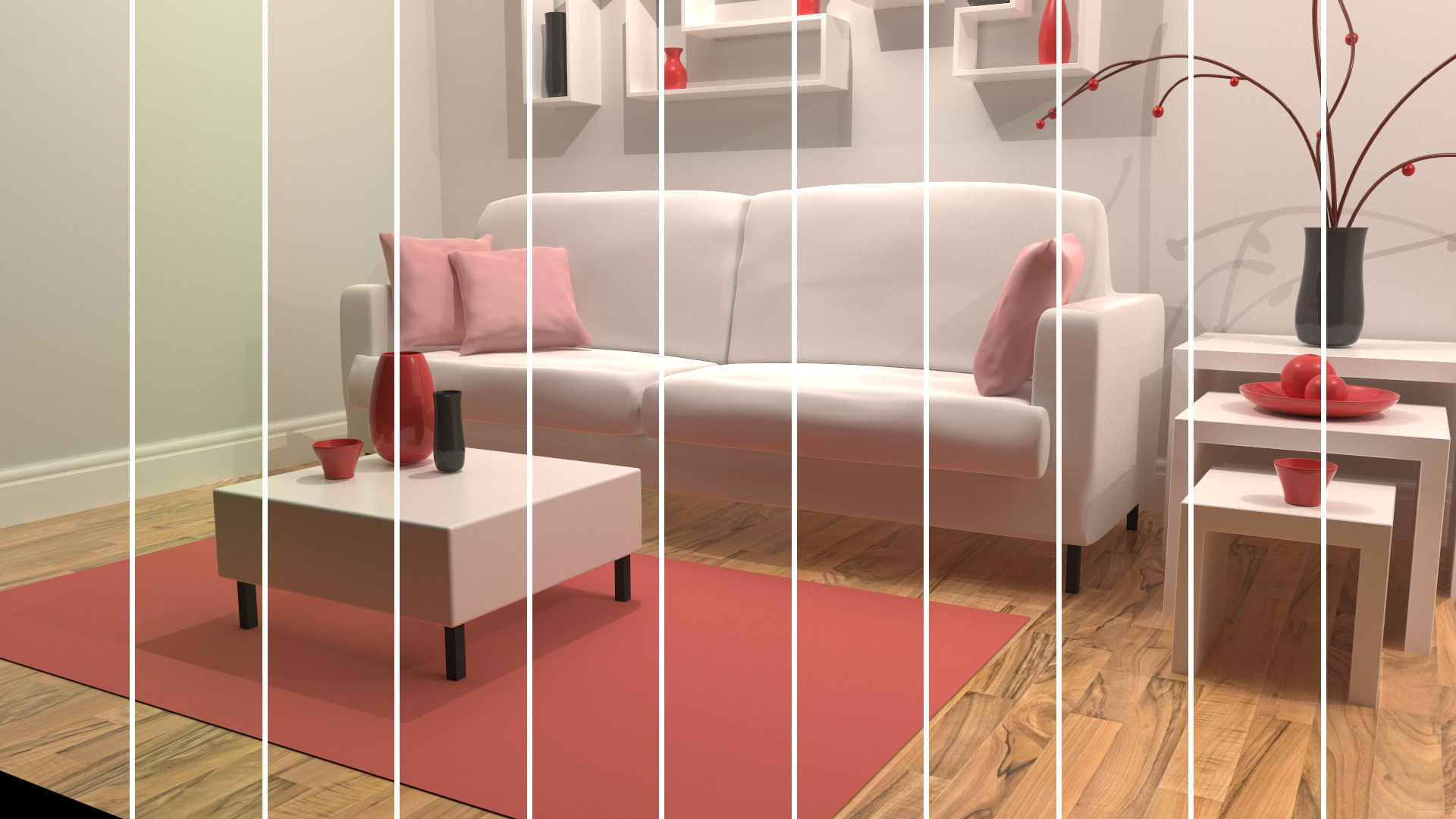}
    \end{overpic}}\\
    1 & 2 & 4 & 8 & 16 & 32 & 64 & 128 & 256 & 512 & 1024 \\
    \multicolumn{11}{c}{Training frames}\\
\end{tabularx}

    \vspace{-3mm}
    \caption{\label{fig:learning}
        From-scratch training of the neural radiance cache.
        We visualize the cache after $1$, $2$, $4$, ..., $1024$ frames.
        Top: to illustrate its training behavior, the radiance cache is visualized directly at the first non-specular vertex.
        Already after the first 64 frames (${\sim0.5}$s) the overall colors are correct and only subtle high-frequency artifacts remain.
        Bottom: using the cache as proposed, the high-frequency artifacts are hidden behind path indirections and the cache is usable starting from the ${\sim8}$th frame (i.e.\ after ${\sim70}$ms).
        This confirms that the cache trains sufficiently fast for the online adaptation to animated content; see the supplementary video for more results.
    }
\end{figure}

Indeed, \autoref{fig:learning} shows that when the radiance cache is trained from scratch, its evolution is gradual and quick at the same time.
Using the cache at the end of paths instead of visualizing it directly filters its approximation error, converging to satisfactory quality in as few as $8$ frames (${\sim70}$ ms); see the supplementary video for more results on animated content.

\subsection{Path Termination}\label{sec:termination}

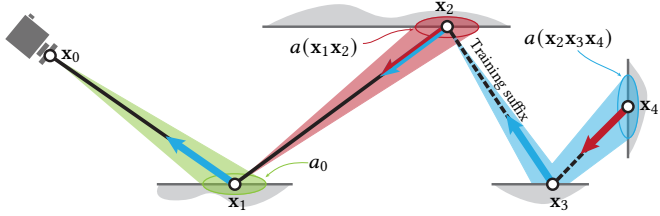
\begin{figure}
    \small
    \begin{overpic}[width=\linewidth]{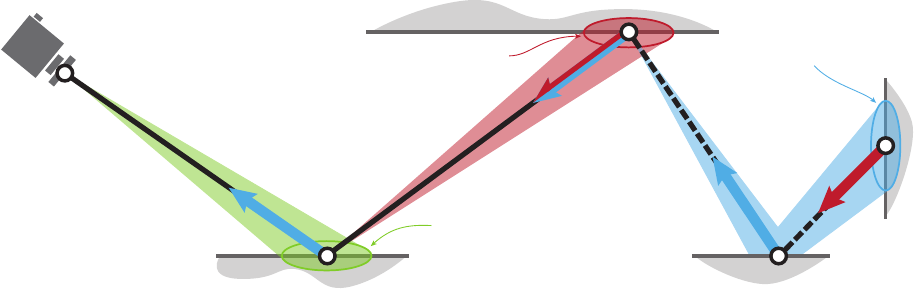}
        \put(8, 23) { $\pos_0$ }
        \put(34, 0) { $\pos_1$ }
        \put(66.2, 30.5) { $\pos_2$ }
        \put(84, 0) { $\pos_3$ }
        \put(98, 15) { $\pos_4$ }
        \put(46.5, 6) { $\AreaHeuristic_0$ }
        \put(43.5, 24) { $\AreaHeuristic(\pos_1\pos_2)$ }
        \put(80, 25.3) { $\AreaHeuristic(\pos_2\pos_3\pos_4)$ }
        \put(71, 25.5) { \scriptsize \rotatebox{-56}{Training suffix} }
    \end{overpic}
    \vspace{-3mm}
    \caption{\label{fig:path-heuristic}
        We terminate our short rendering paths into the neural radiance cache once their scattering interactions blur the signal sufficiently well.
        To this end, we compare the size of the footprint of the path $\AreaHeuristic(\pos_1\pos_2)$ to the size of the directly visible surface in the image plane $\AreaHeuristic_0$.
        The longer training paths are terminated by the same heuristic applied to the vertices of the suffix, i.e.\ we compare $\AreaHeuristic(\pos_2\pos_3\pos_4)$ to $\AreaHeuristic_0$.
    }
\end{figure}

All paths are terminated according to a simple heuristic based on the area-spread of path vertices, illustrated as cones in \autoref{fig:path-heuristic}; we index the camera vertex as $\pos_0$ and the primary vertex as $\pos_1$.
Once the spread becomes large enough to blur away the small-scale inaccuracies of our cache (c.f.\ \autoref{fig:inference-bounce}), we terminate the path.

Following \citet{bekaert2003}, the area spread along the subpath $\pos_1\cdots\pos_n$ can be cheaply approximated as the sum
\begin{align}
    \AreaHeuristic(\pos_1\cdots\pos_n) &= {\left( \sum_{i=2}^n \sqrt{\frac{\|\pos_{i-1} - \pos_i\|^2}{\PdfMC(\diri\,|\,\pos_{i-1}, \diro) \, | \cos{\fsanglei} |}} \right)}^2 \,,
\end{align}
where $\PdfMC$ is the BSDF sampling PDF and $\fsanglei$ is the angle between $\diri$ and the surface normal at $\pos_i$.

To terminate a path, we compare the subpath spread $\AreaHeuristic(\pos_1\cdots\pos_n)$ to the spread at the primary vertex as viewed from the camera, which can be approximated\footnote{By assuming a spherical image plane and ignoring constant factors.} as
\begin{align}
    \AreaHeuristic_0 := \frac{\|\pos_0 - \pos_1\|^2}{4\pi\cos{\fsangle_1}} \,.
\end{align}
That is, we will terminate a path if $\AreaHeuristic(\pos_1\cdots\pos_n) > c \cdot \AreaHeuristic_0$, where $c$ is a hyperparameter that trades variance (longer paths) for bias and speed (shorter paths).
We found $c = 0.01$ to yield satisfactory results.

Lastly, if the path is selected to become a training path, the heuristic will be used once again, this time to terminate the training suffix when $\AreaHeuristic(\pos_n\cdots\pos_m) > c \cdot \AreaHeuristic_0$ is satisfied.
The heuristic is illustrated in \autoref{fig:path-heuristic} for a training path, where the short rendering part of the path ends at vertex ${n=2}$ and the training suffix at ${n=4}$.

\subsection{Amortization in a Real-time Path Tracer}

We target real-time applications, setting ourselves a $16.6$ milliseconds rendering budget in order to achieve a framerate of 60 frames per second.
That budget includes the tracing of paths, shading at every vertex, as well as querying and updating the cache.
In practice, this leaves just a few milliseconds to handle the cache overhead.
We not only tackle this using our proposed fully fused network, described in \autoref{sec:neuralnetwork}, but also in the path tracer integration itself.

We interleave short rendering paths and long training paths by tiling the viewport.
Using a single random offset, we promote one path per tile to be a long training path (see \autoref{fig:path-heuristic}), resulting in a uniform sparse set of training paths in screen space.
This approach of merely prolonging a rendering path to obtain a training path greatly reduces the overhead, as computation is shared between the two.
This contrasts with caching techniques based on probe volumes, which use a separate set of rays to update the cache that do not contribute to the image itself.

Once the tracing is complete, we reconstruct the image by back-propagating the cached radiance from the terminal vertex of each short rendering path.
For training paths, we track two values: the aforementioned rendering radiance and the training radiance.
For every vertex along a training path, we store the training radiance (along with the vertex information) in an array; it will be used as the target to optimize the cache.

We shuffle all training records using a linear congruential generator and distribute them over $s$ training batches of $l$ training records each.
The shuffling ensures that the training batches are not correlated with image regions.
Each training batch is used to perform a single optimization step of the cache.
As we want to ensure a stable work load, we use an adaptive tiling mechanism to match a target training budget;
we select $s=4$ training batches and $l=16384$ records per batch in practice, for a total of $65536$ training records per frame.
We dynamically adjust the tile size
at each frame based on the number of training records generated during the image reconstruction.
Similar to other caching techniques, the cost of training is decoupled from the image resolution, since we use a bounded number of training records.
Lastly, we observe that training our neural radiance cache amounts to a regression over many samples from spatially and temporally nearby locations, i.e.\ a form of path-space denoising.
The variance is thus significantly reduced by replacing one-sample radiance estimates with the cache approximation.

\begin{table}
    \caption{\label{tab:parameters}
        Parameters and their encoding, amounting to 62 dimensions:
        $\PosEnc$ denotes frequency encoding~\citep{mildenhall2020nerf}, $\OneBlob$ denotes one-blob encoding~\citep{mueller2019nis}, $\Spherical$ denotes a conversion to spherical coordinates, normalized to the interval $[0, 1]^2$, and $\Identity$ is the identity.
    }
    \begin{tabularx}{\columnwidth}{rcc}
        \toprule
        Parameter & Symbol & with Encoding \\
        \midrule
        Position & $\pos \in \R^3$ & $\PosEnc(\pos) \in \R^{3\times12}$ \\
        Scattered dir. & $\diro \in \Sphere$ & $\OneBlob(\Spherical(\diro)) \in \R^{2\times4}$ \\
        Surface normal & $\normal(\pos) \in \Sphere$ & $\OneBlob(\Spherical(\normal(\pos))) \in \R^{2\times4}$ \\
        Surface roughness & $r(\pos, \diro) \in \R$ & $\OneBlob\left(1 - e^{-r(\pos, \diro)}\right) \in \R^{4}$ \\
        Diffuse reflectance & $\albedo(\pos, \diro) \in \R^3$ & $\Identity(\albedo(\pos, \diro)) \in \R^3$ \\
        Specular reflectance & $\specAlbedo(\pos, \diro) \in \R^3$ & $\Identity(\specAlbedo(\pos, \diro)) \in \R^3$ \\
        \bottomrule
    \end{tabularx}
\end{table}

\begin{figure*}
    \vspace{-3mm}
    \begin{overpic}[width=0.95\textwidth]{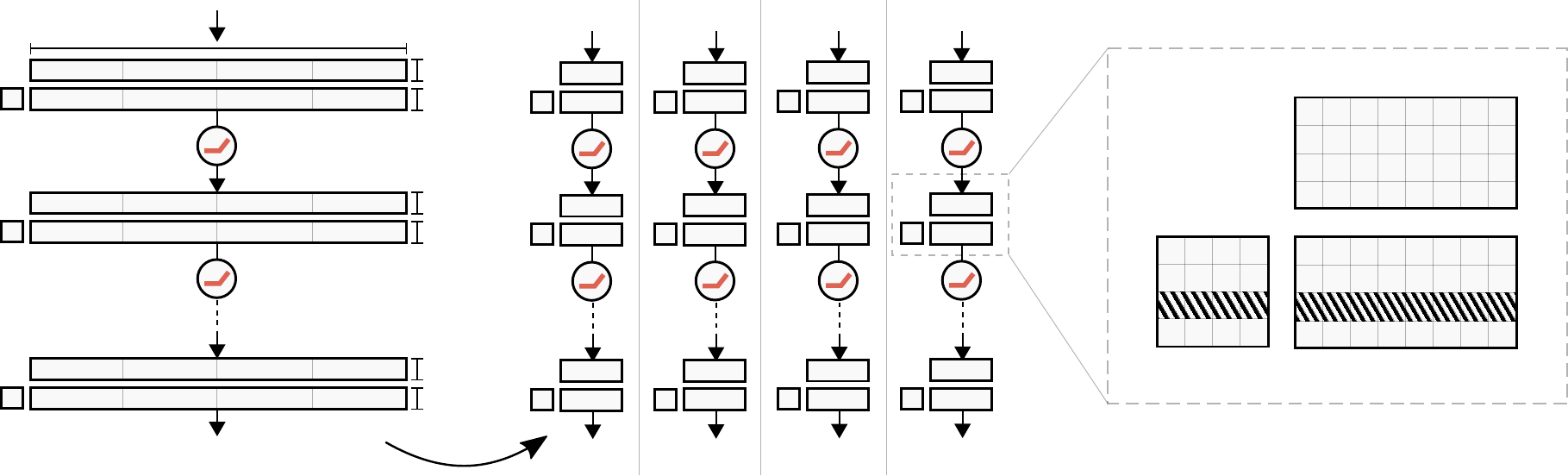}
        \put(0, -3) { \small \textbf{(a)} Batched neural network evaluation }
        \put(11.5, 30.3) { \small Input }
        \put(11, 1) { \small Output }
        \put(18, 28) { \footnotesize Batch size $\batchSize$ }
        \put(15, 20.5) { \footnotesize ReLU }
        \put(26.8, 25.5) { \footnotesize $\inputWidth$ }
        \put(26.8, 23.5) { \footnotesize $\networkWidth$ }
        \put(26.8, 17) { \footnotesize $\networkWidth$ }
        \put(26.8, 15) { \footnotesize $\networkWidth$ }
        \put(26.8, 6.3) { \footnotesize $\networkWidth$ }
        \put(26.8, 4.3) { \footnotesize $\outputWidth$ }
        \put(14.5, 9.5) { \footnotesize $\times\nlayers$ layers }
        \put(33, -3) { \small \textbf{(b)} Distribution of a batch over thread blocks }
        \put(70, -3) { \small \textbf{(c)} Per-thread-block matrix multiplication }
        \put(88.5, 25) { $H_i$ }
        \put(76.1, 6.3) { $W_i$ }
        \put(88, 6.3) { $H_{i+1}'$ }
        \put(79.5, 2) { $H_{i+1}' = W_i \cdot H_i$ }
    \end{overpic}
    \vspace{5mm}
    \caption{
        \textbf{(a)} Evaluating a multi-layer perceptron (MLP) for a large batch of inputs (e.g.\ ${N \approx 2^{21}}$ for a ${1920\times1080}$ frame) amounts to alternating weight-matrix multiplication and element-wise application of the activation function.
        \textbf{(b)} In our fully fused MLP, we parallelize this workload by partitioning the batch into $128$ element wide chunks that are each processed by their own thread block.
        Since our MLP is narrow (${\networkWidth = \inputWidth = 64}$ neurons wide), its weight matrices fit into registers and the intermediate ${64\times128}$ neuron activations fit into shared memory. This is key to the superior performance of the fully fused approach.
        \textbf{(c)} The matrix multiplication performed by each thread block transforms the $i$-th layer $H_i$ into the pre-activated next layer $H_{i+1}'$. It is diced into blocks of ${16\times16}$ elements to match the size of our hardware-accelerated half-precision matrix multiplier (TensorCore).
        Each warp of the thread block computes one ${16\times 128}$ block-row of $H_{i+1}'$ (e.g.\ the striped area) by first loading the corresponding ${16\times 64}$ striped weights from $W_i$ into registers and subsequently multiplying them by all ${64\times 16}$ block-columns of $H_i$.
        Thus, each thread block loads the weight matrix from global memory exactly once (the least possible amount), making multiple passes only over $H_i$ which, however, is located in fast shared memory.
    }\label{fig:fully-fused-illustration}
    \vspace{-2mm}
\end{figure*}

\begin{figure}
    \hspace*{-3mm}\includegraphics[width=1.05\columnwidth]{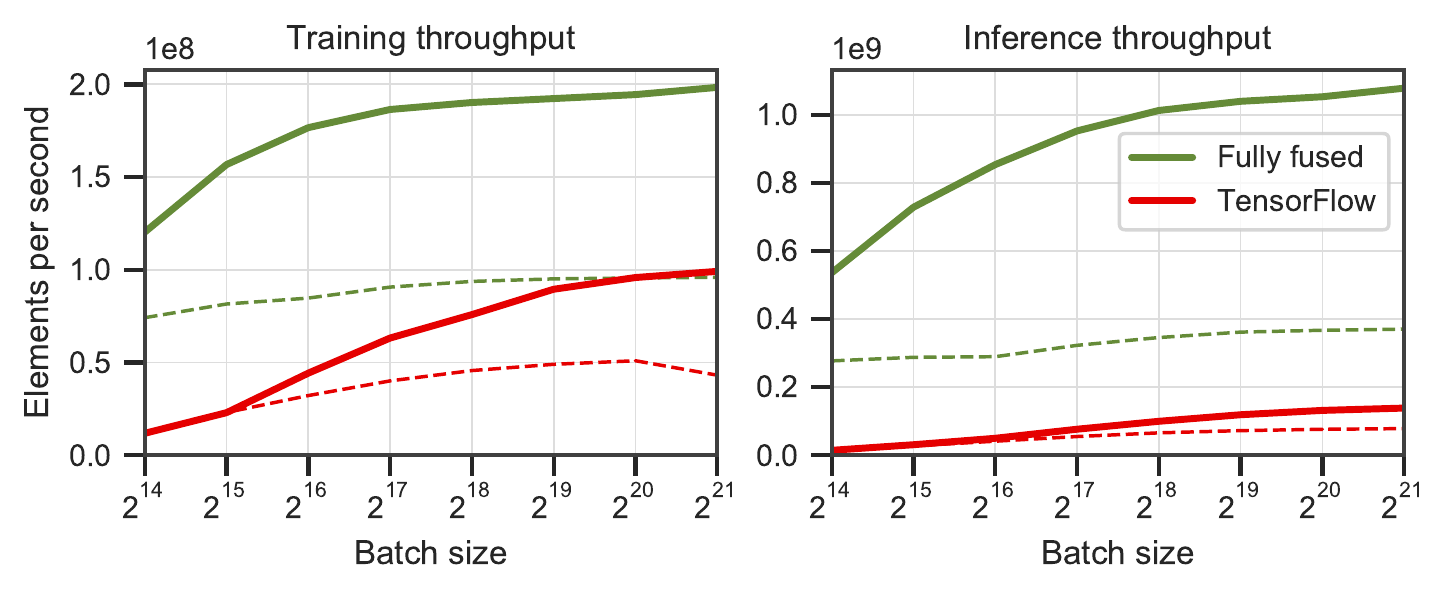}
    \vspace{-6mm}
    \caption{
        Our fully fused neural network outperforms an equivalent XLA-enabled TensorFlow (v2.5.0) implementation.
        Both implementations utilize half precision floating point numbers and TensorCore hardware for matrix multiplication.
        We compare the throughput of training (left) and inference (right) for a 64 (solid line) and a 128 (dashed line) neurons wide multi-layer perceptron.
        The relevant batch sizes for our goal of neural radiance caching are \emph{small} training batches (e.g.\ $2^{14}$ elements) and \emph{large} inference batches (e.g.\ $2^{21}$ elements for evaluating a ${1920 \times 1080}$ frame).
        For these batch sizes, the speed-up over TensorFlow ranges from $5\times$ to $10\times$.
    }\label{fig:us_vs_tensorflow}
    \vspace{-4mm}
\end{figure}

\subsection{Input Encoding}%
\label{sec:encoding}

\citet{Ren:2013:global} showed that solely using the spatio-directional coordinates $(\pos,\diro)$ of the scattered radiance as input to a neural network does not allow it to represent radiance well.
Therefore, the input is augmented by additional parameters that correlate with the scattered radiance:\ the surface normal $\normal$, the surface roughness ${\roughness}$, the diffuse reflectance $\albedo$, and the specular reflectance
$\specAlbedo$.
Being able to exploit such correlations, the neural approximation becomes much more accurate.

It is easier for the network to identify these correlations when they are (nearly) linear.
This is already the case for the diffuse and specular reflectances; we thus input them to the network as-is.
However, the quantities $\pos$, $\diro$, $\normal$, and $\roughness$ have a highly non-linear relation to the scattered radiance.
For these quantities, a well-chosen encoding to a higher-dimensional space can make the relation more linear and thereby make the neural approximation more accurate.\footnote{This is analogous to the ``kernel trick'' that is often employed in machine learning
to make the data linearly separable~\citep{patternrecognition}.}

The extra dimensions do not come for free, as they increase the required memory traffic as well as the cost of the first layer of the neural network.
We thus aim at encoding the quantities $\pos$, $\diro$, $\normal$, and $\roughness$ using \emph{as few as possible} extra dimensions while still profiting from the linearization.

To this end, the one-blob encoding~\citep{mueller2019nis} works well when the scale of the nonlinearities is about the same order of magnitude as the size of the blobs.
This is a good fit for $\diro$, $\normal$, and $\roughness$ as tiny variations in these parameters typically do not change the scattered radiance much.
We thus encode them using a very small number (e.g.\ ${k=4}$) of evenly spaced blobs.

However, tiny changes in the position $\pos$ can cause large variation in the scattered radiance, e.g.\ along shadow and geometric boundaries or in outdoor environments that are much larger than the view frustum.
One-blob encoding is therefore unsuitable for robustly encoding the position within just a few extra dimensions.
Instead, we adopt the frequency encoding from transformer networks~\citep{vaswani2017attention}, introduced to radiance learning by \citet{mildenhall2020nerf}, that leverages a geometric hierarchy of periodic functions to represent a high dynamic range of values in few encoded dimensions.
We use $12$ sine functions, each with frequency $2^d, d \in \{ 0, \ldots, 11 \}$.
To save on dimensions, we found that omitting the cosine terms of the original method does not compromise approximation quality.

In summary, the input of our neural network is a concatenation of the following:
the frequency-encoded position $\pos$, the one-blob encoded parameters $\diro$, $\normal$, and $\roughness$, and the raw diffuse albedo $\albedo$ and specular reflectance $\specAlbedo$; see \autoref{tab:parameters} for a detailed breakdown.
This results in a total of $62$ input dimensions to the neural network, which we pad to $64$ for compatibility with the hardware matrix-multiplication accelerator.
We pad with a value of $1$, which allows the network to implicitly learn a bias term (the corresponding columns of the first weight matrix) even though our architecture lacks explicit biases.

\section{Fully Fused Neural Networks}%
\label{sec:neuralnetwork}

We implemented our neural network from scratch in a GPU programming language in order to take full advantage of the GPU memory hierarchy.
In \autoref{fig:us_vs_tensorflow}, we compare the performance of this implementation to TensorFlow (v2.5.0)~\citep{tensorflow2015-whitepaper}, which we outperform by almost an order of magnitude.

To understand where this dramatic speedup comes from, we examine the bottleneck of evaluating a fully connected neural network like ours.
The \emph{computational} cost of such a neural network scales quadratically with its width, whereas its \emph{memory traffic} scales linearly.
Modern GPUs have vastly larger computational throughput than they have memory bandwidth, though, meaning that for \emph{narrow} neural networks like ours, the linear memory traffic is the bottleneck.
The key to improving performance is thus to minimize traffic to slow ``global'' memory (VRAM and high-level caches) and to fully utilize fast on-chip memory (low-level caches, ``shared'' memory, and registers).

Our fully fused approach does precisely this: we implement the \emph{entire} neural network as a single GPU kernel that is designed such that the only slow global memory accesses are reading and writing the network inputs and outputs.
Furthermore, implementing the kernel from scratch as opposed to building it out of existing frameworks allows us to specifically tailor the implementation to the network architecture and the GPU that we use.

\autoref{fig:fully-fused-illustration} illustrates how the fully fused approach is mapped to the memory hierarchy.
Using CUDA terminology: a given batch of input vectors is partitioned into block-column segments that are processed by a single thread block each (\autoref{fig:fully-fused-illustration}(b)).
The thread blocks independently evaluate the network by alternating between weight-matrix multiplication and element-wise application of the activation function.
By making the thread blocks small enough such that all intermediate neuron activations fit into on-chip shared memory, traffic to slow global memory is minimized.
This is the key advantage of the fully fused approach and stands in contrast to typical implementations of general matrix multiplication.

Within a matrix multiplication (\autoref{fig:fully-fused-illustration}(c)), each warp of the thread block computes the matrix product of a single block-row (striped area).
In our case, the striped weights in $\Params_i$ are few enough to fit into the registers of the warp and can thus be re-used for every block of $H_{i+1}'$ that the warp computes, yielding an additional performance gain.
Furthermore, since each warp loads a distinct block-row of the weight matrix, the entire thread block loads the weight matrix from global memory exactly once, which cannot be reduced further.

The only possible remaining reduction of global memory traffic is thus to minimize the \emph{number} of thread blocks by making them as large as fits into shared memory.
On our hardware (NVIDIA RTX 3090) and with our $64$-neurons-wide network, this sweet-spot is met when each thread block processes $128$ elements of the batch.
Each thread block thus computes matrix products of a $64\times64$ weight matrix with a $64\times128$ chunk of the data.

\begin{figure}
    \vspace{-2mm}
    \small
    \setlength{\tabcolsep}{1.0pt}%
    \renewcommand{\arraystretch}{0.8}%
    \setlength{\fboxrule}{1.5pt}%
    \vspace{0.2mm}
    \centering
    Visualization of factored neural radiance cache at primary vertex
    \begin{overpic}[width=1.0\columnwidth,trim={0px 50px 0px 200px},clip]{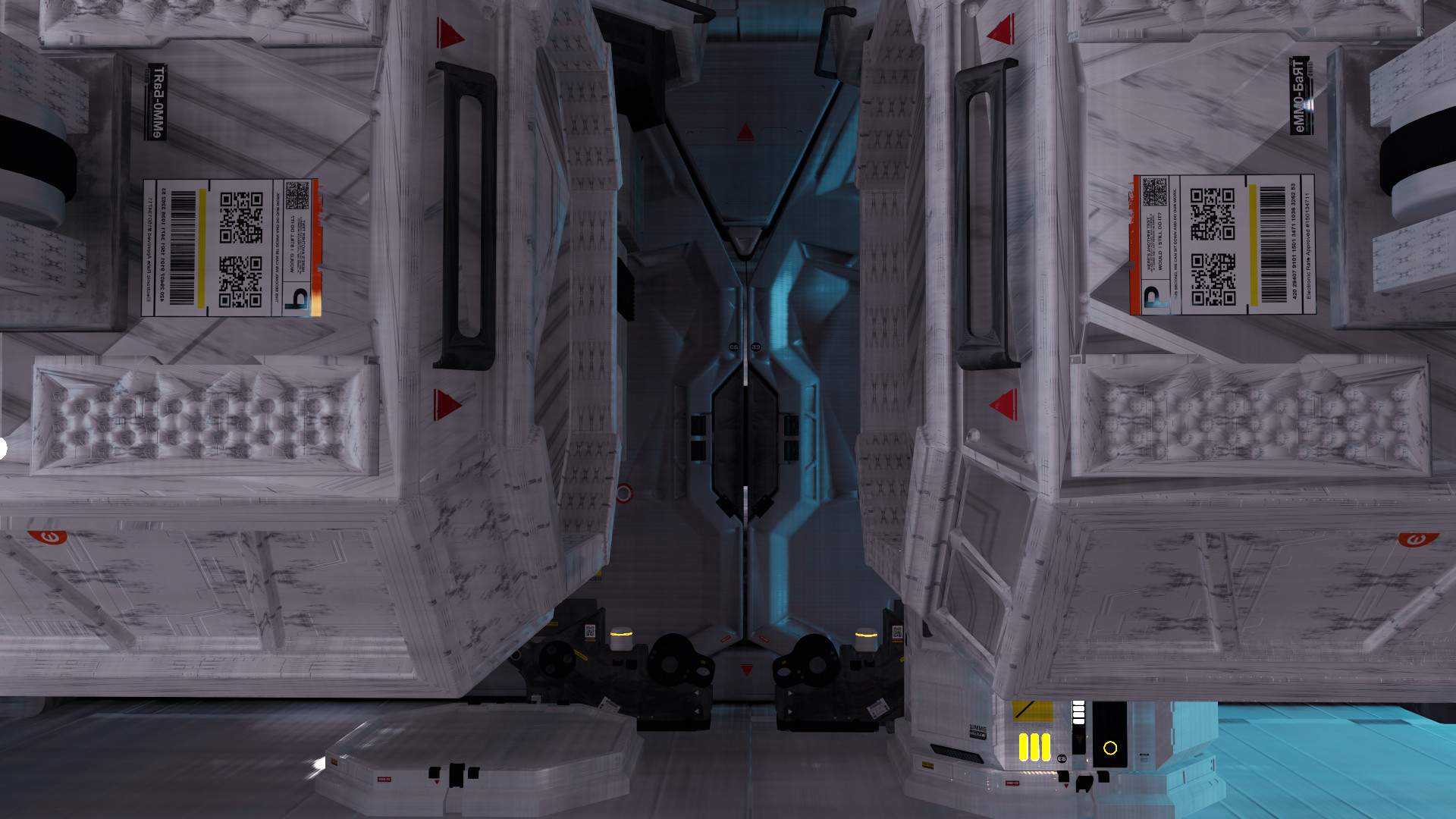}
        \put(84,36.7) { \makebox(0,0){\tikz\draw[red,ultra thick] (0,0) rectangle (0.18\linewidth, 0.125\linewidth);} }
    \end{overpic}\\
    \begin{tabularx}{\columnwidth}{cccc}%
        \hspace{-1.0pt}\fcolorbox{red}{red}{\includegraphics[width=0.232\columnwidth,trim={1450px 630px 120px 190px},clip]{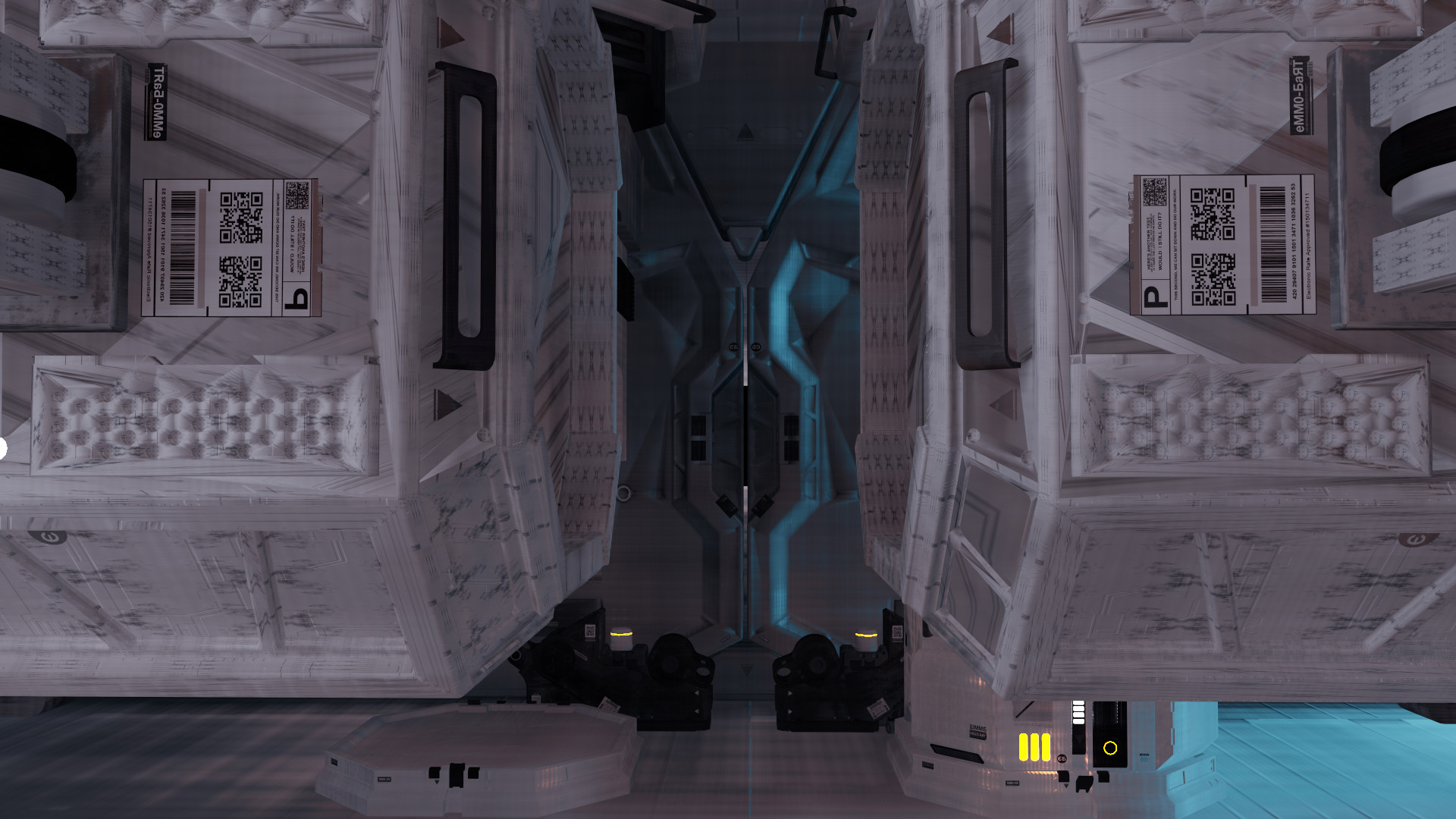}} &
        \fcolorbox{red}{red}{\includegraphics[width=0.232\columnwidth,trim={1450px 630px 120px 190px},clip]{images/factorization/measure_seven/enabled.jpg}} &
        \fcolorbox{red}{red}{\includegraphics[width=0.232\columnwidth,trim={1450px 630px 120px 190px},clip]{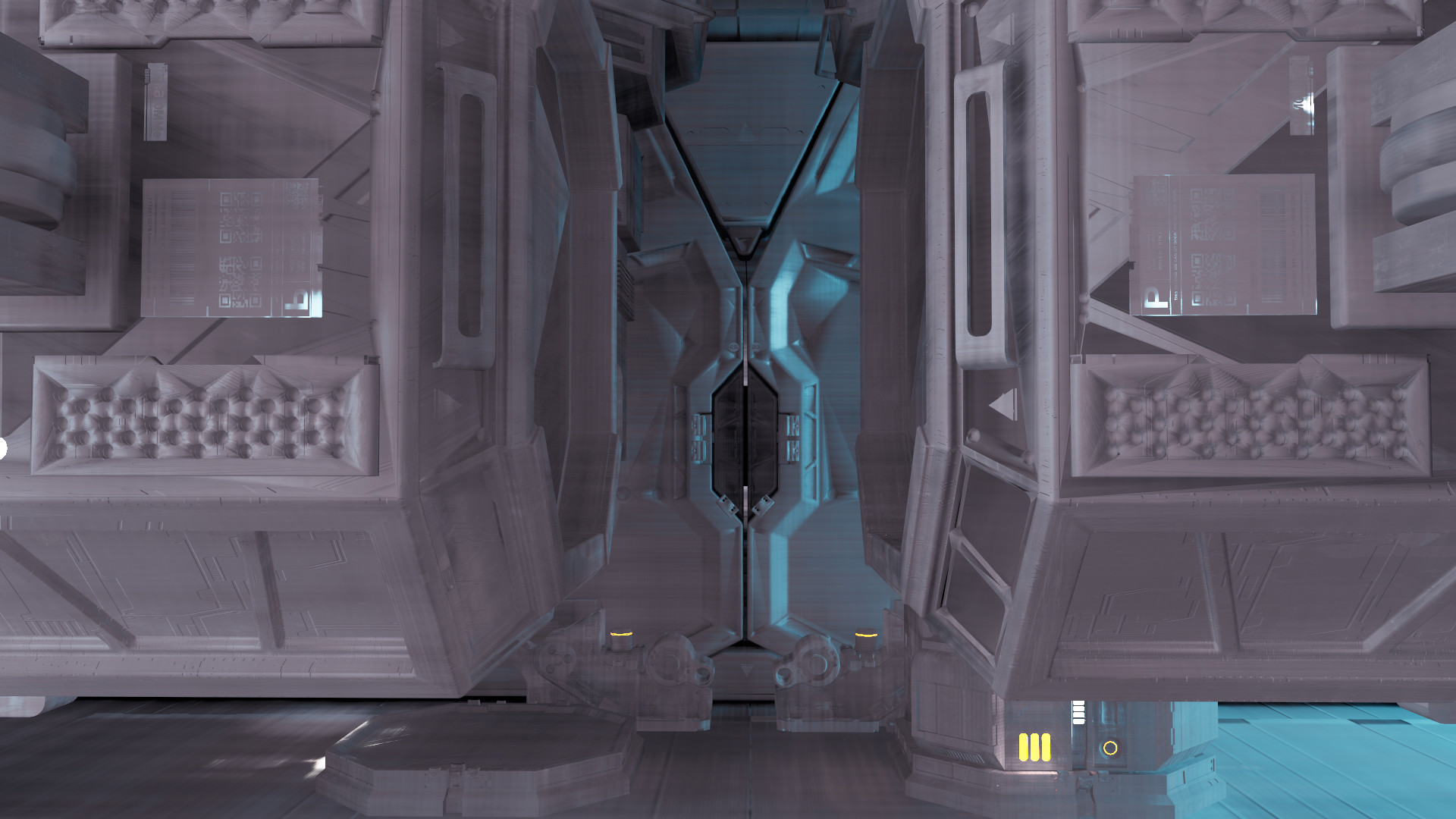}} &
        \fcolorbox{red}{red}{\includegraphics[width=0.232\columnwidth,trim={1450px 630px 120px 190px},clip]{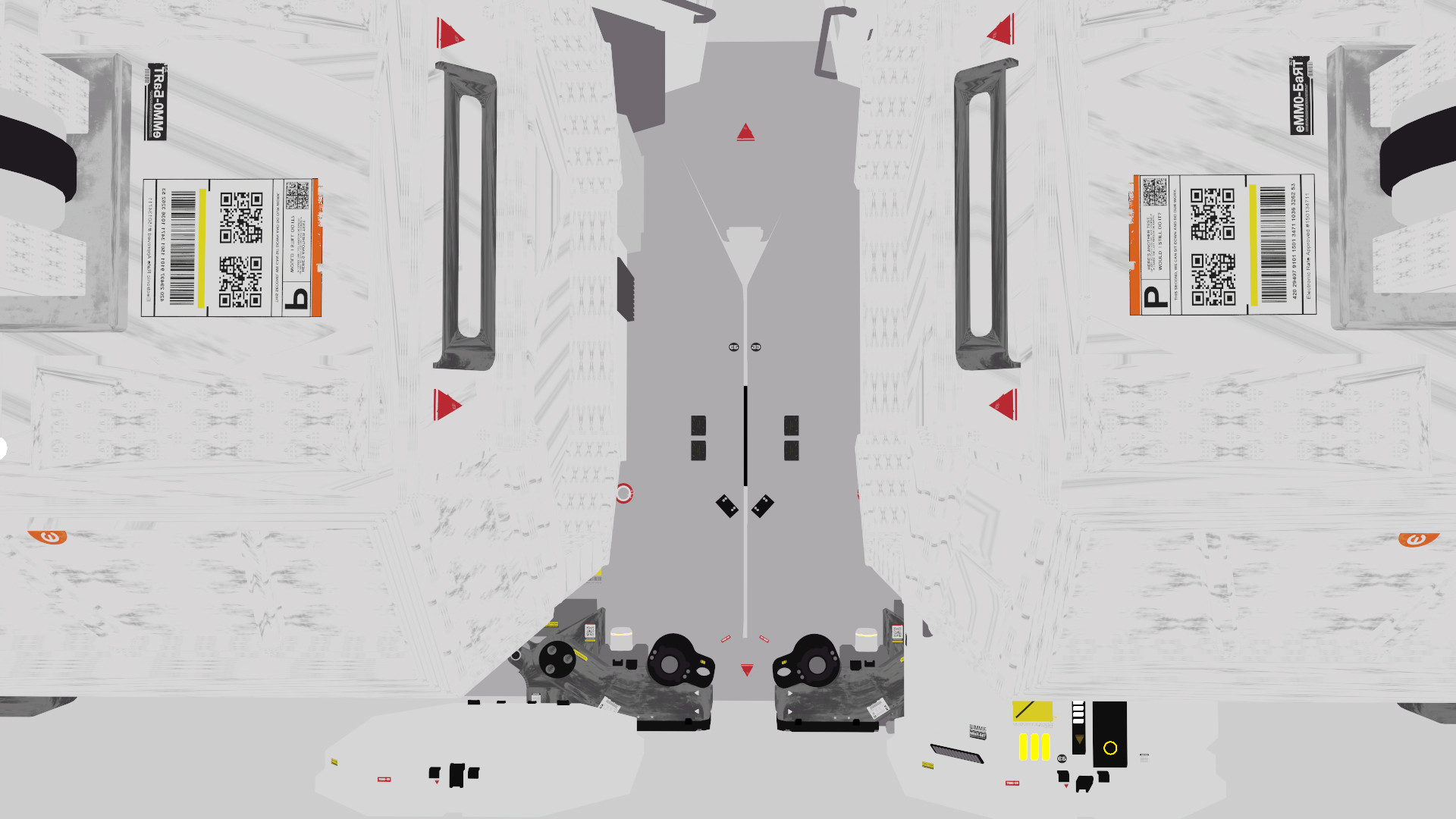}} \\[0.6mm]
        Radiance cache & \multicolumn{3}{c}{Radiance cache $\,\,\,\,=\,\,\,\,$ Prediction $\,\,\,\,\,\,\times\,\,\,\,\,\,$ Reflectance $\,\,\,$} \\
        \cmidrule(lr){1-1}
        \cmidrule(lr){2-4}
        Direct prediction & \multicolumn{3}{c}{Factorization}
    \end{tabularx}
    \vspace{-1mm}
    \caption{\label{fig:factorization}
        Reflectance factorization leads to more accurate textured colors and lets the neural prediction focus on complementary detail such as glossy highlights.
        Note that sharp texture detail is also present when predicting the product, because the reflectance is input to the network in any case.
    }
    \vspace{-2mm}
\end{figure}

\paragraph{Training the fully fused neural network}

For training, the forward and backward passes admit the same matrix multiplication structure as the previously discussed inference pass.
However, they require additional global-memory traffic, because intermediate activations and their gradients must be written out for backpropagation.
Furthermore, additional matrix multiplications are necessary to turn the results of backpropagation into the gradients of the weight matrices.
We compute these additional matrix multiplications using the general matrix multiplication (GEMM) routines of the CUTLASS template library (in split-k mode) as we were unable to produce a faster implementation ourselves.

All these additional operations make training slower than inference by a factor of roughly $4\times$--$5\times$; see \autoref{fig:us_vs_tensorflow}.

\begin{figure}
    \vspace{-3mm}
    \hspace*{-3mm}\begin{overpic}[width=1.05\columnwidth]{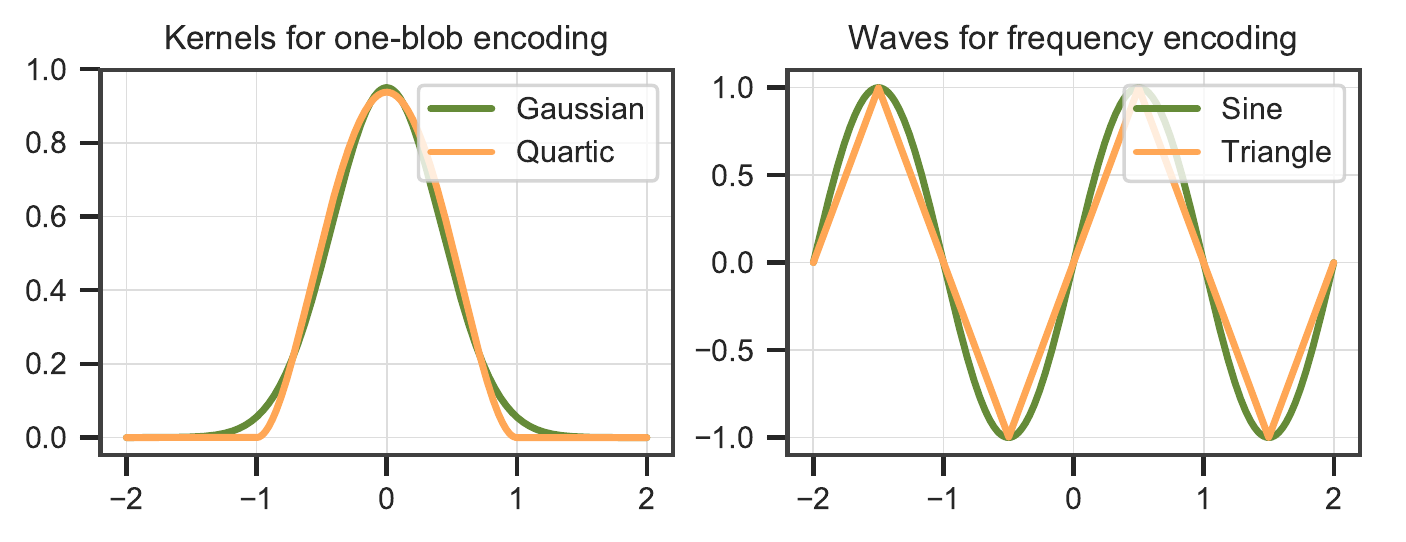}
        \put(9, -3.5) { $\mathrm{quartic}(x) := \frac{15}{16} (1 - x^2)^2$ }
        \put(54, -3.5) { $\mathrm{tri}(x) := 2\,| x\bmod 2 - 1 | - 1$ }
    \end{overpic}
    \vspace{-1mm}
    \caption{\label{fig:cheap_primitives}
        To avoid expensive mathematical operations, we replace the Gaussian kernel of the one-blob encoding by a quartic kernel and the sine function in the frequency encoding by a triangle wave.
        With these replacements, the cost per frame is reduced by $0.25$ms with no visible loss of quality.
    }
    \vspace{-2mm}
\end{figure}

\section{Practical Considerations}\label{sec:practical}

\paragraph{Architecture}
Our fully fused neural network architecture (see \autoref{fig:fully-fused-illustration}) comprises of seven fully connected layers.
The five hidden layers have 64 neurons each with ReLU activation functions.
The output layer reduces the 64 dimensions to three RGB values.
None of the layers has a bias vector, as biases did not result in any measurable quality benefit and omitting them makes the fully fused implementation simpler and more efficient.
Note that the neural network is shallow enough for vanishing gradients not to be a problem.
Hence there is no need for residual layers using skip links to help training, which we confirmed experimentally.

\begin{figure*}
    \vspace{-2.5mm}
    \setlength{\fboxrule}{10pt}%
\setlength{\insetvsep}{20pt}%
\setlength{\tabcolsep}{-0.1pt}%
\renewcommand{\arraystretch}{1}%
\footnotesize%
\hspace*{0.5mm}\begin{tabular}{lcccccc}
    &Reference
    &Path tracing
    &PT+NRC (Ours)
    &PT+ReSTIR
    &PT+ReSTIR+NRC (Ours)
    &Reference
    \\%
    \setInset{A}{red}{700}{200}{244}{120}%
    \setInset{B}{orange}{1300}{795}{326}{160}%
    \hspace{-4mm}\rotatebox{90}{\hspace{-17mm}\ZeroDay}%
    &\addBeautyCrop{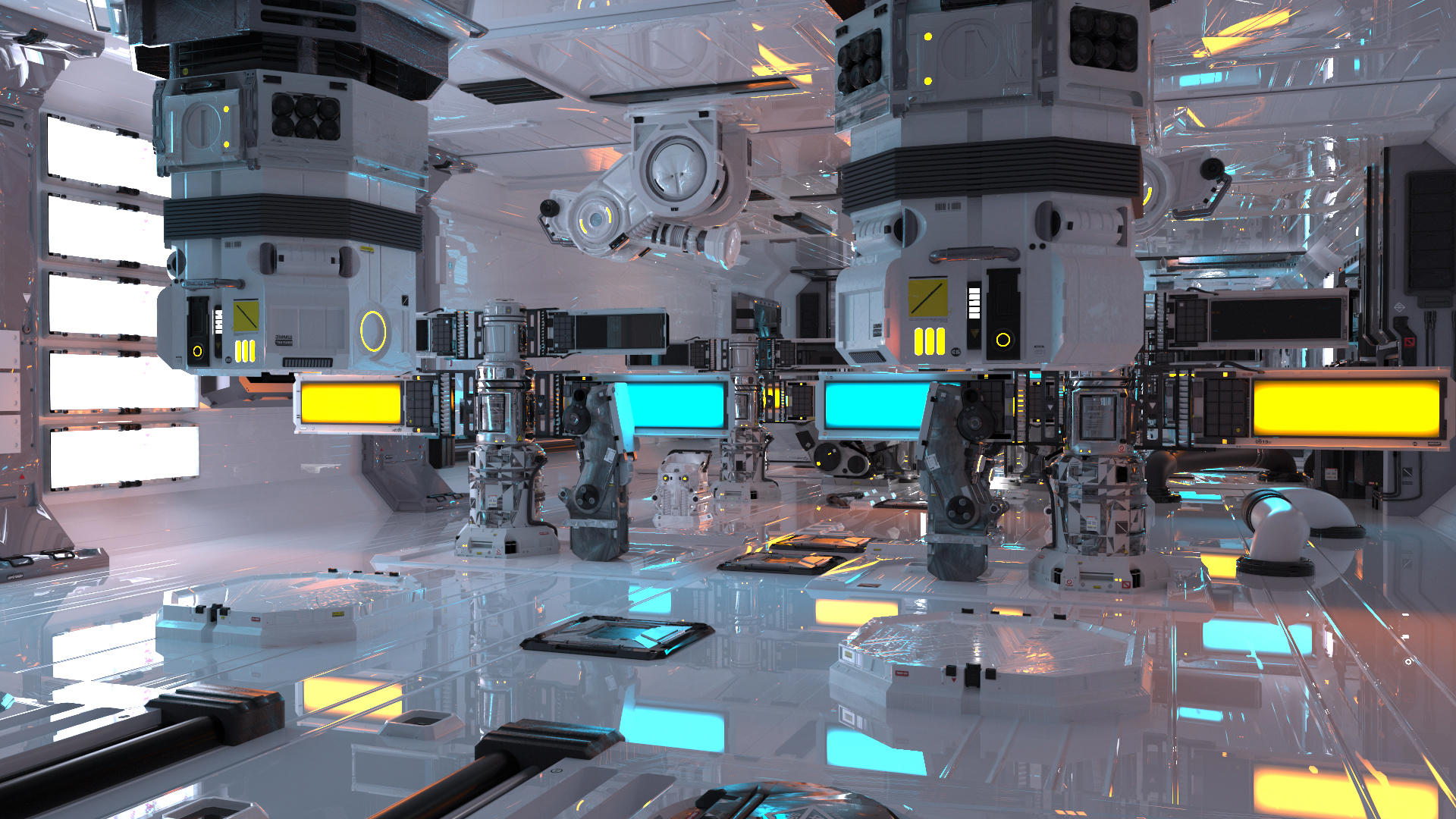}{0.263}{1920}{1080}{0}{0}{1920}{1080}%
    &\addInsets{figures/fig_main_result/measure_seven/pt.jpg}%
    &\addInsets{figures/fig_main_result/measure_seven/nrc.jpg}%
    &\addInsets{figures/fig_main_result/measure_seven/restir.jpg}%
    &\addInsets{figures/fig_main_result/measure_seven/restir+nrc.jpg}%
    &\addInsets{figures/fig_main_result/measure_seven/ref.jpg}%
    \\%
    & MRSE / Framerate:
    &20.52 / 96.6 fps%
    &5.07 / 126 fps%
    &17.79 / 70.8 fps%
    &\textbf{3.55} / 87.7 fps%
    &%
    \\%
    & \FLIP{}:
    &0.47%
    &0.30%
    &0.43%
    &\textbf{0.25}%
    &%
    \\%
    \setInset{A}{red}{300}{990}{163}{80}%
    \setInset{B}{orange}{1150}{640}{326}{160}%
    \hspace{-4mm}\rotatebox{90}{\hspace{-14mm}\Attic}%
    &\addBeautyCrop{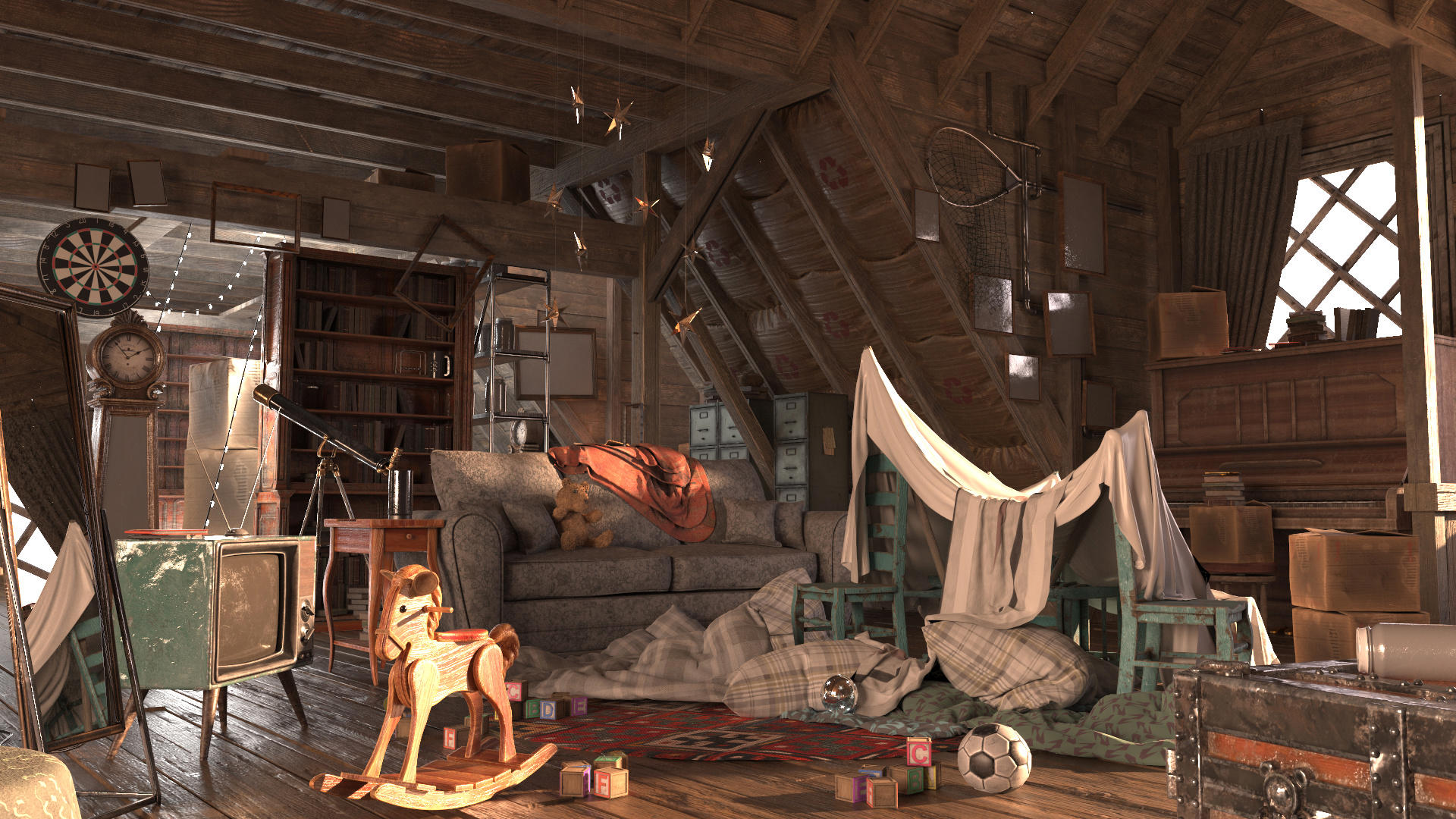}{0.263}{1920}{1080}{0}{0}{1920}{1080}%
    &\addInsets{figures/fig_main_result/attic/pt.jpg}%
    &\addInsets{figures/fig_main_result/attic/nrc.jpg}%
    &\addInsets{figures/fig_main_result/attic/restir.jpg}%
    &\addInsets{figures/fig_main_result/attic/restir+nrc.jpg}%
    &\addInsets{figures/fig_main_result/attic/ref.jpg}%
    \\%
    & MRSE / Framerate:
    &17.29 / 111 fps%
    &10.81 / 103 fps%
    &27.69 / 76.7 fps%
    &\textbf{3.01} / 72.6 fps%
    &%
    \\%
    & \FLIP{}:
    &0.46%
    &0.38%
    &0.28%
    &\textbf{0.21}%
    &%
    \\%
    \setInset{A}{red}{319}{531}{244}{120}%
    \setInset{B}{orange}{900}{650}{489}{240}%
    \hspace{-4mm}\rotatebox{90}{\hspace{-14mm}\BistroExterior}%
    &\addBeautyCrop{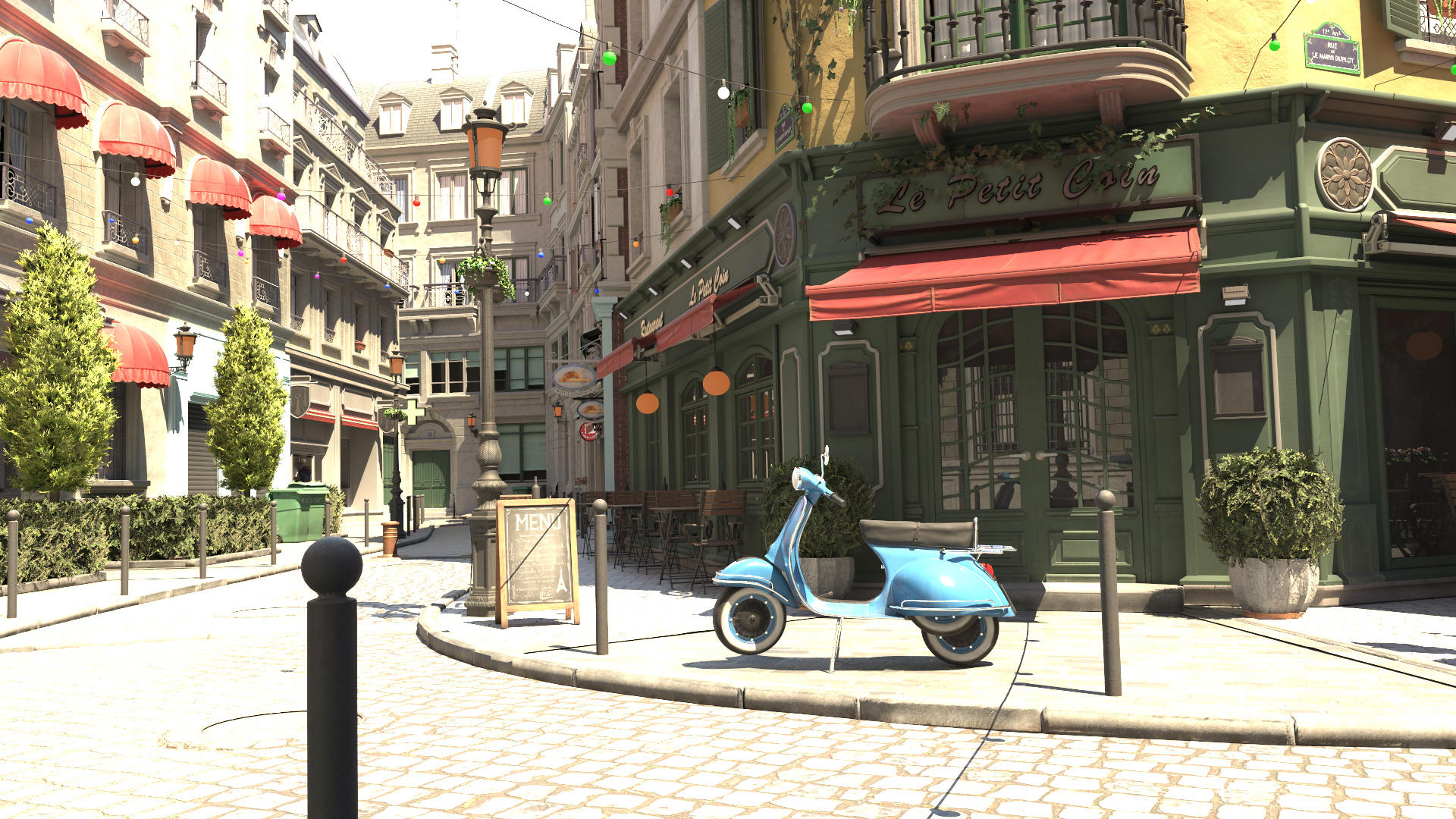}{0.263}{1920}{1080}{0}{0}{1920}{1080}%
    &\addInsets{figures/fig_main_result/exterior/pt.jpg}%
    &\addInsets{figures/fig_main_result/exterior/nrc.jpg}%
    &\addInsets{figures/fig_main_result/exterior/restir.jpg}%
    &\addInsets{figures/fig_main_result/exterior/restir+nrc.jpg}%
    &\addInsets{figures/fig_main_result/exterior/ref.jpg}%
    \\%
    & MRSE / Framerate:
    &11.71 / 105 fps%
    &3.43 / 98.8 fps%
    &10.99 / 73.3 fps%
    &\textbf{1.47} / 69.6 fps%
    &%
    \\%
    & \FLIP{}:
    &0.63%
    &0.52%
    &0.39%
    &\textbf{0.28}%
    &%
    \\%
    \setInset{A}{red}{50}{660}{163}{80}%
    \setInset{B}{orange}{1250}{600}{326}{160}%
    \hspace{-4mm}\rotatebox{90}{\hspace{-18mm}\Classroom}%
    &\addBeautyCrop{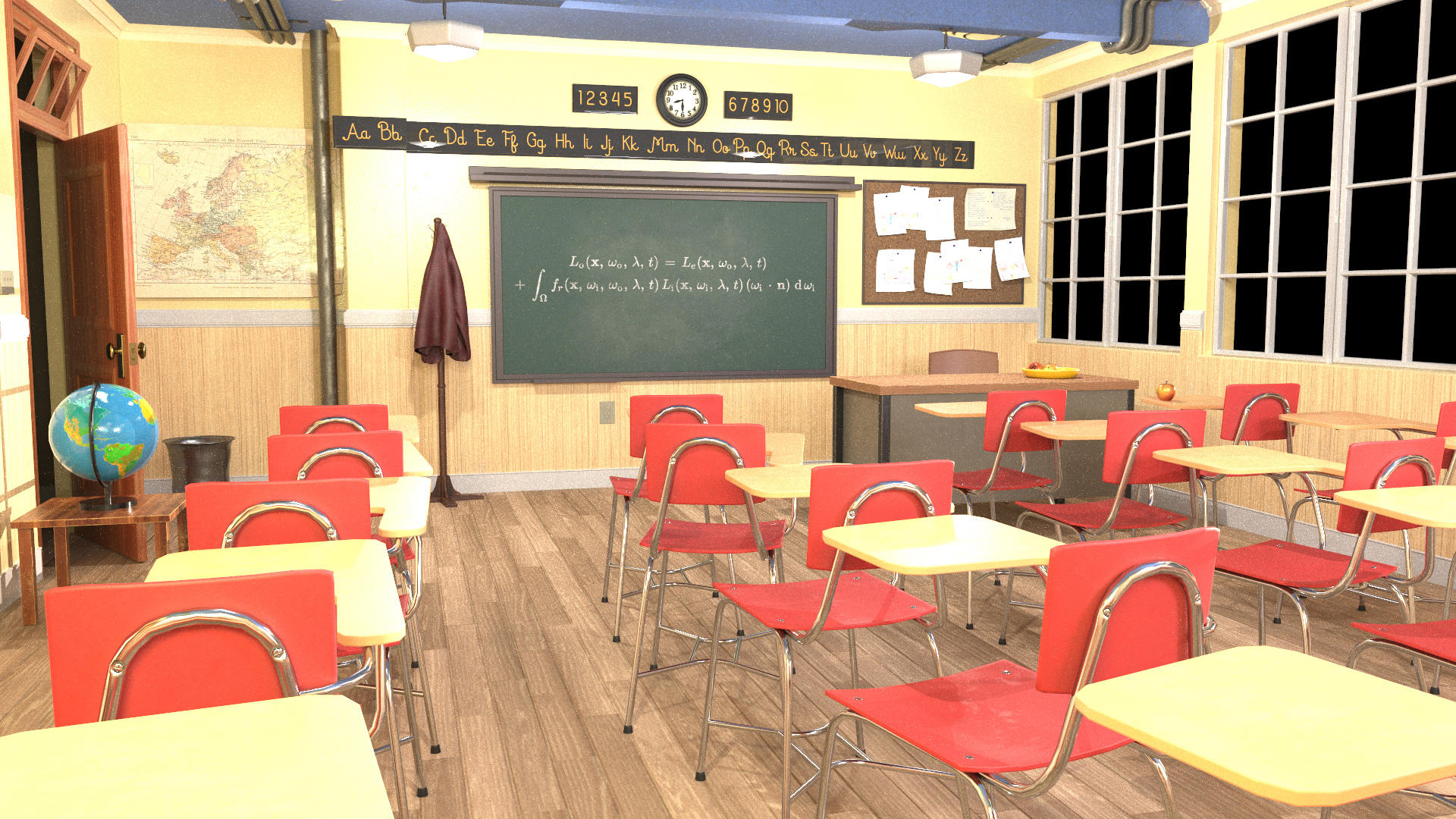}{0.263}{1920}{1080}{0}{0}{1920}{1080}%
    &\addInsets{figures/fig_main_result/classroom/pt.jpg}%
    &\addInsets{figures/fig_main_result/classroom/nrc.jpg}%
    &\addInsets{figures/fig_main_result/classroom/restir.jpg}%
    &\addInsets{figures/fig_main_result/classroom/restir+nrc.jpg}%
    &\addInsets{figures/fig_main_result/classroom/ref.jpg}%
    \\%
    & MRSE / Framerate:
    &380.97 / 70.7 fps%
    &\textbf{7.76} / 87.9 fps%
    &336.11 / 54.9 fps%
    &8.20 / 66.7 fps%
    &%
    \\%
    & \FLIP{}:
    &0.60%
    &0.47%
    &0.41%
    &\textbf{0.26}%
    &%
    \\%
    \setInset{A}{red}{1000}{460}{326}{160}%
    \setInset{B}{orange}{1220}{750}{244}{120}%
    \hspace{-4mm}\rotatebox{90}{\hspace{-18mm}\LivingRoom}%
    &\addBeautyCrop{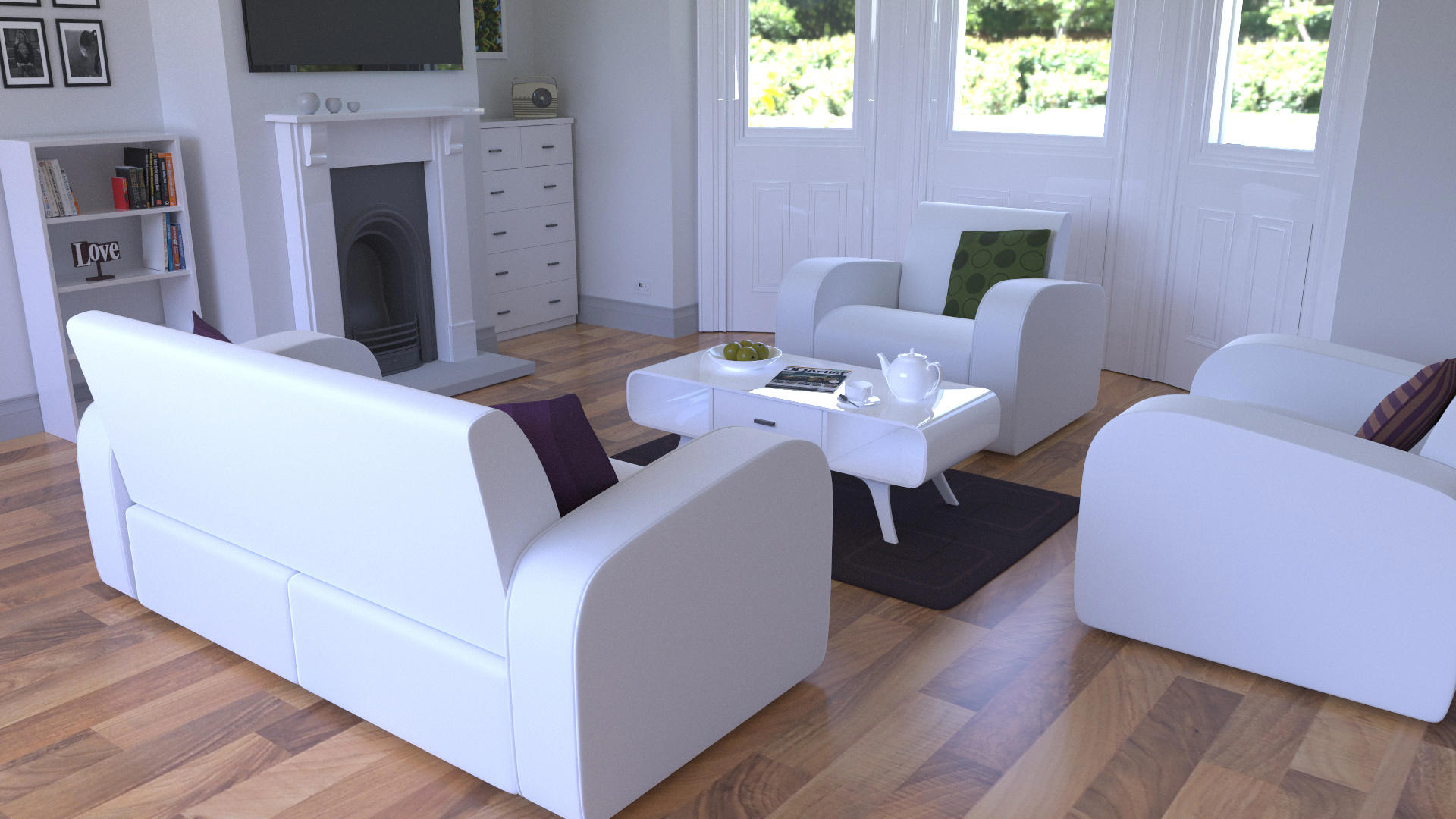}{0.263}{1920}{1080}{0}{0}{1920}{1080}%
    &\addInsets{figures/fig_main_result/living_room/pt.jpg}%
    &\addInsets{figures/fig_main_result/living_room/nrc.jpg}%
    &\addInsets{figures/fig_main_result/living_room/restir.jpg}%
    &\addInsets{figures/fig_main_result/living_room/restir+nrc.jpg}%
    &\addInsets{figures/fig_main_result/living_room/ref.jpg}%
    \\%
    & MRSE / Framerate:
    &52.82 / 186 fps%
    &5.67 / 172 fps%
    &62.04 / 121 fps%
    &\textbf{1.31} / 111 fps%
    &%
    \\%
    & \FLIP{}:
    &0.75%
    &0.44%
    &0.55%
    &\textbf{0.21}%
    &%
    \\%
    \setInset{A}{red}{1250}{50}{244}{120}%
    \setInset{B}{orange}{1350}{660}{326}{160}%
    \hspace{-4mm}\rotatebox{90}{\hspace{-17mm}\PinkRoom}%
    &\addBeautyCrop{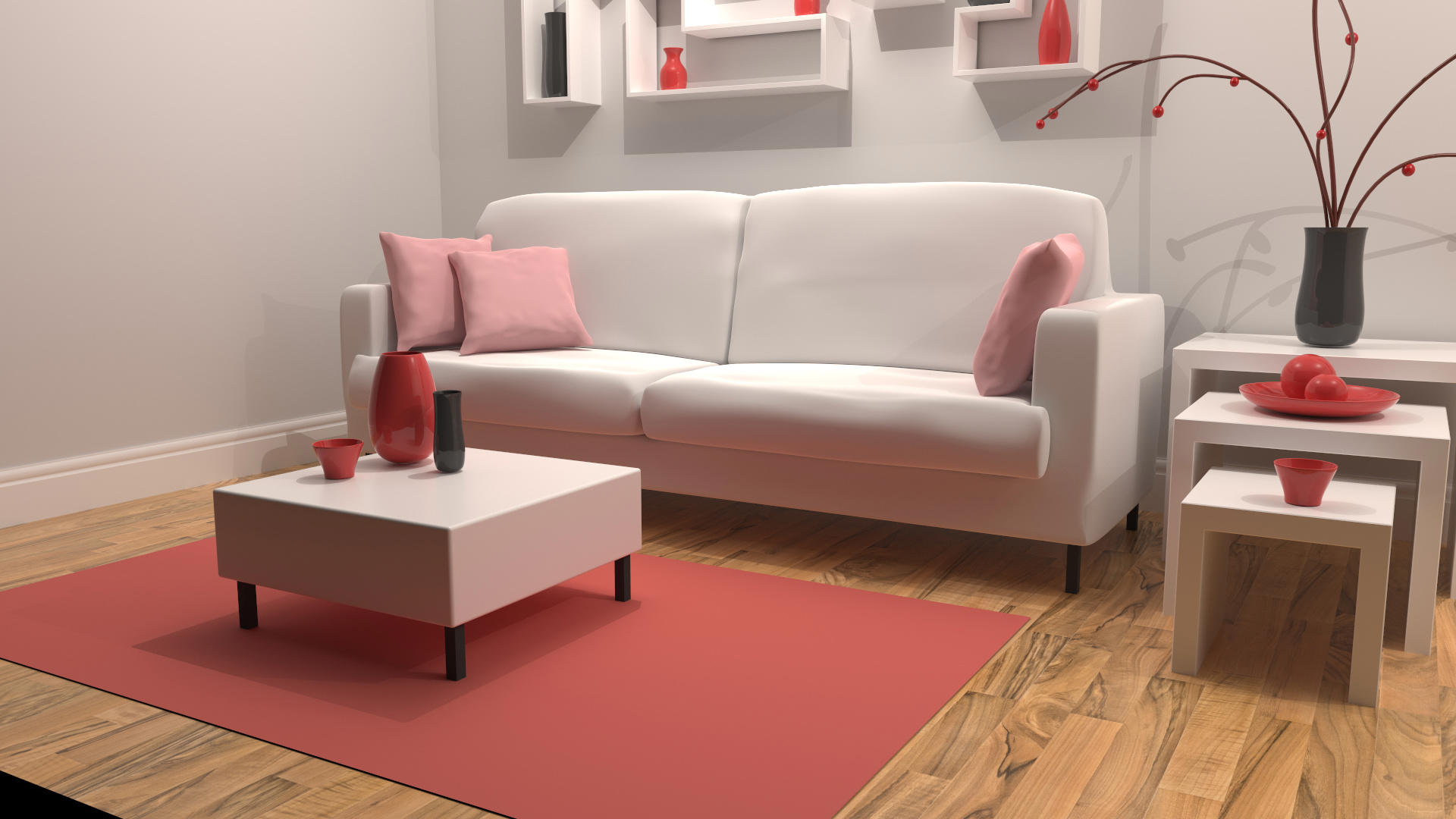}{0.263}{1920}{1080}{0}{0}{1920}{1080}%
    &\addInsets{figures/fig_main_result/pink_room/pt.jpg}%
    &\addInsets{figures/fig_main_result/pink_room/nrc.jpg}%
    &\addInsets{figures/fig_main_result/pink_room/restir.jpg}%
    &\addInsets{figures/fig_main_result/pink_room/restir+nrc.jpg}%
    &\addInsets{figures/fig_main_result/pink_room/ref.jpg}%
    \\%
    & MRSE / Framerate:
    &7.80 / 275 fps%
    &0.91 / 203 fps%
    &7.54 / 147 fps%
    &\textbf{0.77} / 124 fps%
    &%
    \\%
    & \FLIP{}:
    &0.47%
    &0.28%
    &0.47%
    &\textbf{0.27}%
    &%
    \\%
\end{tabular}

    \vspace{-2.5mm}
    \caption{\label{fig:main_result}
        We demonstrate the benefit of our neural radiance cache (NRC) in 1 spp real-time renderings with varied material and lighting complexity.
        From left to right: our baseline is unbiased path tracing with Russian roulette and next-event estimation driven by a light BVH~\citep{moreau19lightbvh}.
        Then, we add spatiotemporal reservoir resampling (ReSTIR)~\citep{bitterli20spatiotemporal} for low-variance direct and NRC for low-variance indirect illumination.\@
        Individually, these complementary techniques excel in their respective domains (e.g.\ ReSTIR in the directly lit \BistroExterior{} and NRC in the glossy \ZeroDay{} scene), but their combination unlocks the biggest improvement.
        Together, the three techniques reduce the mean relative squared error (MRSE) of path tracing by $1$--$2$ orders of magnitude while incurring a comparatively little performance loss thanks to the drastic path shortening of NRC.\@
        In all scenes, the combined technique exceeds 60 frames per second at a resolution of ${1920\times1080}$.
        We also report the perceptually based \protect\FLIP{} metric that is more robust to outliers (``fireflies'').
    }
\end{figure*}

\paragraph{Reflectance factorization}
To improve textured colors reproduction, we multiply the network output by the sum of the diffuse albedo and specular reflectance $\albedo(\pos,\diro) + \specAlbedo(\pos,\diro)$.
For Lambertian materials, this amounts to irradiance factorization~\citep{Ward:1988:caching}, and the network is effectively tasked with learning irradiance as opposed to reflected radiance.
However, even in our highly non-Lambertian scenes, the above factorization is helpful.
The factorization is \emph{not} necessary to recover sharp detail---$\albedo(\pos,\diro)$ and $\specAlbedo(\pos,\diro)$ are input to the network in any case---but it helps recover colors while letting the cache focus on complementary details; see \autoref{fig:factorization}.

\paragraph{High-performance primitives for encoding}
The one-blob and frequency encodings rely on primitives that are computationally expensive: Gaussian kernels and trigonometric functions.
We thus replace the primitives with approximations that are far cheaper to evaluate.
Specifically, we replace the Gaussian with a quartic kernel and the sine function with a triangle wave as illustrated in \autoref{fig:cheap_primitives}, reducing the cost per frame by $0.25$ ms with no visible loss of quality.

\paragraph{Relative loss}
To facilitate effective training, we use the relative $\Loss^2$ loss that admits unbiased gradient estimates when the training signal---the reflected radiance $\reflectedRadiance(\pos, \diro)$---is noisy~\citep{Lehtinen:2018}.
The loss is normalized by the neural prediction:
\begin{align}
    \Loss^2 \big(\reflectedRadiance(\pos, \diro), \cachedReflectedRadiance(\pos, \diro; \Params_t)\big) := \frac{{\big(\reflectedRadiance(\pos, \diro) - \cachedReflectedRadiance(\pos, \diro; \Params_t)\big)}^2}{{\StopGradient\big(\cachedReflectedRadiance(\pos, \diro; \Params_t)\big)}^2+\epsilon} \,,
    \label{eq:relative-l2}
\end{align}
where $\epsilon = 0.01$ and $\StopGradient(\,\cdot\,)$ denotes that its argument is treated as a constant in the optimization, i.e.\ no gradient is propagated back.
Furthermore, for spectral values of $\reflectedRadiance(\pos, \diro)$, we normalize the loss of each color channel by the squared \emph{luminance} across the spectrum.

\paragraph{Optimizer}
The choice of optimizer is crucial to effectively leverage the little training data we have per frame.
To this end, we compared multiple first-order optimizers, i.e.\ stochastic gradient descent (SGD), Adam~\citep{KingmaB14}, and Novograd~\citep{novograd}) and found that Adam converges in the fewest iterations while having an overhead of practically zero.

We also investigated a second order optimizer, Shampoo~\citep{gupta2018shampoo,anil2020second}, which converges with slightly fewer iterations than Adam.
However, the $0.3$ milliseconds per-frame overhead of our optimized implementation did not justify its benefit in our experiments.
We thus use Adam in all results.

\section{Results and Discussion}%
\label{sec:results}

We implemented all components of our neural radiance cache in CUDA, i.e.\ input encoding, the fully fused network, and the optimizer, the source code of which we release publically~\citep{tiny-cuda-nn}.
The radiance cache is integrated into a path tracer implemented in Direct3D 12 using the Falcor rendering framework~\citep{Benty20} with which we generated all results in this paper.

All images were rendered at a resolution of ${1920\times1080}$ on a high-end desktop machine (i9 9900k and RTX 3090).
For each image, we report the mean relative squared error (MRSE)~\citep{Rousselle:2011} or its decomposition into relative bias (rBias) and variance (rVar) to aid the reader in gauging the improved Monte Carlo efficiency.
When applicable, we also list the perceptually based \FLIP{} metric~\citep{Andersson:2020:flip} that is more robust to outliers (``fireflies'').
Our reference images were created using \emph{ReSTIR-enabled} path tracing to ensure that we only measure the bias caused by radiance caching and not that of ReSTIR.\@

\paragraph{Real-time rendering}
In \autoref{fig:main_result}, we utilize neural radiance caching (NRC) to reduce indirect illumination noise of a path tracer.
By combining NRC with complementary direct-lighting techniques, we get global illumination with both low noise \emph{and} little bias in real-time.

Our baseline is an unbiased path tracer with Russian roulette and next-event estimation driven by a light BVH~\citep{moreau19lightbvh}, to which we add screen-space spatiotemporal reservoir resampling (ReSTIR)~\citep{bitterli20spatiotemporal}.
While this algorithm has low variance in its direct lighting estimates (PT+ReSTIR column), it suffers from noisy estimates of \emph{indirect} lighting, which can be remedied by terminating paths into our cache (PT+ReSTIR+NRC column).
Shortening the paths in this way not only results in lower variance but sometimes also in \emph{higher} framerates due to the low overhead of querying and training the fully fused network.

Compared with path tracing, the combination of ReSTIR and NRC reduces the MRSE by $1$--$2$ orders of magnitude while having a comparatively small impact on performance.
In all scenes, the combined technique exceeds 60 frames per second.

We measure our speedup in \autoref{tab:equal_quality} by letting the baseline converge to equal MRSE as a single rendered frame of our method.
The average speedup across the test scenes is $13.6\times$.

\begin{table}
    \caption{\label{tab:equal_quality}
        Time to converge to equal MRSE.
    }%
    \vspace{-2mm}
    \setlength{\tabcolsep}{3.5pt}%
\renewcommand{\arraystretch}{0.75}%
\small%
\hspace*{-1mm}\begin{tabularx}{\columnwidth}{cccrrc}%
    \toprule
    Scene & Method &Frames&Time&MRSE&Speedup\\
    \midrule
    \multirow{2}{*}{\Attic{}}
    & PT+ReSTIR &7&92.5 ms&2.739&\multirow{2}{*}{$\mathbf{6.6\times}$}\\
    & PT+ReSTIR+NRC &1&14.0 ms&2.727&\\
    \\[-1.5mm]
    \multirow{2}{*}{\BistroExterior{}}
    & PT+ReSTIR &8&110.7 ms&1.498&\multirow{2}{*}{$\mathbf{7.6\times}$}\\
    & PT+ReSTIR+NRC &1&14.6 ms&1.407&\\
    \\[-1.5mm]
    \multirow{2}{*}{\Classroom{}}
    & PT+ReSTIR &145&2625 ms&5.882&\multirow{2}{*}{$\mathbf{172.7\times}$}\\
    & PT+ReSTIR+NRC &1&15.2 ms&5.847\\
    \\[-1.5mm]
    \multirow{2}{*}{\LivingRoom{}}
    & PT+ReSTIR &53&431.5 ms&1.379&\multirow{2}{*}{$\mathbf{49.6\times}$}\\
    & PT+ReSTIR+NRC &1&8.7 ms&1.376&\\
    \\[-1.5mm]
    \multirow{2}{*}{\PinkRoom{}}
    & PT+ReSTIR &10&66.9 ms&0.769&\multirow{2}{*}{$\mathbf{8.4\times}$}\\
    & PT+ReSTIR+NRC &1&8.0 ms&0.765&\\
    \\[-1.5mm]
    \multirow{2}{*}{\ZeroDay{}}
    & PT+ReSTIR &5&69.4 ms&3.799&\multirow{2}{*}{$\mathbf{6.1\times}$}\\
    & PT+ReSTIR+NRC &1&11.3 ms&3.430&\\
    \midrule
    \multirow{2}{*}{Average}
        & PT+ReSTIR &16.6&154.1 ms&2.037&\multirow{2}{*}{$\mathbf{13.6\times}$}\\
        & PT+ReSTIR+NRC &1&11.3 ms&1.941&\\
    \bottomrule
\end{tabularx}

\end{table}

\begin{figure}
    \setlength{\tabcolsep}{1pt}%
\renewcommand{\arraystretch}{0.75}%
\small%
\hspace*{-1mm}\begin{tabular}{lccc}%
    &\multicolumn{2}{c}{PT+ReSTIR+NRC at 256 spp} & PT+ReSTIR\\
    \cmidrule(lr){2-3}
    \cmidrule(lr){4-4}
    & no Self-training & Self-training & Reference \\
    \hspace{-2.5mm}\rotatebox{90}{\hspace{7mm}\ZeroDay}%
    &\includegraphics[width=0.33\linewidth, trim=300 0 350 0, clip]{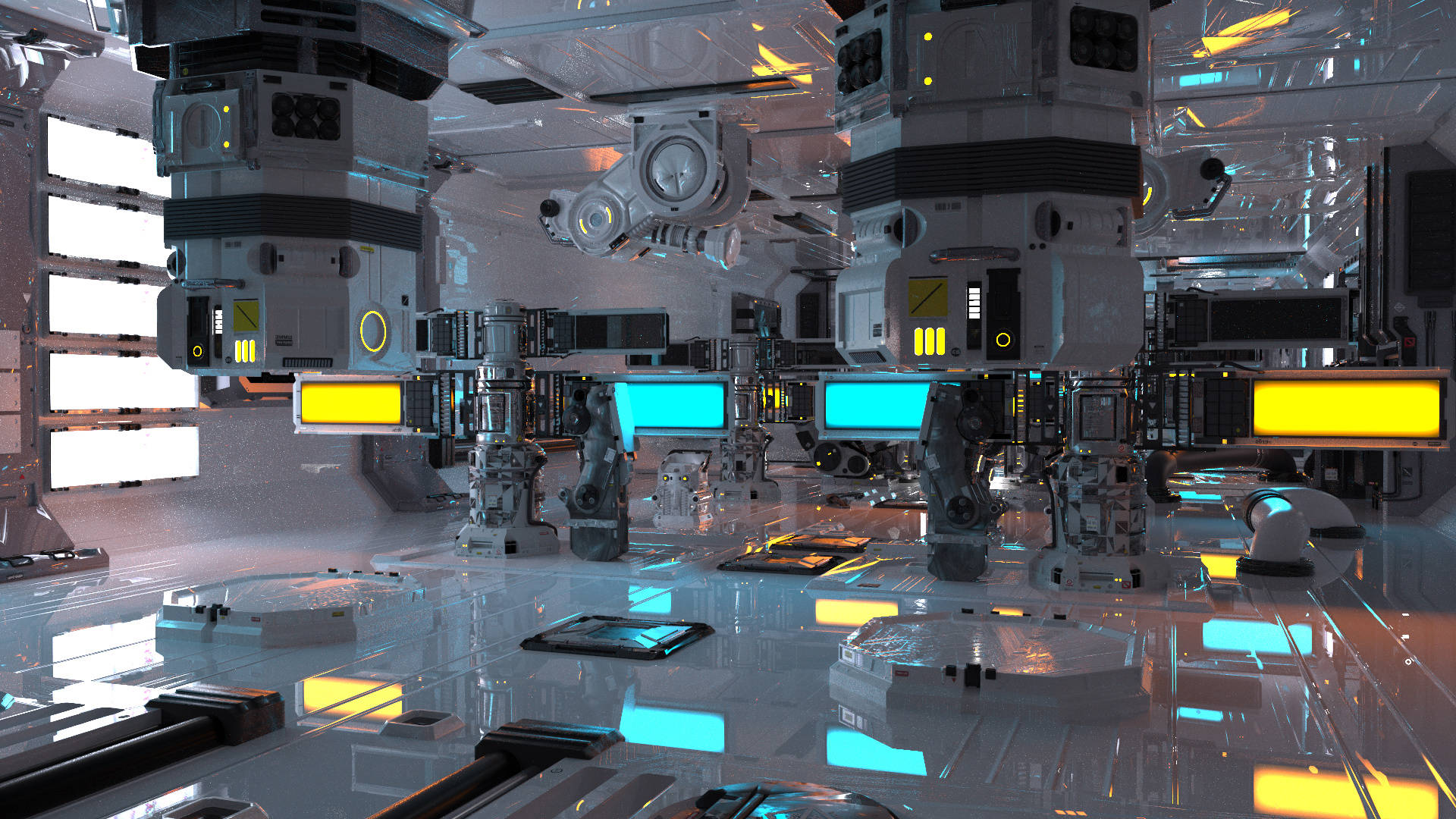}%
    &\includegraphics[width=0.33\linewidth, trim=300 0 350 0, clip]{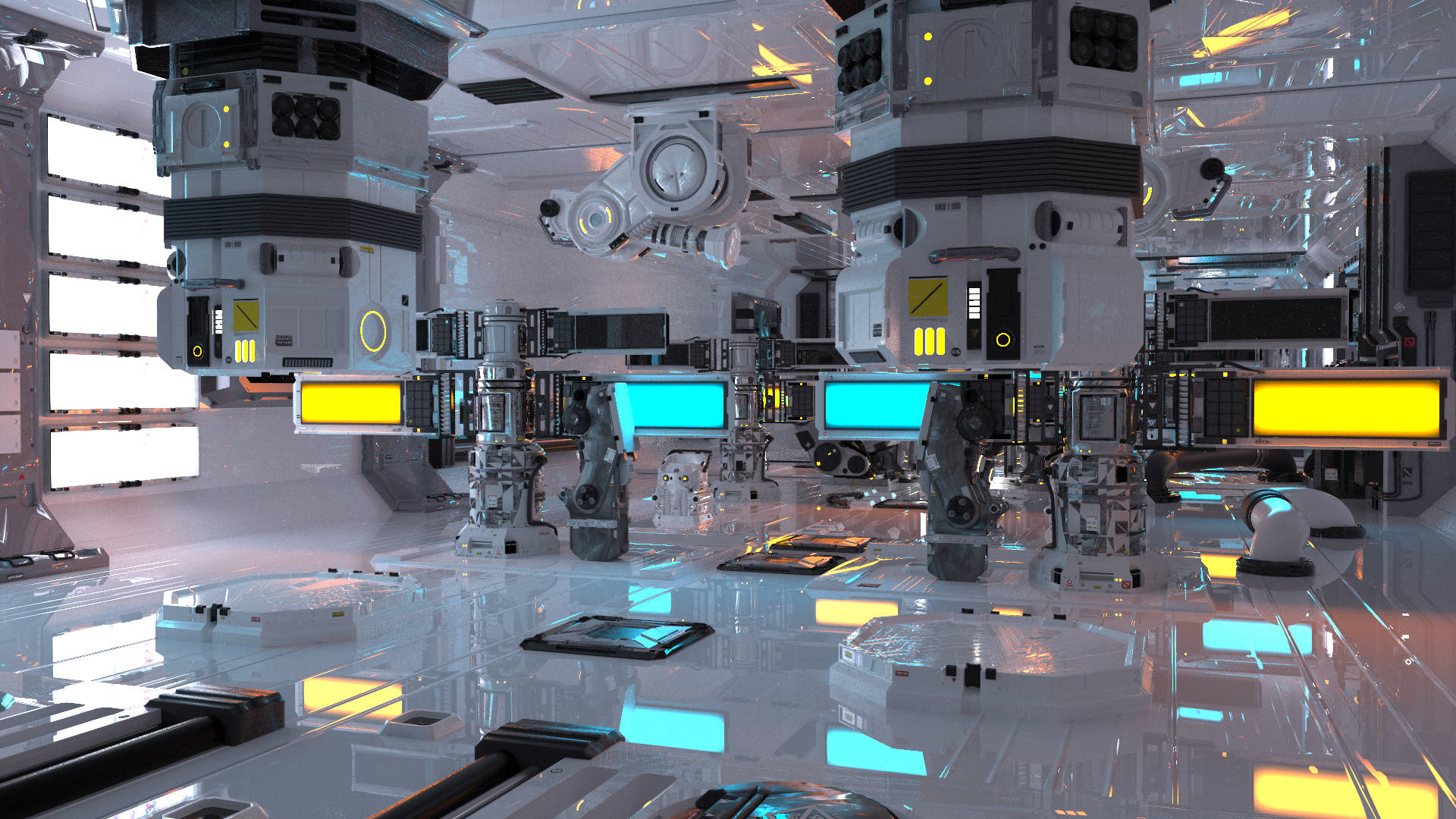}%
    &\includegraphics[width=0.33\linewidth, trim=300 0 350 0, clip]{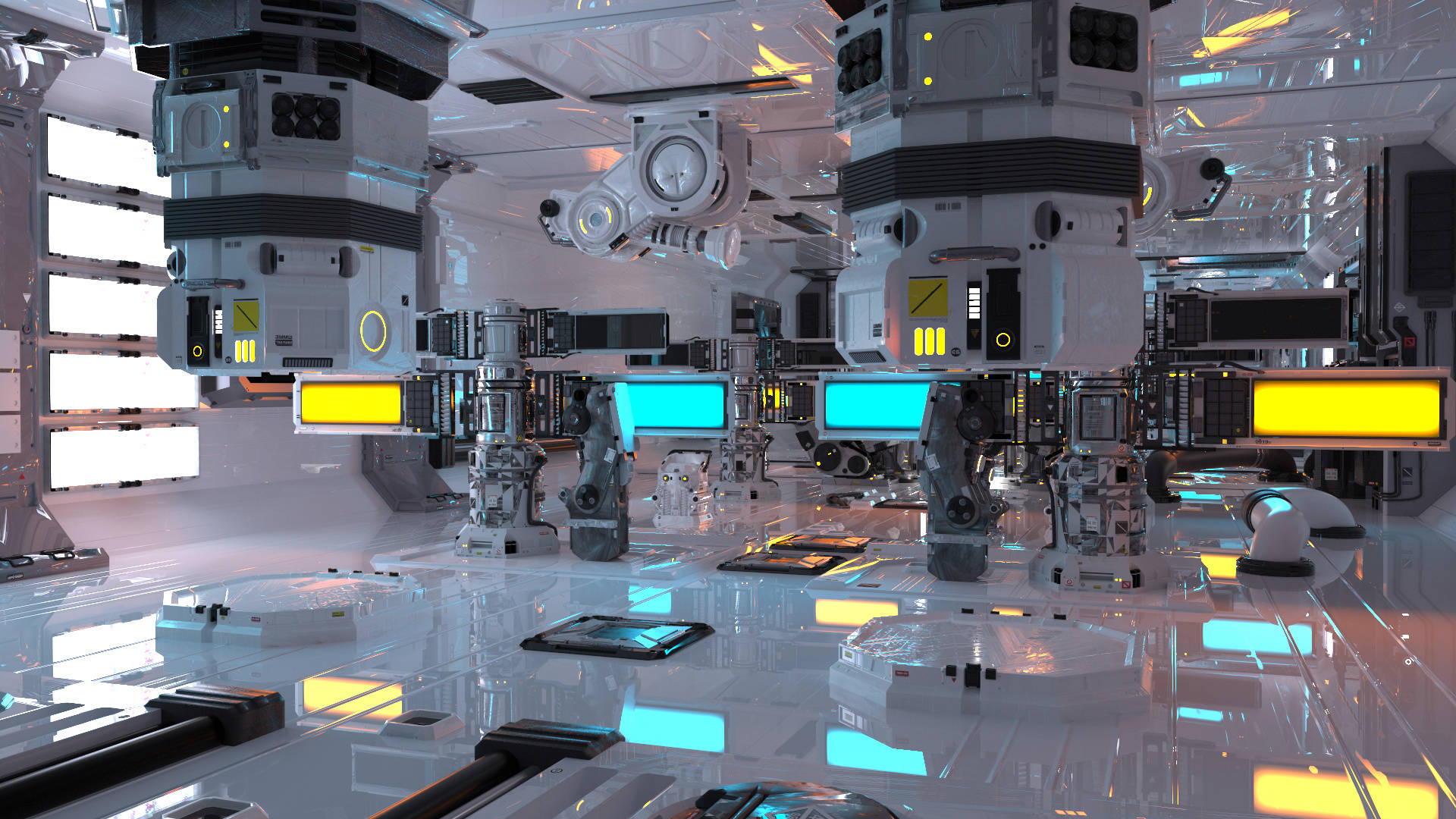}%
    \\%
    &
    rBias\textsuperscript{2}: 0.096&%
    rBias\textsuperscript{2}: 0.006&%
    \\%
    \hspace{-2.5mm}\rotatebox{90}{\hspace{5mm}\LivingRoom}%
    &\includegraphics[width=0.33\linewidth, trim=300 0 350 0, clip]{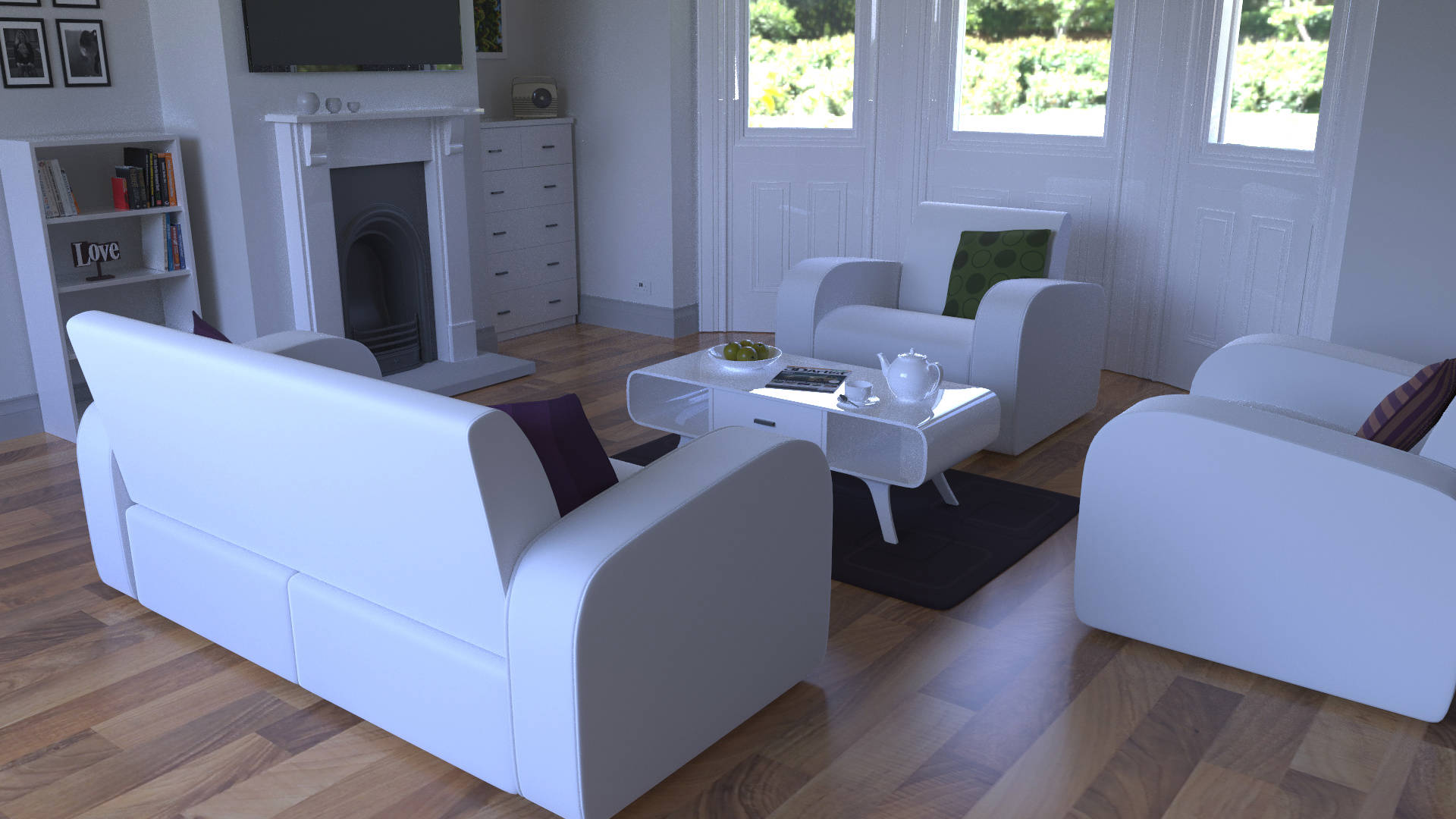}%
    &\includegraphics[width=0.33\linewidth, trim=300 0 350 0, clip]{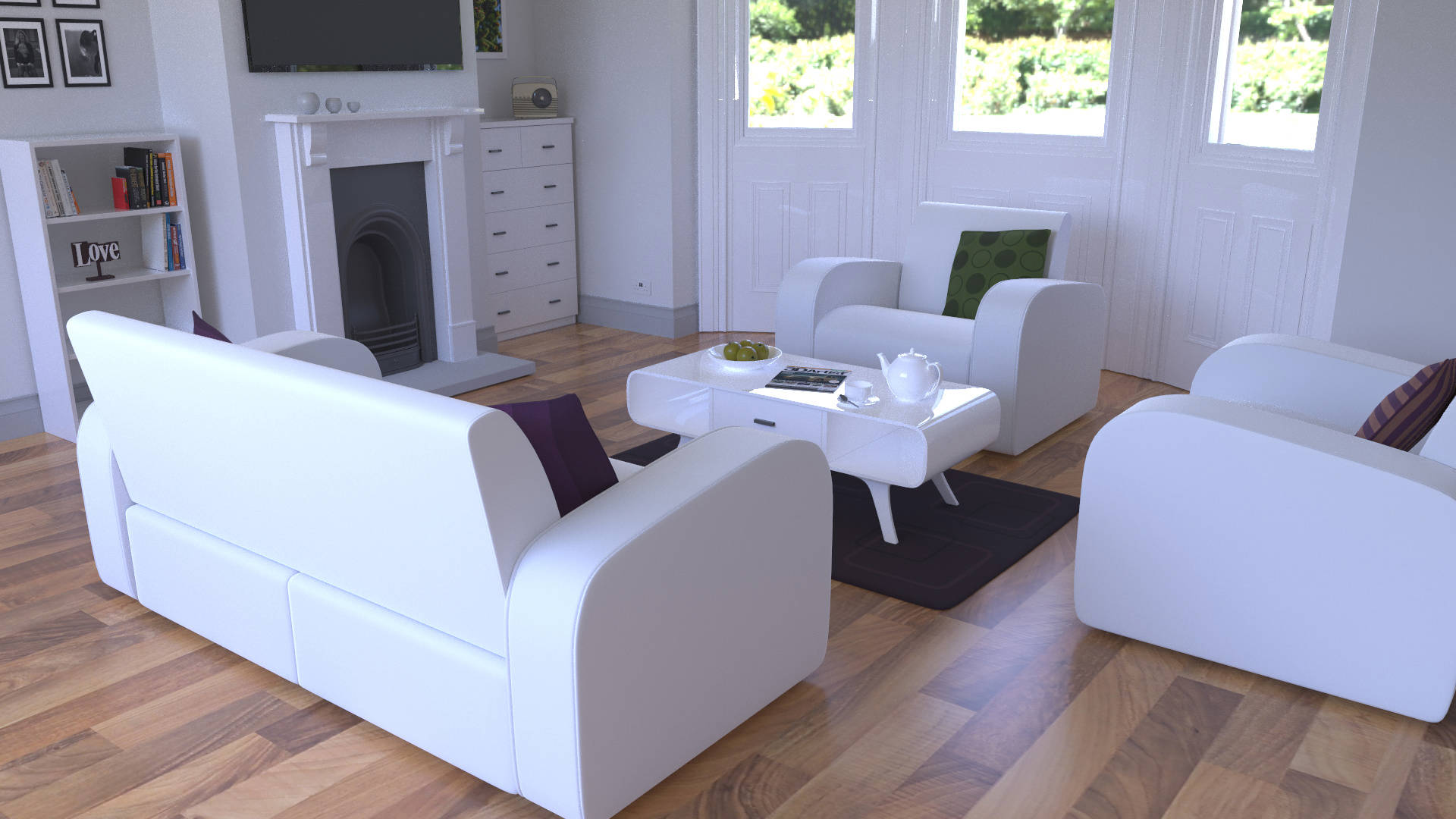}%
    &\includegraphics[width=0.33\linewidth, trim=300 0 350 0, clip]{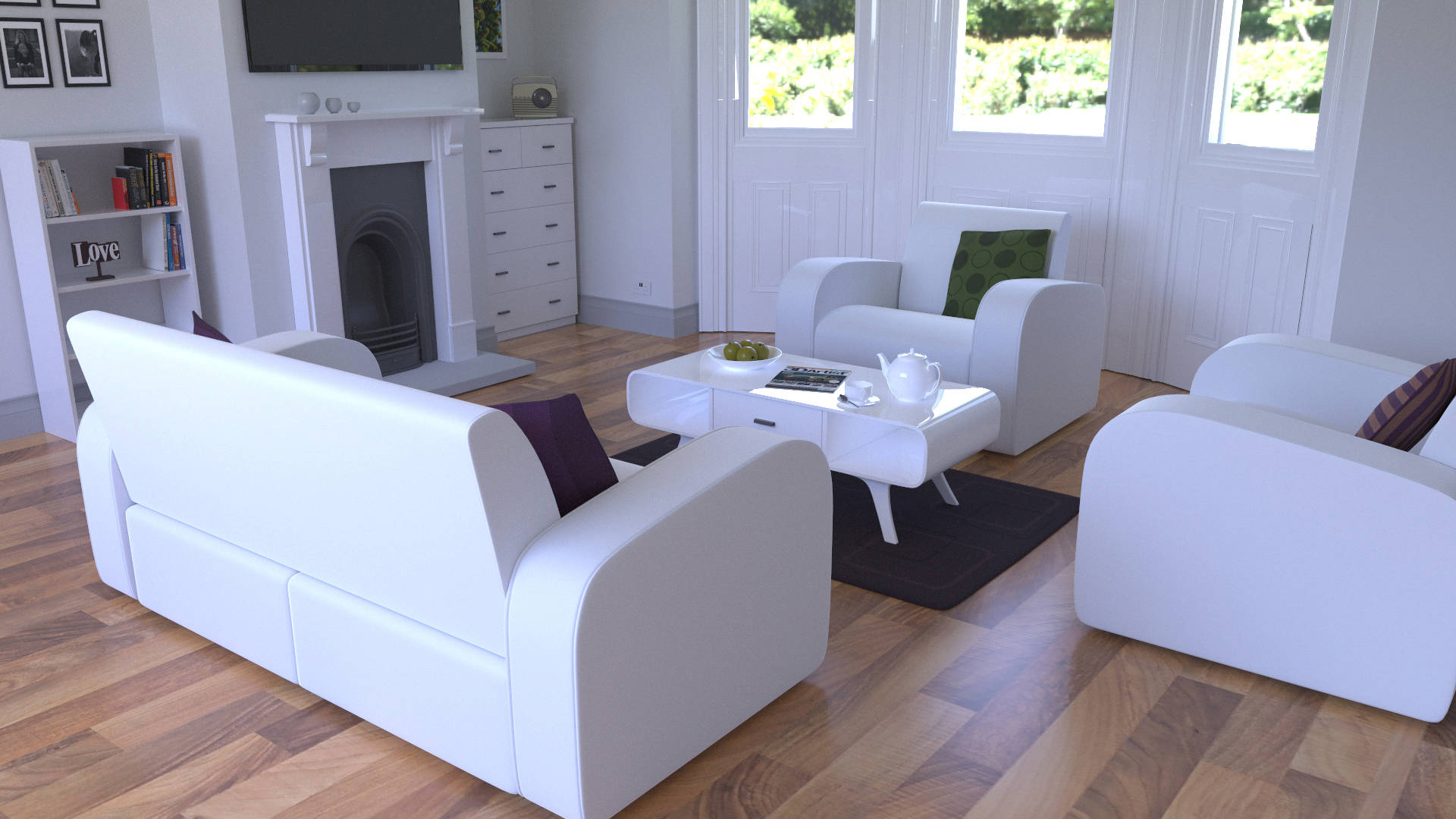}%
    \\%
    &
    rBias\textsuperscript{2}: 0.118&%
    rBias\textsuperscript{2}: 0.002&%
    \\%
    \hspace{-2.5mm}\rotatebox{90}{\hspace{6mm}\PinkRoom}%
    &\includegraphics[width=0.33\linewidth, trim=300 0 350 0, clip]{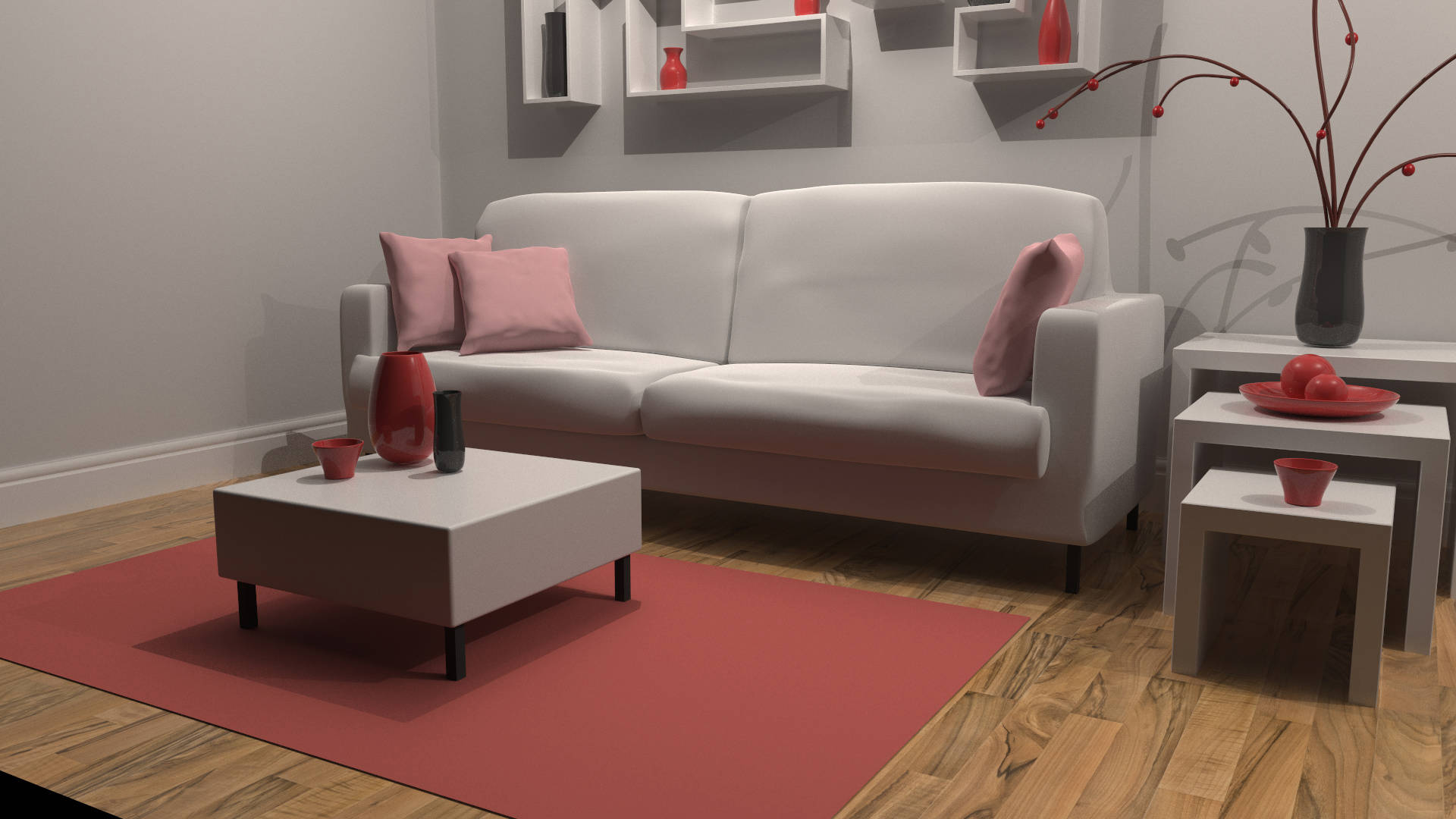}%
    &\includegraphics[width=0.33\linewidth, trim=300 0 350 0, clip]{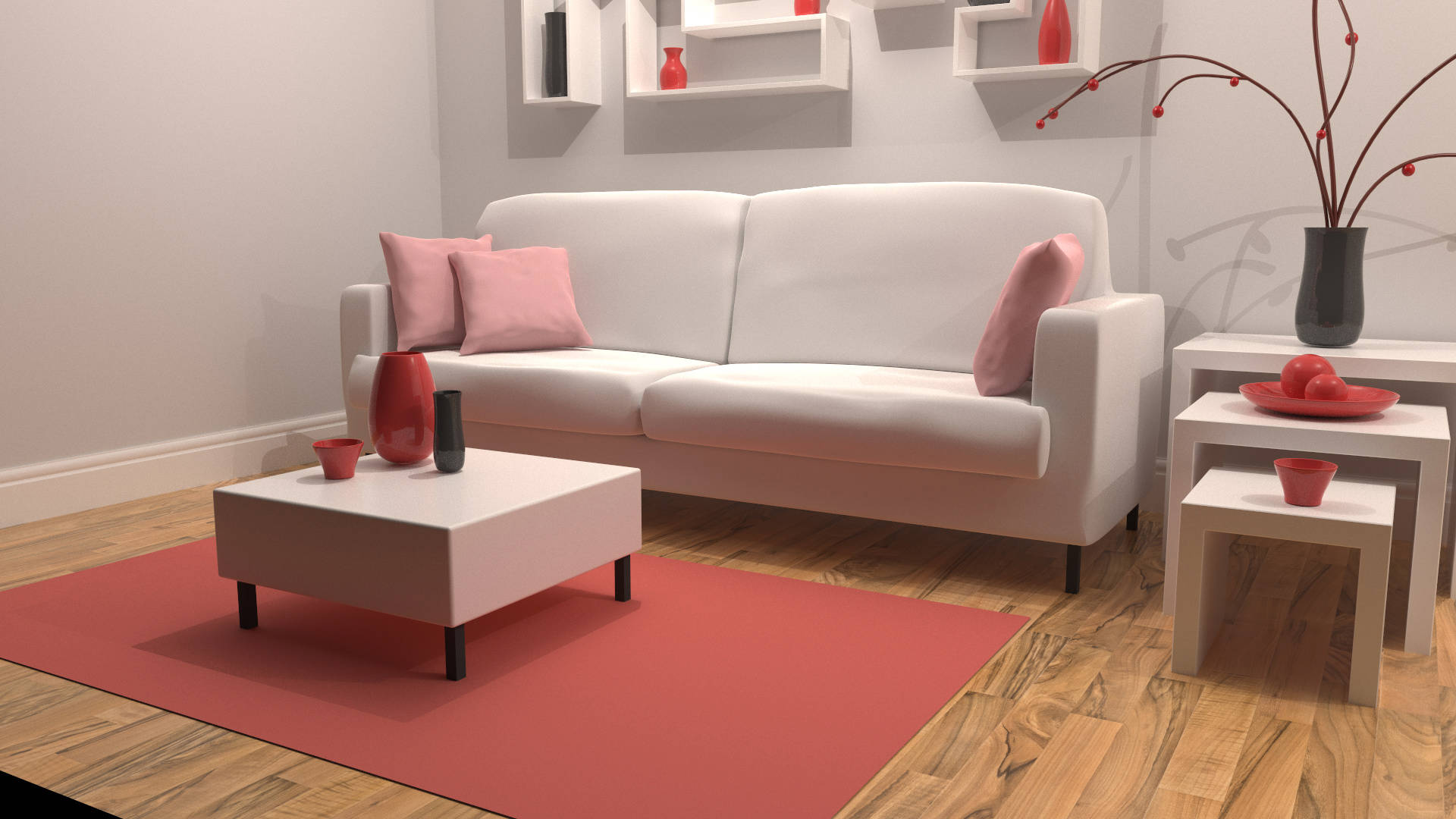}%
    &\includegraphics[width=0.33\linewidth, trim=300 0 350 0, clip]{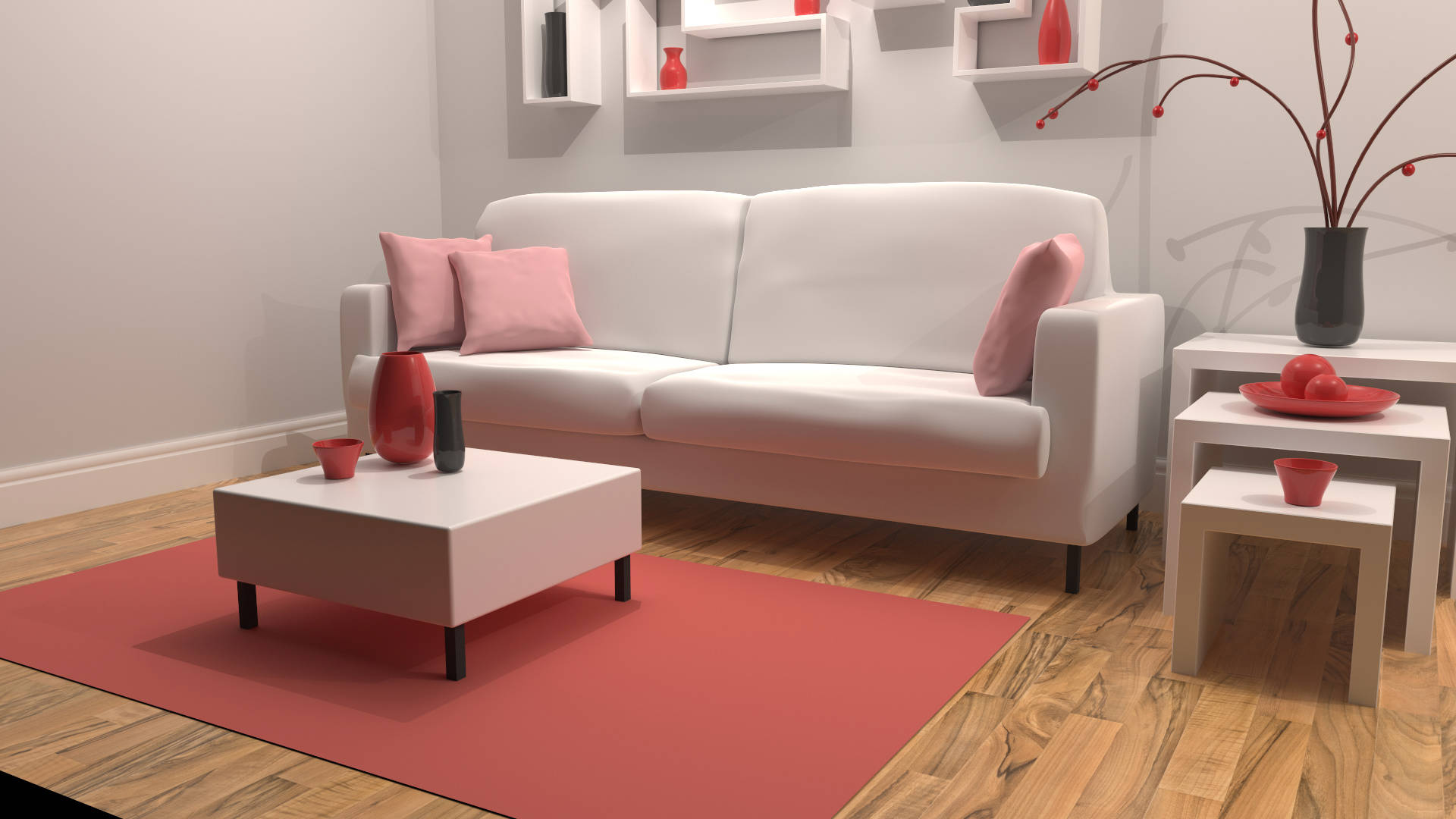}%
    \\%
    &
    rBias\textsuperscript{2}: 0.137&%
    rBias\textsuperscript{2}: 0.002&%
    \\%
\end{tabular}

    \vspace{-2mm}
    \caption{\label{fig:qlearning}
        Self-training enables efficient learning of indirect radiance that is not captured by short paths: compare the left and middle images, both of which trace paths of the same length (including the sparse set of unbiased training paths described in \autoref{sec:practical}).
        On the right, we provide a reference image obtained by tracing paths that are truncated with Russian roulette.
    }
\end{figure}

\paragraph{Self-training}
We compare the performance of self-training to training relying on pure path-tracing.
Terminating training paths with our termination heuristic, we set the tail contribution either to black (path-traced training) or to the radiance prediction at the last vertex (self-training).
As shown in \autoref{fig:qlearning}, the self-trained solution accurately captures multi-bounce light transport.
The computational overhead of self-training amounts to the cost of querying the radiance cache one additional time for each of the few training paths, which amounts roughly to a $1$\% overall overhead---a small amount compared with the cost that is saved by not having to trace longer paths to learn global illumination.

\begin{figure*}
    \vspace{-2mm}
    \setlength{\fboxrule}{5pt}%
\setlength{\insetvsep}{10pt}%
\setlength{\tabcolsep}{-0.5pt}%
\renewcommand{\arraystretch}{1}%
\small%
\begin{tabular}{lcccc}
    &$\cachedReflectedRadiance$ at heuristic hit
    &$\cachedReflectedRadiance$ at 1\textsuperscript{st} non-specular hit
    &$\cachedReflectedRadiance$ at heuristic hit
    &Reference
    \\%
    \setInset{A}{red}{1100}{359}{360}{120}%
    \setInset{B}{orange}{360}{795}{480}{160}%
    \hspace{-4mm}\rotatebox{90}{\hspace{-17mm}\ZeroDay}%
    &\addBeautyCrop{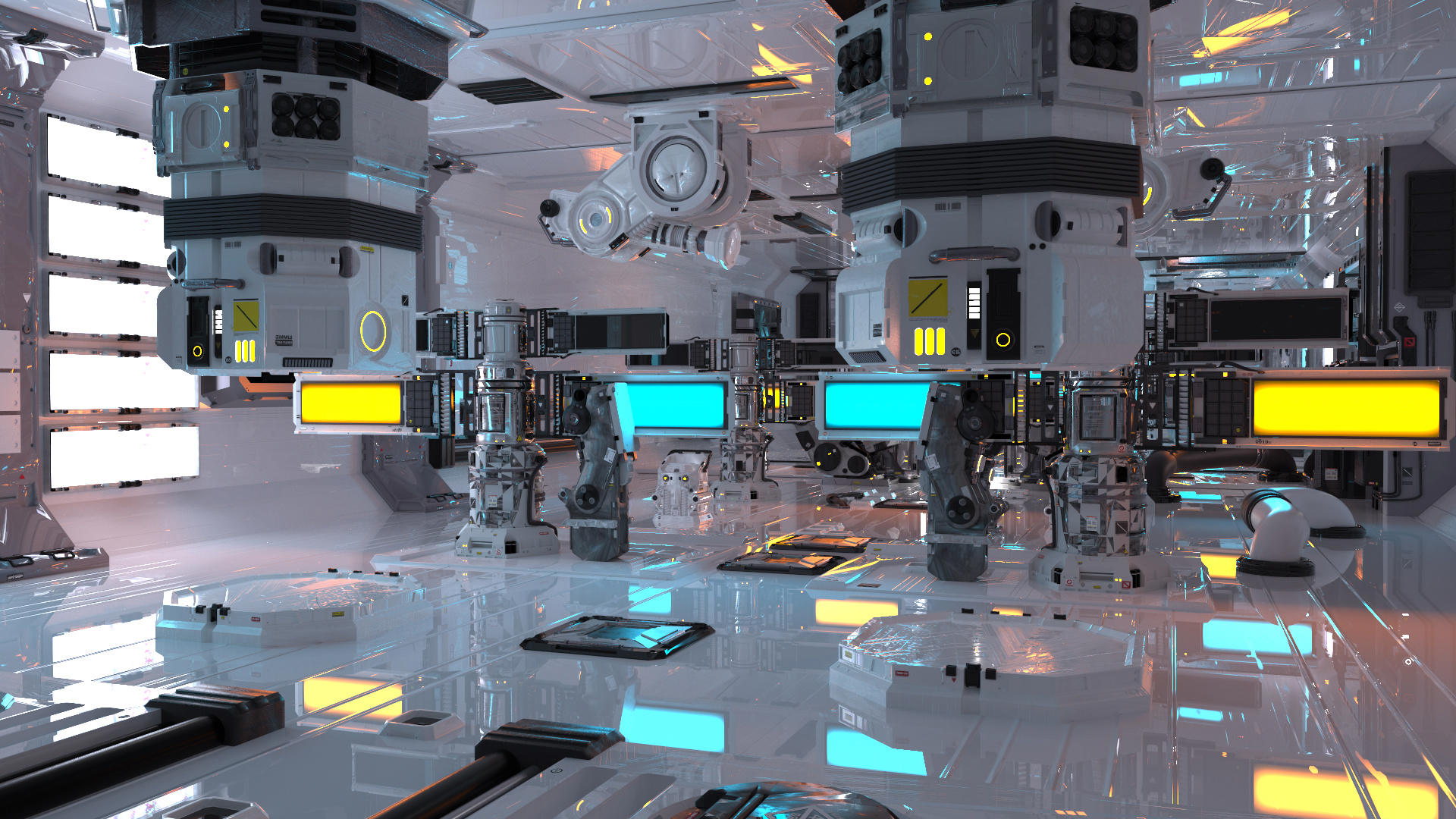}{0.293}{1920}{1080}{0}{0}{1920}{1080}%
    &\addInsets{figures/fig_inference_bounce/measure_seven/Ls_1st_vertex.jpg}%
    &\addInsets{figures/fig_inference_bounce/measure_seven/Ls_heuristic.jpg}%
    &\addInsets{figures/fig_inference_bounce/measure_seven/Reference.jpg}%
    \\%
    &rBias\textsuperscript{2}:
    &0.3433%
    &0.0014%
    &%
    \\%
    &\FLIP:
    &0.1948%
    &0.0439%
    &%
    \\%
    \setInset{A}{red}{319}{531}{360}{120}%
    \setInset{B}{orange}{840}{720}{720}{240}%
    \hspace{-4mm}\rotatebox{90}{\hspace{-15.5mm}\BistroExterior}%
    &\addBeautyCrop{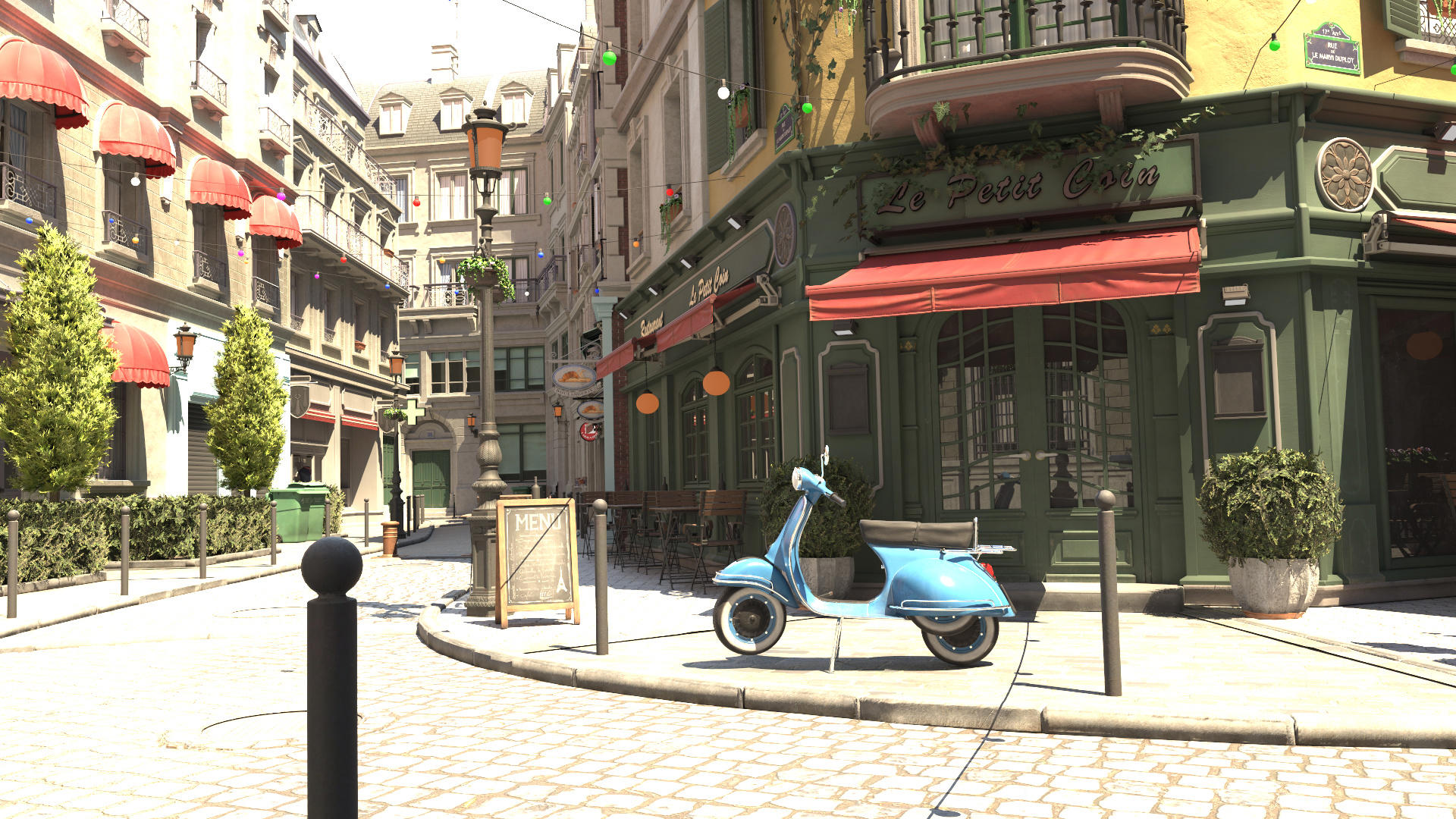}{0.293}{1920}{1080}{0}{0}{1920}{1080}%
    &\addInsets{figures/fig_inference_bounce/exterior/Ls_1st_vertex.jpg}%
    &\addInsets{figures/fig_inference_bounce/exterior/Ls_heuristic.jpg}%
    &\addInsets{figures/fig_inference_bounce/exterior/Reference.jpg}%
    \\%
    &rBias\textsuperscript{2}:
    &0.8447%
    &0.0016%
    &%
    \\%
    &\FLIP:
    &0.2290%
    &0.0401%
    &%
    \\%
    \setInset{A}{red}{710}{580}{240}{80}%
    \setInset{B}{orange}{1135}{640}{480}{160}%
    \hspace{-4mm}\rotatebox{90}{\hspace{-15.5mm}\Attic}%
    &\addBeautyCrop{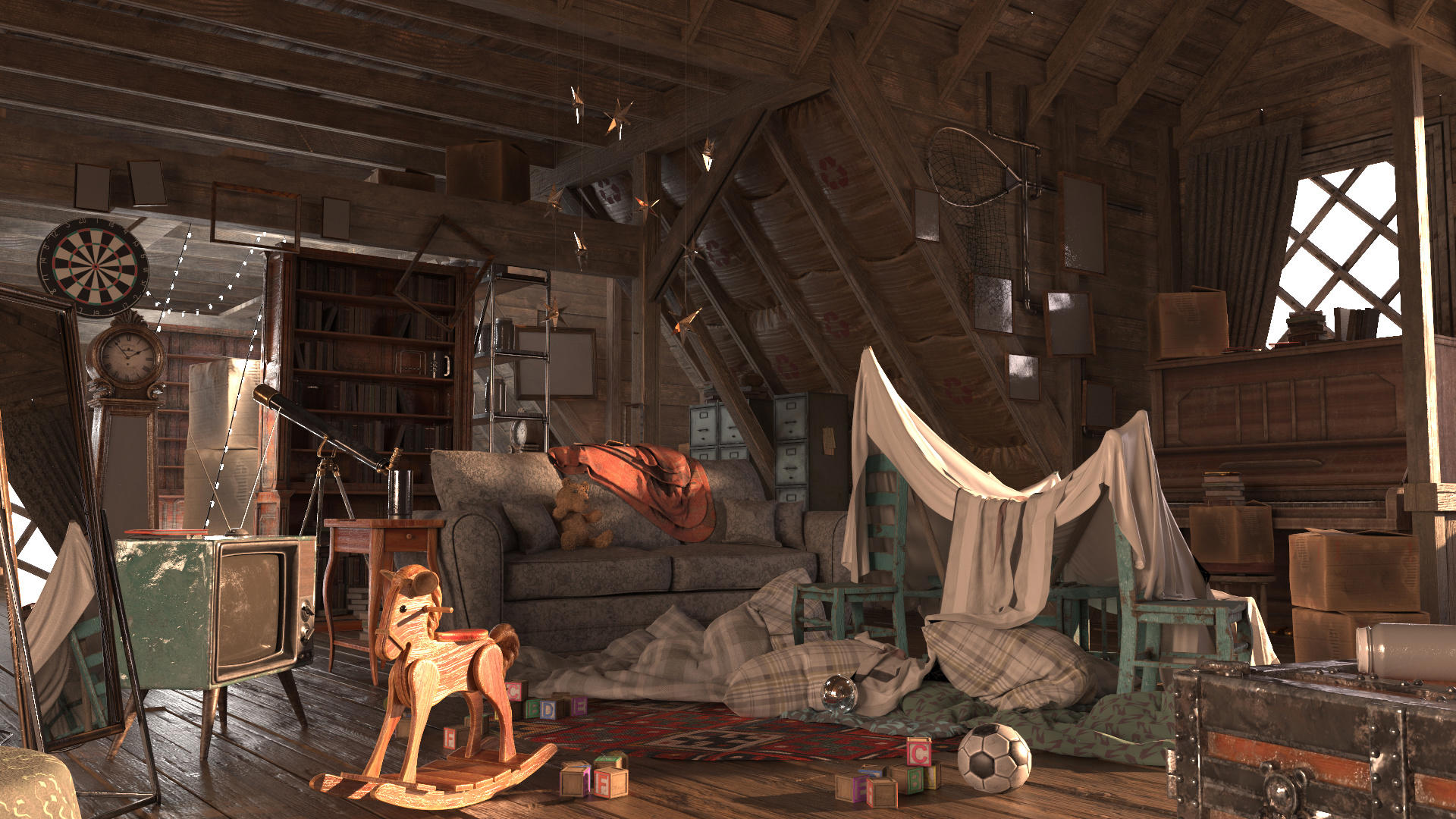}{0.293}{1920}{1080}{0}{0}{1920}{1080}%
    &\addInsets{figures/fig_inference_bounce/attic/Ls_1st_vertex.jpg}%
    &\addInsets{figures/fig_inference_bounce/attic/Ls_heuristic.jpg}%
    &\addInsets{figures/fig_inference_bounce/attic/Reference.jpg}%
    \\%
    &rBias\textsuperscript{2}:
    &0.1104%
    &0.0017%
    &%
    \\%
    &\FLIP:
    &0.1885%
    &0.0479%
    &%
    \\%
\end{tabular}
    \vspace{-3mm}
    \caption{\label{fig:inference-bounce}
        Converged renderings when querying the cache at the first non-specular vertex, or according to the path termination heuristic.
        We measure accuracy using the objective relative square bias metric, and the perceptual \protect\FLIP{} metric~\citep{Andersson:2020:flip}.
        The \ZeroDay{} scene features complex area lighting, glossy materials, and high-order indirect illumination due to the high albedo of surfaces.
        The \BistroExterior{} scene features small geometry and shadow details in a relatively large environment, highlighting the local adaptation afforded by the encoding of the network inputs.
        The \Attic{} scene is an interesting test case for light leakage as it features a fairly high geometric complexity, including thin elements such as the tarp.
        In all three scenes, employing the termination heuristic leads to a high quality result that closely resembles the reference image.
        In particular, the heuristic recovers contact shadows and local ambient occlusion, whereas other real-time caching techniques typically require an additional screen-space ambient occlusion (SSAO) pass to recover these details.
    }
\end{figure*}

\paragraph{Quality of the cache}
In \autoref{fig:inference-bounce}, we study the quality of the neural radiance cache by visualizing it at the first non-specular vertex.
By being agnostic to the materials and geometry of the underlying scenes, the cache handles a wide range of visual phenomena.
For example, it handles complex glossy transport in the \ZeroDay{} scene, shadow detail at a distance in the \BistroExterior{} scene, and thin geometry without light leakage, such as the tarp in the \Attic{} scene.
The cache also performs well at capturing the overall color of almost every region in each scene.

The limitations of the cache are twofold.
First, the cache does not capture sharp detail very well if that detail is absent from the inputs to the network (e.g.\ a sharp contact shadow or caustic).
And second, the cache exhibits subtle axis-aligned stripes that are a byproduct of the frequency encoding~\citep{mildenhall2020nerf}.
Since we rely on the frequency encoding to handle a spatial detail at scale (e.g.\ the accurate far-away shadows in the \BistroExterior{} scene), we cannot simply switch to a different encoding.
All other encodings that we tried either exhibited worse artifacts or did not scale well.

Thus, to mask away the remaining cache inaccuracies, we defer its evaluation according to the termination heuristic from \autoref{sec:termination}.
Since the heuristic is based on the path footprint, the cache artifacts are observed through rough reflections and therefore averaged out.
The heuristic also leads to fewer cache queries in concavities, where contact-shadow inaccuracies could become an issue.
The insets in \autoref{fig:inference-bounce} confirm this observation: combined with the termination heuristic, the cache produces images that are difficult to distinguish from the ground truth, both visually and numerically.

\begin{figure*}
    \vspace{-2mm}
    \setlength{\tabcolsep}{1pt}%
\renewcommand{\arraystretch}{0.75}%
\small%
\hspace*{-1.5mm}\begin{tabular}{lccc@{\hskip 3mm}ccc}%
    &\multicolumn{3}{c}{Frostbite diffuse} & \multicolumn{3}{c}{Lambertian diffuse (exposure $\times$0.5)}\\
    \cmidrule(lr){2-4}
    \cmidrule(lr){5-7}
    & DDGI & NRC & Reference & DDGI & NRC & Reference \\
    \hspace{-2.5mm}\rotatebox{90}{\hspace{6mm}\ZeroDay}%
    &\includegraphics[width=0.161\linewidth]{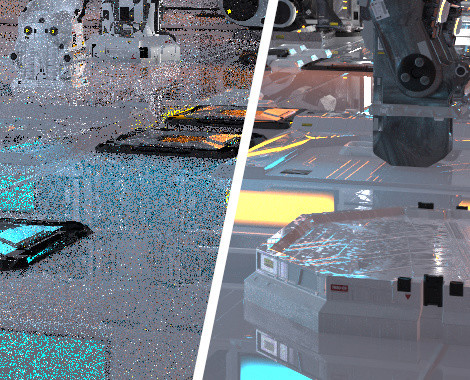}%
    &\includegraphics[width=0.161\linewidth]{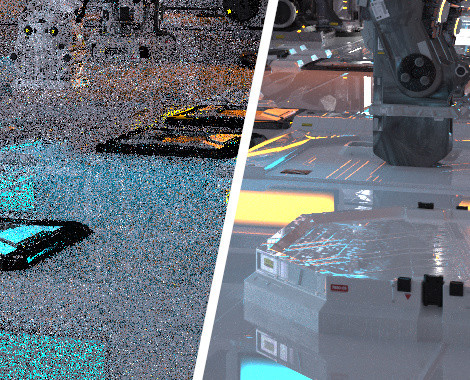}%
    &\includegraphics[width=0.161\linewidth]{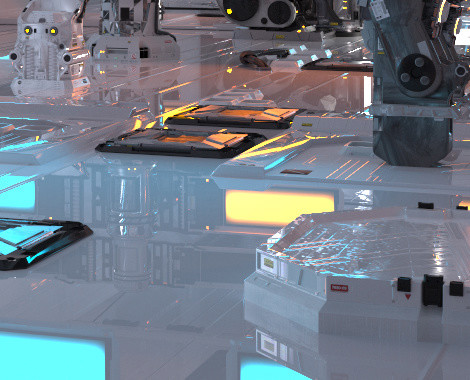}%
    &\includegraphics[width=0.161\linewidth]{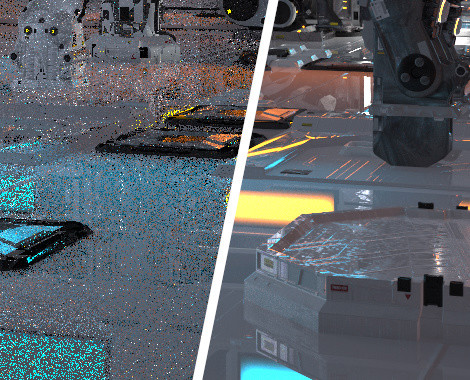}%
    &\includegraphics[width=0.161\linewidth]{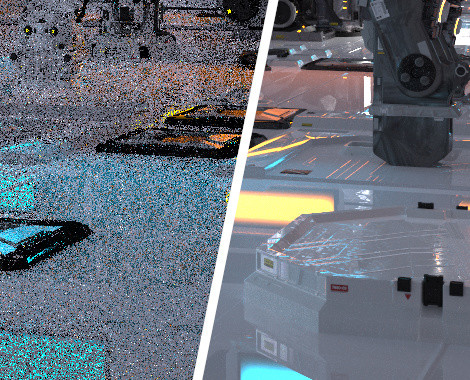}%
    &\includegraphics[width=0.161\linewidth]{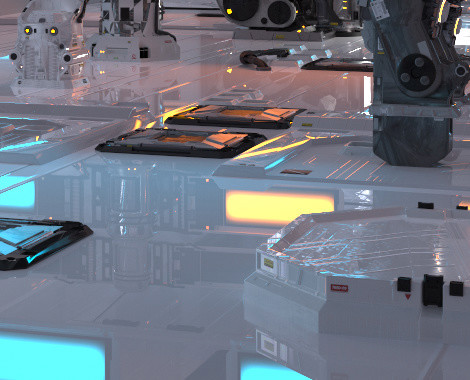}%
    \\%
    &
    rVar: 1.917, rBias\textsuperscript{2}: 0.057&%
    rVar: 3.757, rBias\textsuperscript{2}: 0.006&%
    &%
    rVar: 1.187, rBias\textsuperscript{2}: 0.118&%
    rVar: 3.349, rBias\textsuperscript{2}: 0.005&%
    \\%
    &
    96.7 fps&%
    87.7 fps&%
    &%
    96.8 fps&%
    86.3 fps&%
    \\%
    \hspace{-2.5mm}\rotatebox{90}{\hspace{5mm}\LivingRoom}%
    &\includegraphics[width=0.161\linewidth]{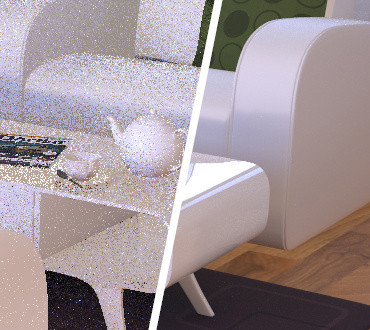}%
    &\includegraphics[width=0.161\linewidth]{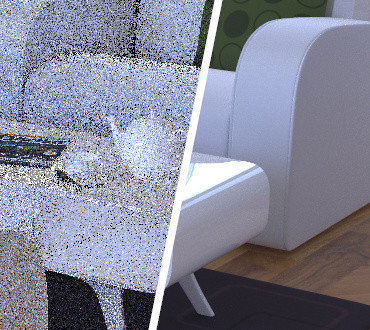}%
    &\includegraphics[width=0.161\linewidth]{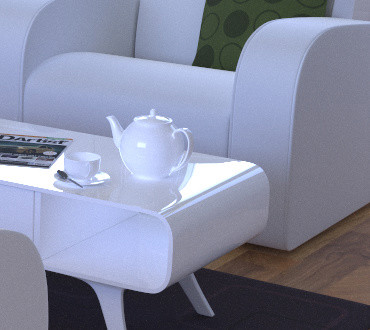}%
    &\includegraphics[width=0.161\linewidth]{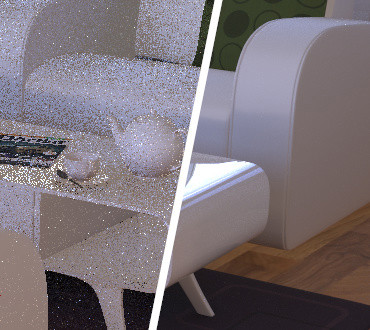}%
    &\includegraphics[width=0.161\linewidth]{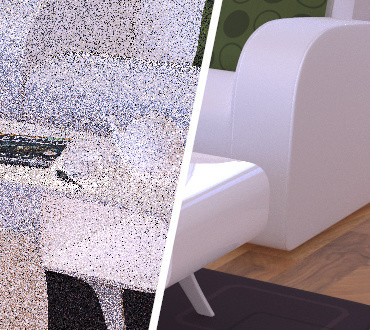}%
    &\includegraphics[width=0.161\linewidth]{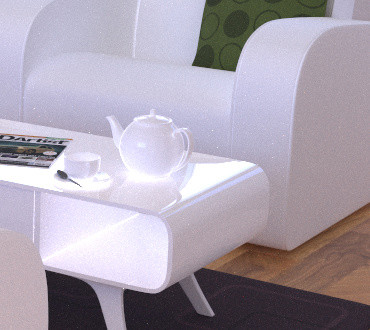}%
    \\%
    &
    rVar: 0.589, rBias\textsuperscript{2}: 0.174&%
    rVar: 0.950, rBias\textsuperscript{2}: 0.003&%
    &%
    rVar: 0.178, rBias\textsuperscript{2}: 0.264&%
    rVar: 0.522, rBias\textsuperscript{2}: 0.006&%
    \\%
    &
    134 fps&%
    111 fps&%
    &%
    134 fps&%
    111 fps&%
    \\%
    \hspace{-2.5mm}\rotatebox{90}{\hspace{7mm}\PinkRoom}%
    &\includegraphics[width=0.161\linewidth]{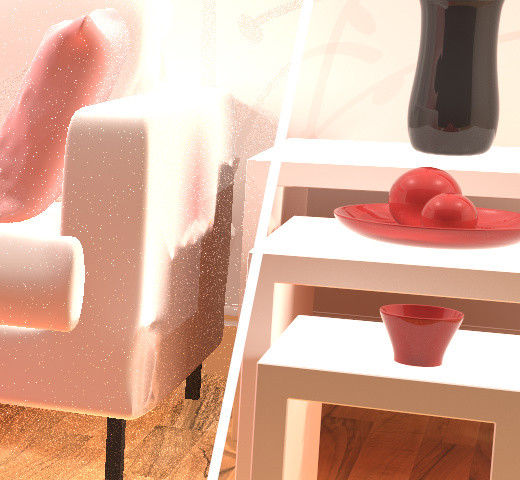}%
    &\includegraphics[width=0.161\linewidth]{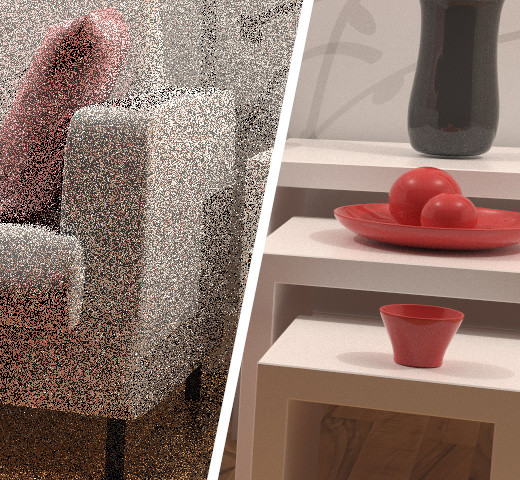}%
    &\includegraphics[width=0.161\linewidth]{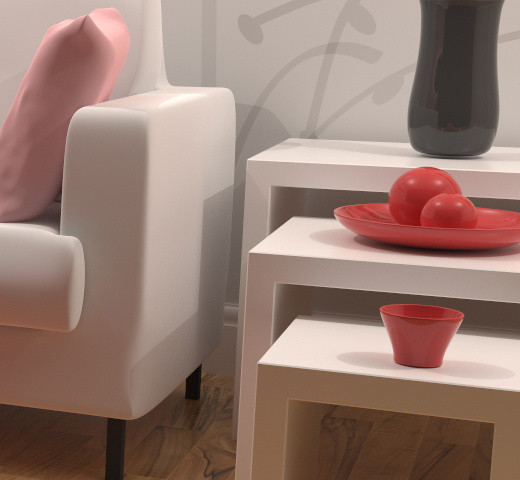}%
    &\includegraphics[width=0.161\linewidth]{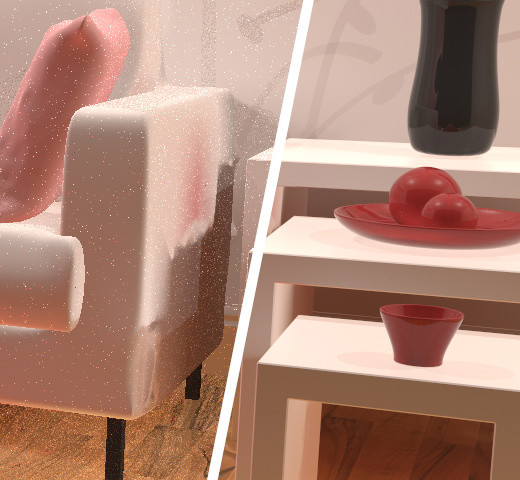}%
    &\includegraphics[width=0.161\linewidth]{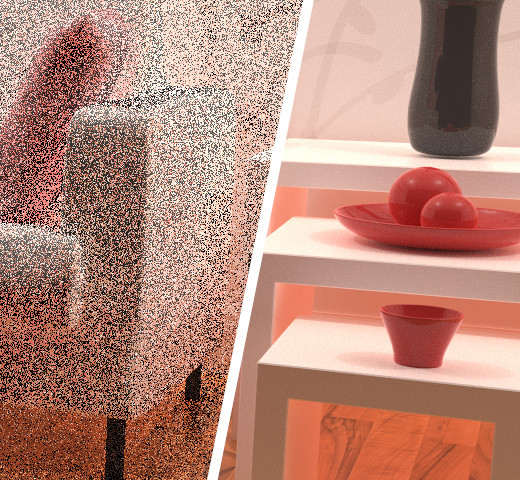}%
    &\includegraphics[width=0.161\linewidth]{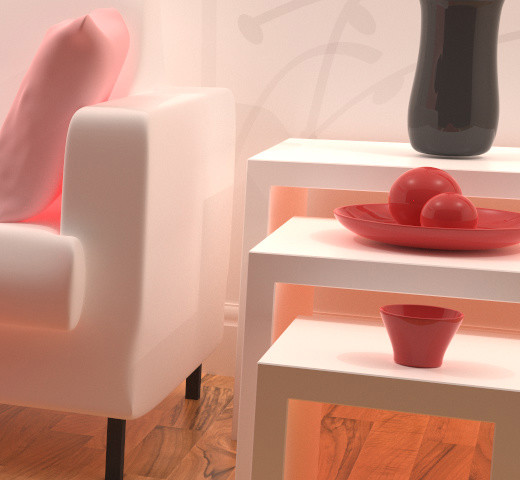}%
    \\%
    &
    rVar: 0.394, rBias\textsuperscript{2}: 3.836&%
    rVar: 1.087, rBias\textsuperscript{2}: 0.002&%
    &%
    rVar: 0.232, rBias\textsuperscript{2}: 0.542&%
    rVar: 1.297, rBias\textsuperscript{2}: 0.008&%
    \\%
    &
    141 fps&%
    124 fps&%
    &%
    144 fps&%
    124 fps&%
    \\%
\end{tabular}
    \vspace{-2mm}
    \caption{\label{fig:ddgi}
        Comparison of NRC with DDGI under challenging indirect illumination.\@ Both methods use ReSTIR for direct illumination.
        Since DDGI caches radiance as an extension of irradiance probes, its implementation assumes a Lambertian diffuse BRDF model and recovers detail through surface normal and albedo modulation.
        As our scenes were modeled using the Frostbite model~\citep{lagarde14frostbite}, we show results with both the Frostbite (left) and the Lambertian (right) model.\@
        DDGI performs at its best on the Lambertian material that it was designed for, whereas NRC---being agnostic to the BSDF---performs similarly in all situations.
        For each method, we show results at 1spp (left trapezoid) and 1024spp (converged, right trapezoid) to allow for the visual inspection of the bias-variance trade-off.
        We also report relative bias (rBias) and variance (rVar).
        DDGI achieves the least variance and runs fastest at the cost of sometimes incurring strong bias, e.g.\ around problematic geometry in the \PinkRoom{}.
        NRC has much lower bias, however incurring a moderately higher variance and cost.
        Note that contact shadows and ambient occlusion are implicitly built into NRC via the path termination heuristic, whereas DDGI lacks these visual features, necessitating an additional render pass in practice.
    }
\end{figure*}

\paragraph{Comparison with dynamic diffuse global illumination (DDGI)}

In \autoref{fig:ddgi} we compare the neural cache to DDGI~\cite{Majercik:2019:ddgi}.
DDGI is a modern extension of irradiance probes, relying on modulation by the surface normal and albedo to approximate the scattered radiance.
As a consequence, DDGI works best on Lambertian materials, which is why we show results using both a Lambertian diffuse BSDF model as well as the more physically based Frostbite model~\citep{lagarde14frostbite} that our scenes were modeled with.

DDGI makes an aggressive trade-off for performance and low noise: paths are terminated into the irradiance probes at their \emph{first} diffuse interaction (as opposed to glossy or specular), which is frequently the primary vertex.
Consequently, DDGI is on one hand very performant (paths are short) and has little noise, but on the other hand lacks ambient occlusion and can expose visible bias (e.g.\ in the \PinkRoom{}) due to its limited spatial resolution.

With NRC, we make the opposite trade-off.
We minimize bias at the cost of slightly reduced performance and (sometimes much) more noise.
In contrast to DDGI, our neural representation makes no assumptions about the underlying material model, and our path termination criterion helps to avoid the remaining inaccuracies of NRC while also recovering ambient occlusion.
The larger cost of our model could thus be offset by the cost of the separate ambient occlusion pass that DDGI requires.\@

\begin{table}
    \caption{\label{tab:timings}
        Breakdown of rendering cost by component.
    }%
    \vspace{-2mm}
    \setlength{\tabcolsep}{4pt}%
\renewcommand{\arraystretch}{0.75}%
\small%
\hspace*{-1mm}\begin{tabularx}{\columnwidth}{ccrrr}%
    \toprule
    Scene & Method &Trace \& shade&Query&Training\\
    \midrule
    \multirow{3}{*}{\Attic{}}
    & PT+ReSTIR &12.96 ms&---&---\\
    & PT+ReSTIR+DDGI &11.56 ms&0.64 ms&1.78 ms\\
    & PT+ReSTIR+NRC &10.88 ms&1.66 ms&1.12 ms\\
    \\[-1.5mm]
    \multirow{3}{*}{\BistroExterior{}}
    & PT+ReSTIR &13.75 ms&---&---\\
    & PT+ReSTIR+DDGI &12.71 ms&0.65 ms&1.68 ms\\
    & PT+ReSTIR+NRC &11.96 ms&1.38 ms&1.11 ms\\
    \\[-1.5mm]
    \multirow{3}{*}{\Classroom{}}
    & PT+ReSTIR &18.06 ms&---&---\\
    & PT+ReSTIR+DDGI &12.93 ms&0.59 ms&1.65 ms\\
    & PT+ReSTIR+NRC &12.28 ms&1.70 ms&1.11 ms\\
    \\[-1.5mm]
    \multirow{3}{*}{\LivingRoom{}}
    & PT+ReSTIR &8.32 ms&---&---\\
    & PT+ReSTIR+DDGI &5.68 ms&0.52 ms&0.99 ms\\
    & PT+ReSTIR+NRC &5.82 ms&1.85 ms&1.11 ms\\
    \\[-1.5mm]
    \multirow{3}{*}{\PinkRoom{}}
    & PT+ReSTIR &6.73 ms&---&---\\
    & PT+ReSTIR+DDGI &5.56 ms&0.52 ms&0.89 ms\\
    & PT+ReSTIR+NRC &5.36 ms&1.54 ms&1.12 ms\\
    \\[-1.5mm]
    \multirow{3}{*}{\ZeroDay{}}
    & PT+ReSTIR &13.89 ms&---&---\\
    & PT+ReSTIR+DDGI &8.34 ms&0.54 ms&1.21 ms\\
    & PT+ReSTIR+NRC &8.67 ms&1.41 ms&1.09 ms\\
    \midrule
    \multirow{3}{*}{Average}
        & PT+ReSTIR &12.29 ms&---&---\\
        & PT+ReSTIR+DDGI &9.46 ms&0.58 ms&1.37 ms\\
        & PT+ReSTIR+NRC &9.16 ms&1.59 ms&1.11 ms\\
    \bottomrule
\end{tabularx}
\end{table}

\paragraph{Performance breakdown}
In \autoref{tab:timings}, we analyze the time spent in DDGI and NRC in more detail.
We break down the rendering cost into (i) path tracing \& shading, (ii) querying the cache/DDGI, and (iii) training the cache/DDGI.

Compared to the PT+ReSTIR baseline, the low path-tracing cost of DDGI and NRC arises from tracing fewer rays: in addition to Russian roulette, DDGI terminates its paths when a diffuse lobe is sampled, and NRC terminates each path according to its footprint.
On average, both methods reduce the cost of path tracing by a similar amount of roughly $2.8$ ms per frame ($25$\%).
However, this improvement is partially offset by the overhead of querying and training or updating the respective caches.

As expected, querying DDGI is faster than querying our neural network.
However, given the reputation of neural networks to be expensive, the difference is smaller than what might be expected:
A full-frame DDGI query costs on average $0.58$ ms, whereas the neural radiance cache costs $1.59$ ms.
Both methods are thus well within reasonable cost for real-time settings.

Most interestingly, training the neural radiance cache is \emph{cheaper} than training the DDGI volume ($1.11$ ms vs.\ $1.37$ ms on average).
This has two reasons.
First, NRC is very data efficient, using only ${2^{16} = 65536}$ training records per frame.
On the other hand, DDGI in our $16\times16\times16$ probe-grid configuration traces ${16^3 \cdot 256 = 1048576}$ update rays per frame---more than $10$ times as many as NRC.\@
Second, the few training paths that are traced for NRC share their first couple of vertices with the paths that need to be traced for rendering anyway, further saving cost.

\section{Discussion and Future Work}%
\label{sec:discussion-and-future-work}

\paragraph{Precomputation}
While we do not perform any precomputation, it could be incorporated by e.g.\ pre-training a good initial state of the neural network or by utilizing a low-overhead meta-learning technique~\citep{Hospedales:2020:metalearning}.
In our design, we consider precomputation as strictly optional and merely to enhance the performance when possible.
In fact, as the neural radiance cache rapidly learns the current situation (8 frames are sufficient, see \autoref{fig:learning}),
precomputation can be only of limited benefit.
Still, it is of interest to explore the utility of static sets of neural network weights and to determine the bounds of the domain of validity of a fixed set of neural network weights.

\paragraph{Cache artifacts}
While we were able to suppress high-frequency temporal flickering using an exponential moving average over the optimized network weights, subtle low-frequency scintillation remains.
Additionally, the frequency encoding causes distracting axis-aligned oscillations throughout space.
These artifacts are imperceptible when using the neural radiance cache at non-primary path vertices, but in some use cases (e.g.\ when noise is not an option), using the cache at the primary vertex would be desirable.
To this end future work is needed to stabilize the prediction in a visually pleasing manner.

\paragraph{Additional network inputs.}
Input encodings alone are not sufficient to learn a detailed, high-frequency representation of features in the scattered radiance that correlate poorly with all of the network inputs.
Examples of such features are shadows and caustics, which are unrelated to the local surface attributes that we can easily pass to the network.
Shadows and caustics are therefore learned at a much slower rate---or not at all, if the network is too small or the radiance estimates are too noisy.
It is thus very interesting to think about additional, simple-to-compute network inputs that correlate well with such features.

\paragraph{Offline rendering}
While our neural cache design focuses on real-time rendering, we believe its use of self-training would be beneficial in offline scenarios.
Indeed, it allows to capture high-order indirect illumination without the costly tracing and shading of long paths;
this could be particularly useful to tackle path-length limitations of batch rendering~\cite{Burley:2018:design}.

\begin{figure}
    \vspace{-2mm}
    \setlength{\fboxrule}{1pt}%
\setlength{\tabcolsep}{1pt}%
\renewcommand{\arraystretch}{1}%
\small%
\centering%
\begin{tabular}{cccccccc}%
    Path tracing & + NRC (Ours) & Reference
    \\%
    \multicolumn{3}{c}{\setlength{\fboxsep}{1.5pt}\begin{overpic}[width=\linewidth]{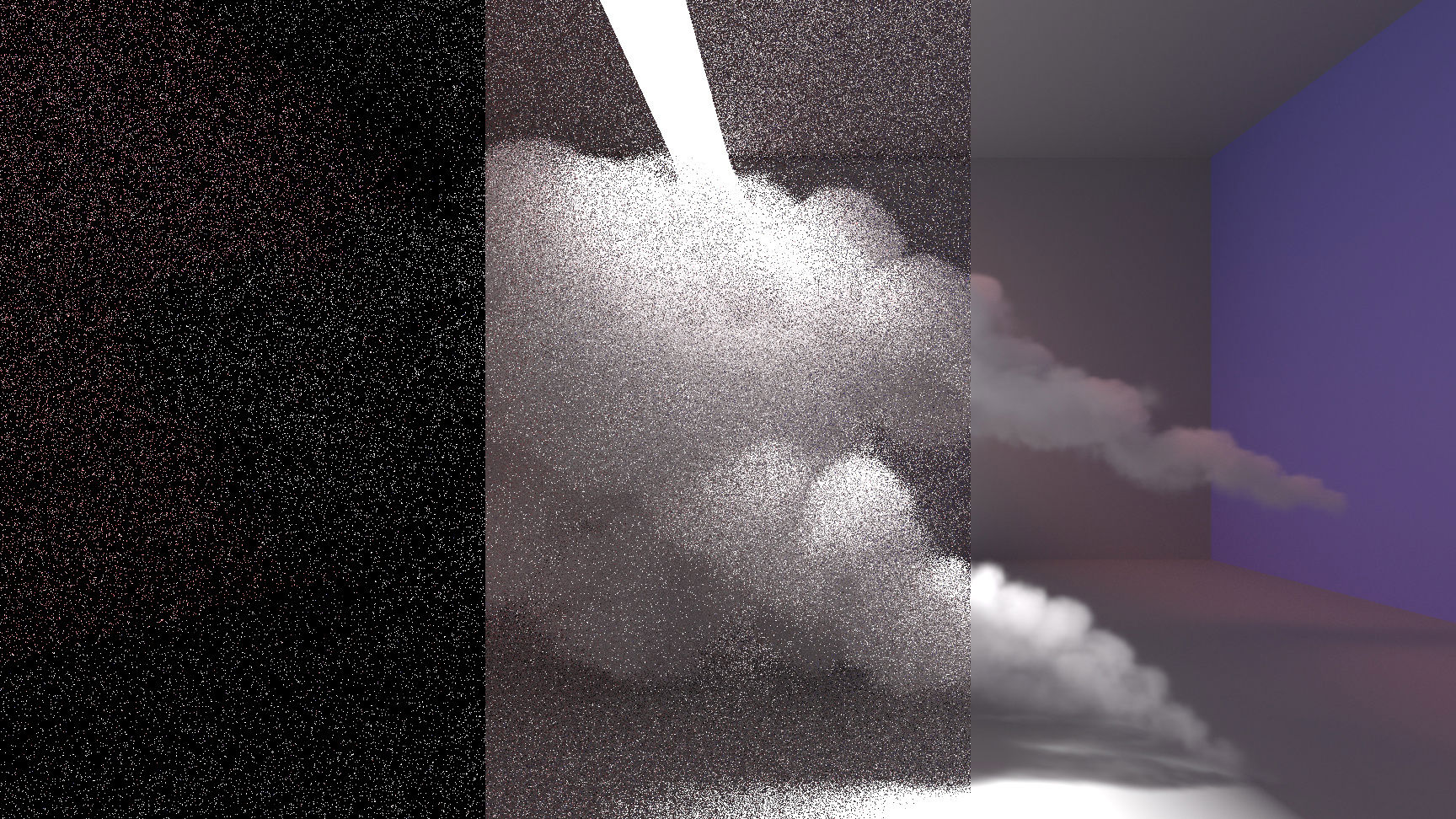}
        \put(35.879629629629626, 44.675925925925924){\makebox(0,0){\tikz\draw[red,ultra thick] (0,0) rectangle (0.11458333333333333\linewidth, 0.06875\linewidth);}}
        \put(72.33796296296296, 12.268518518518517){\makebox(0,0){\tikz\draw[orange,ultra thick] (0,0) rectangle (0.11458333333333333\linewidth, 0.06875\linewidth);}}
    \end{overpic}}
    \\%
    \fcolorbox{red}{red}{\includegraphics[width=0.32\linewidth]{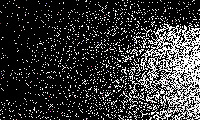}} &
    \fcolorbox{red}{red}{\includegraphics[width=0.32\linewidth]{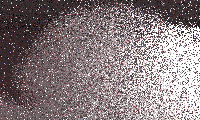}} &
    \fcolorbox{red}{red}{\includegraphics[width=0.32\linewidth]{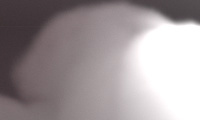}}
    \\%
    \fcolorbox{orange}{orange}{\includegraphics[width=0.32\linewidth]{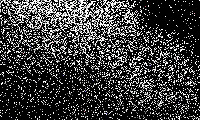}} &
    \fcolorbox{orange}{orange}{\includegraphics[width=0.32\linewidth]{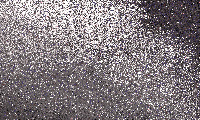}} &
    \fcolorbox{orange}{orange}{\includegraphics[width=0.32\linewidth]{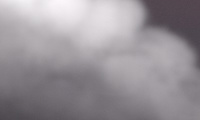}}
    \\%
    Path tracing & + NRC (Ours) & Reference
    \\%
            14.6 / 116 fps&
            2.62 / 125 fps&
        \\%
\end{tabular}

    \caption{
        Neural radiance caching also works in volumetric rendering.
        We render at 1 spp and report MRSE (left number) and frames per second.
        Illumination enters the room through a narrow slit in the ceiling and the volume has an isotropic phase function.
        There is one single neural radiance cache for the scene.
        For volume queries, undefined parameters (surface roughness, normal, albedo, and specular coefficients) are simply set to default constants.
    }\label{fig:volumes}
\end{figure}

\paragraph{Volumes}
We note that our neural cache parameterization is not tied to a surface representation and can thus also be used in volumetric rendering.
A straightforward implementation yields promising results (see \autoref{fig:volumes}), but an in-depth investigation needs to be carried out.

\begin{figure}[t]
    \vspace{-2mm}
    \setlength{\fboxrule}{1pt}%
\setlength{\tabcolsep}{1pt}%
\renewcommand{\arraystretch}{1}%
\small%
\centering%
\begin{tabular}{cccccccc}%
    Path tracing & + ReSTIR & + NRC (Ours) & Reference
    \\%
    \multicolumn{4}{c}{\setlength{\fboxsep}{1.5pt}\begin{overpic}[width=\linewidth]{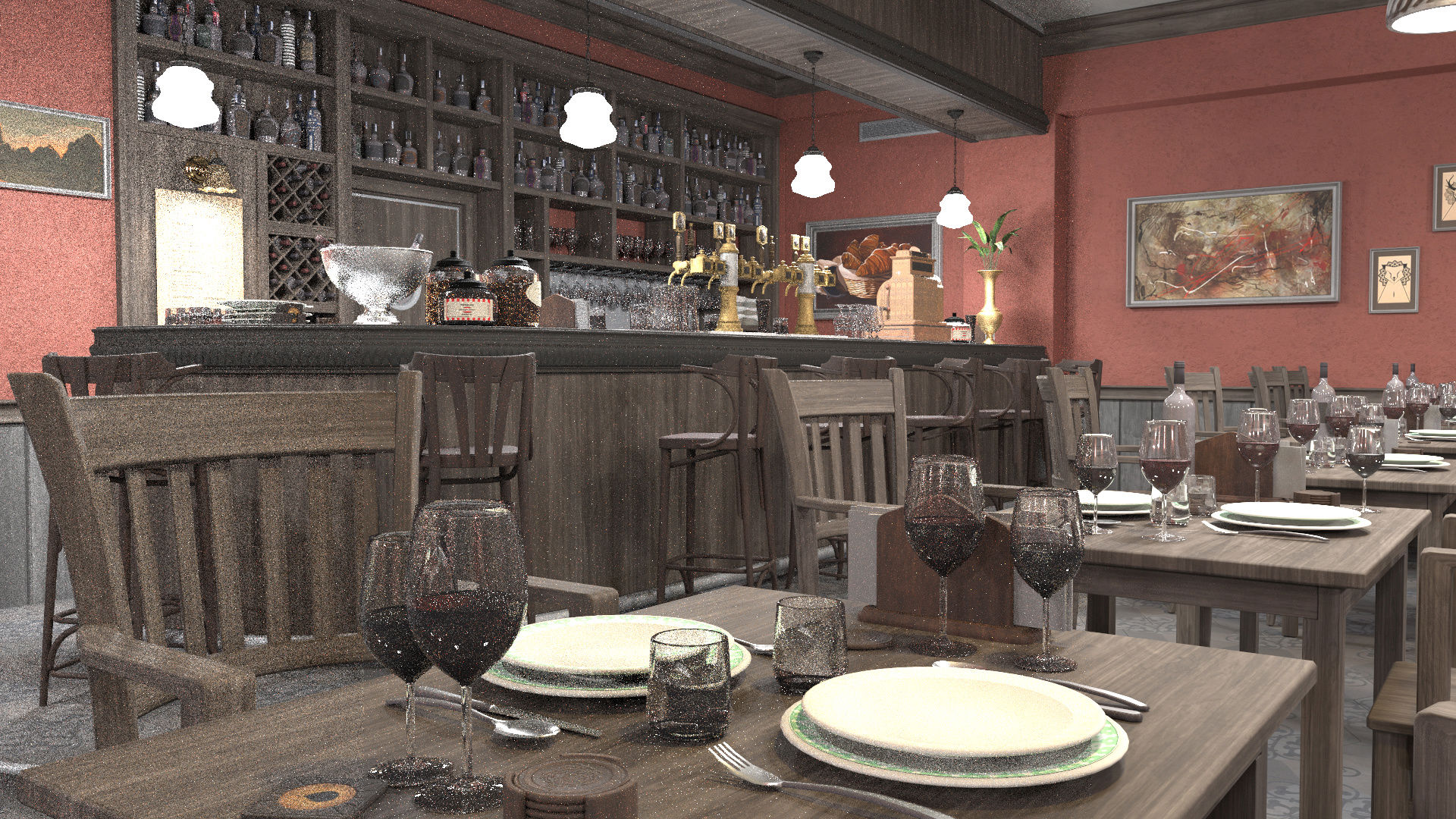}
        \put(28.645833333333332, 36.979166666666664){\makebox(0,0){\tikz\draw[red,ultra thick] (0,0) rectangle (0.1546875\linewidth, 0.061875\linewidth);}}
        \put(44.27083333333333, 35.9375){\makebox(0,0){\tikz\draw[orange,ultra thick] (0,0) rectangle (0.10312500000000001\linewidth, 0.041249999999999995\linewidth);}}
        \put(65.10416666666666, 20.833333333333332){\makebox(0,0){\tikz\draw[yellow,ultra thick] (0,0) rectangle (0.10312500000000001\linewidth, 0.041249999999999995\linewidth);}}
    \end{overpic}}
    \\%
    \fcolorbox{red}{red}{\includegraphics[width=0.236\linewidth]{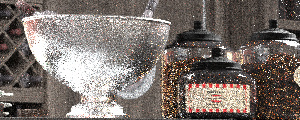}} &
    \fcolorbox{red}{red}{\includegraphics[width=0.236\linewidth]{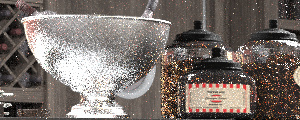}} &
    \fcolorbox{red}{red}{\includegraphics[width=0.236\linewidth]{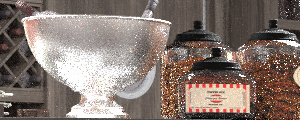}} &
    \fcolorbox{red}{red}{\includegraphics[width=0.236\linewidth]{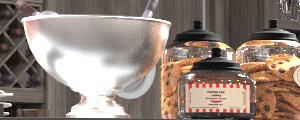}}
    \\%
    \fcolorbox{orange}{orange}{\includegraphics[width=0.236\linewidth]{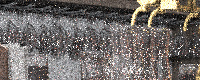}} &
    \fcolorbox{orange}{orange}{\includegraphics[width=0.236\linewidth]{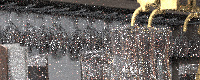}} &
    \fcolorbox{orange}{orange}{\includegraphics[width=0.236\linewidth]{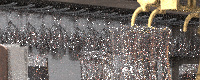}} &
    \fcolorbox{orange}{orange}{\includegraphics[width=0.236\linewidth]{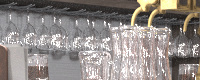}}
    \\%
    \fcolorbox{yellow}{yellow}{\includegraphics[width=0.236\linewidth]{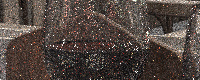}} &
    \fcolorbox{yellow}{yellow}{\includegraphics[width=0.236\linewidth]{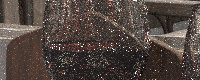}} &
    \fcolorbox{yellow}{yellow}{\includegraphics[width=0.236\linewidth]{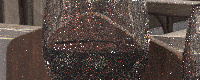}} &
    \fcolorbox{yellow}{yellow}{\includegraphics[width=0.236\linewidth]{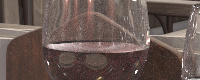}}
    \\%
    Path tracing & + ReSTIR & + NRC (Ours) & Reference
    \\%
            2.79 / 192 ms&
            2.45 / 320 ms&
            2.10 / 379 ms&
        \\%
\end{tabular}
    \vspace{-1mm}
    \caption{
        32 spp rendering of the Bistro scene, of which we report MRSE (left number) and render time.
        The improvement by neural radiance caching is marginal, because it does not address complex, branching specular transport.
        To alleviate this, we believe that a combination with path guiding as well as an improved path termination criterion should be investigated in the future.
    }\label{fig:bistro_interior}
\end{figure}

\paragraph{Path guiding}
The main source of noise in our results is the indirect use of the cache.
While the indirection results in decreased bias, it also leads to increased variance due to the (hemi-)spherical sampling, even though the cache approximation itself is noise-free.
One could consider bringing neural importance sampling~\cite{mueller2019nis} to real-time applications;
as with our neural radiance cache, the cost appears prohibitive at first, yet applying similar principles may prove successful.

\paragraph{Improved path termination}
More accurate approximations of the anisotropic area spreads, such as covariance tracing~\citep{Belcour:2013:COV} or bundle coherence~\citep{meng15granular}, could be used. While we did not encounter specific failure modes of the isotropic approximation of Bekaert et al., scenes with strong anisotropic lighting effects would likely benefit from the aforementioned methods.

An additional challenge that goes beyond the choice of area-spread approximation is the lack of path termination in long, branching, specular chains of interactions.
Consequently, our cache currently provides little benefit when transport is dominated by dielectric materials such as glass, as seen in \autoref{fig:bistro_interior}.
An improvement of the termination heuristic in such cases, as well as a more accurate cache to resolve sharp specular details, is of high interest.

\begin{figure}[t]
    \vspace{-2mm}
    \setlength{\fboxrule}{1pt}%
\setlength{\tabcolsep}{1pt}%
\renewcommand{\arraystretch}{1}%
\small%
\centering%
\begin{tabular}{cccccccc}%
    Denoised PT & + ReSTIR & + NRC (Ours) & Reference
    \\%
    \multicolumn{4}{c}{\setlength{\fboxsep}{1.5pt}\begin{overpic}[width=\linewidth]{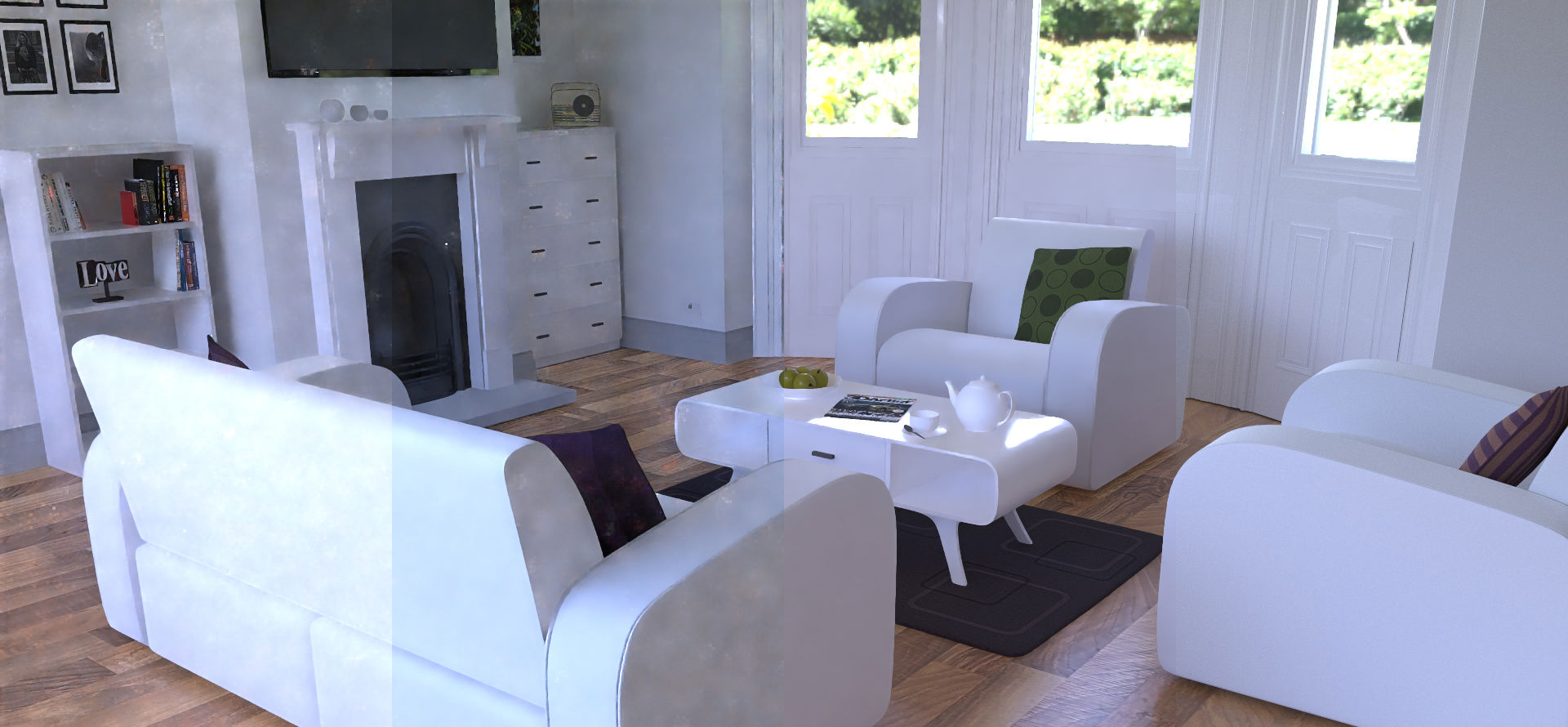}
        \put(8.854166666666668, 33.85416666666667){\makebox(0,0){\tikz\draw[red,ultra thick] (0,0) rectangle (0.10312500000000001\linewidth, 0.041249999999999995\linewidth);}}
        \put(33.85416666666667, 39.583333333333336){\makebox(0,0){\tikz\draw[orange,ultra thick] (0,0) rectangle (0.10312500000000001\linewidth, 0.041249999999999995\linewidth);}}
        \put(65.10416666666666, 6.7708333333333375){\makebox(0,0){\tikz\draw[yellow,ultra thick] (0,0) rectangle (0.10312500000000001\linewidth, 0.041249999999999995\linewidth);}}
    \end{overpic}}
    \\%
    \fcolorbox{red}{red}{\includegraphics[width=0.236\linewidth]{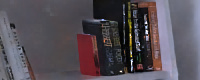}} &
    \fcolorbox{red}{red}{\includegraphics[width=0.236\linewidth]{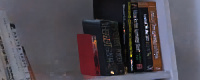}} &
    \fcolorbox{red}{red}{\includegraphics[width=0.236\linewidth]{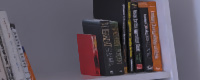}} &
    \fcolorbox{red}{red}{\includegraphics[width=0.236\linewidth]{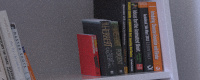}}
    \\%
    \fcolorbox{orange}{orange}{\includegraphics[width=0.236\linewidth]{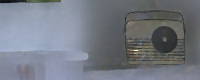}} &
    \fcolorbox{orange}{orange}{\includegraphics[width=0.236\linewidth]{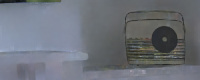}} &
    \fcolorbox{orange}{orange}{\includegraphics[width=0.236\linewidth]{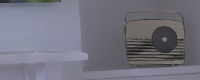}} &
    \fcolorbox{orange}{orange}{\includegraphics[width=0.236\linewidth]{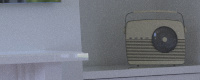}}
    \\%
    \fcolorbox{yellow}{yellow}{\includegraphics[width=0.236\linewidth]{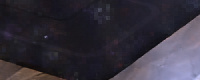}} &
    \fcolorbox{yellow}{yellow}{\includegraphics[width=0.236\linewidth]{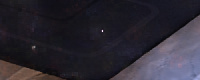}} &
    \fcolorbox{yellow}{yellow}{\includegraphics[width=0.236\linewidth]{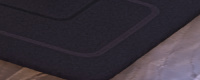}} &
    \fcolorbox{yellow}{yellow}{\includegraphics[width=0.236\linewidth]{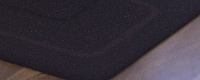}}
    \\%
    Denoised PT & + ReSTIR & + NRC (Ours) & Reference
    \\%
            0.0440 / 59 fps&
            0.0268 / 46 fps&
            0.0105 / 45 fps&
        \\%
\end{tabular}
    \vspace{-1mm}
    \caption{
        The \LivingRoom{} scene from the teaser image at 1 spp, passed through a deep-learning based real-time denoiser~\citep{hasselgren2020denoising}.
        Despite being trained on datasets with noise characteristics much different from our algorithm, the denoiser produces the cleanest results on it, achieving the lowest relative squared bias (left number).
        Even with denoising enabled, the framerate of our method stays well above 30.
    }\label{fig:restir-and-denoising}
\end{figure}

\paragraph{Denoising}
Our neural radiance cache could be considered a path-space denoiser, as it effectively performs a regression over spatio-temporal samples to produce noise-free approximations.
In \autoref{fig:restir-and-denoising}, we demonstrate that this path-space denoising \emph{complements} existing screen-space techniques.
This preliminary experiment employed an off-the-shelf denoiser that was trained on data not representative of our method, so we expect a tighter coupling of our cache and the denoiser to offer potential for further improvements.

\section{Conclusion}

We have introduced a real-time neural radiance caching technique for path-traced global illumination.
It can handle dynamic content while providing predictable performance and resource consumption,
which is enabled by our fully fused neural networks that achieve \emph{generalization via online adaptation}.
While the necessary performance requires a lot of engineering, robustness comes as a collateral.
The resulting high rendering quality makes up for the cost of the neural radiance cache, and could be further improved through orthogonal variance reduction techniques such as by path guiding.

The neural radiance cache is a rather different approach to real-time rendering than previous techniques.
It could be characterized as wasteful in terms of \emph{compute}---some neurons have little impact on the output, yet their contribution is still evaluated.
Competing techniques with sophisticated data structures could be characterized as wasteful in terms of \emph{memory}---the memory is never used in its entirety as queries access only small (random) neighborhoods.

The neural radiance cache employs fixed function hardware (the GPU tensor cores), and heavily relies on cheap computation, instead of costly memory accesses.
This efficiency is reflected in the timings of \autoref{tab:timings}, where our neural approach is only twice as expensive as irradiance probes, despite requiring a comparatively massive amount of compute (both implementations are reasonably well optimized).
We posit this compute efficiency is the key ingredient of the neural cache robustness.
This paradigm---cheap compute---appears to be worth investigating~\citep{DS-Accelerators}, and we hope it inspires further experiments in applications where compute is typically considered a precious commodity.

\begin{acks}

The article is dedicated to our dear friend and colleague Jaroslav K{\v{r}}iv{\'a}nek.
The authors thank Nikolaus Binder for his early contributions to the high performance neural networks.
Zander Majercik assisted us for the DDGI comparison, and Simon Kallweit contributed to the Falcor integration.

\end{acks}

\bibliography{bibliography}
\bibliographystyle{ACM-Reference-Format}

\end{document}